\documentclass[a4paper,11pt]{article}
\pdfoutput=1

\usepackage{jheppub}
\usepackage[T1]{fontenc}

\usepackage{bbm}
\usepackage{bm}
\usepackage{mathrsfs}
\usepackage{mathtools}
\usepackage{physics}

\allowdisplaybreaks
\renewcommand{\arraystretch}{1.2}

\let\de=\partial
\let\eps=\epsilon

\let\n=\nu
\let\om=\omega
\let\t=\theta
\let\x=\xi
\let\w=\wedge

\let\Res\relax
\DeclareMathOperator*{\Res}{Res}

\newcommand\im{\text{i}}% imaginary unit
\newcommand\gr[1]{\mathrm{#1}}% font for group names such as SU(2)
\newcommand\La{\mathscr{L}}% Lagrangian symbol
\newcommand{\R}{\mathbb{R}}% real numbers
\newcommand{\C}{\mathbb{C}}% complex numbers
\newcommand{\vek}[1]{\bm{#1}}% spatial vector in bold
\newcommand{\he}[1]{{#1}^\dagger}% Hermitian conjugate
\newcommand{\Sum}[2]{\smashoperator{\sum_{\substack{#1\\(#2)}}}}
\renewcommand{\Omega}{0}% vacuum state

\newcommand\exotic{quarton}

\title{Nonrelativistic effective field theories\\ with enhanced symmetries and soft behavior}

\preprint{TUM-HEP-1381/21}

\author[a]{Martin A. Mojahed}
\author[b]{and Tom\'{a}\v{s} Brauner}

\affiliation[a]{Physik-Department T70, Technische Universit\"{a}t M\"{u}nchen,\\
James-Franck-Straße 1, D-85748 Garching, Germany}
\affiliation[b]{Department of Mathematics and Physics, University of Stavanger,\\
N-4036 Stavanger, Norway}

\emailAdd{martin.mojahed@tum.de}
\emailAdd{tomas.brauner@uis.no}

\abstract{We systematically explore the landscape of nonrelativistic effective field theories with a local $S$-matrix and enhanced symmetries and soft behavior. The exploration is carried out using both conventional quantum field theory methods based on symmetry arguments, and recently developed on-shell recursion relations. We show that, in contrary to relativistic theories, enhancement of the soft limit of scattering amplitudes in nonrelativistic theories is generally not a byproduct of symmetry alone, but requires additional low-energy data. Sufficient conditions for enhanced scattering amplitudes can be derived by combining symmetries and dispersion relations of the scattered particles. This has direct consequences for the infrared dynamics that different types of nonrelativistic Nambu-Goldstone bosons can exhibit. We then use a bottom-up soft bootstrap approach to narrow down the landscape of nonrelativistic effective field theories that possess a consistent low-energy $S$-matrix. We recover two exceptional theories of a complex Schr\"odinger-type scalar, namely the $\C P^1$ nonlinear sigma model and the Schr\"odinger-Dirac-Born-Infeld theory. Moreover, we use soft recursion to prove a no-go theorem ruling out the existence of other exceptional Schr\"odinger-type theories. We also prove that all exceptional theories of a single real scalar with a linear dispersion relation are necessarily Lorentz-invariant. Soft recursion allows us to obtain some further general bounds on the landscape of nonrelativistic effective theories with enhanced soft limits. Finally, we present a novel theory of a complex scalar with a technically natural quartic dispersion relation. Altogether, our work represents the first step of a program to extend the developments in the study of scattering amplitudes to theories without Lorentz invariance.}

\begin{document} 
\maketitle
\flushbottom

%%%%%%%%%%%%%%%%%%%%%%%%%%%%%%%%%%%%%%%%%%%%%%%

\section{Introduction}
\label{sec:intro}

Effective field theory (EFT) is a general framework that encodes the dynamics of the degrees of freedom present in a physical system below a given energy scale. The fundamental principle of EFT is to include in the action all possible terms allowed by symmetry principles up to a certain order in one or more parameters such as momentum or mass. The relevance of different terms is determined through a power-counting scheme. Irrelevant terms are suppressed by powers of the ultraviolet cutoff and have dimensionless Wilson coefficients that encode effects from the physics beyond this cutoff. 

EFT is particularly powerful for physical systems with an ordered ground state, where the low-energy dynamics is dominated by Nambu-Goldstone (NG) modes of the symmetry spontaneously broken by the order parameter. If the system is invariant under spatial rotations and spacetime translations, then the NG modes stemming from spontaneously broken global symmetries can be classified into two different families. These families are referred to as type $A_m$ and type $B_{2m}$~\cite{Griffin2015a}, where $m$ is a positive integer. A type $A_m$ NG mode is described by a real scalar field with dispersion relation $\omega^2\propto\boldsymbol{p}^{2m}$. The well-known NG bosons that arise in relativistic systems with spontaneously broken global symmetries belong to the type $A_1$ subfamily. On the other hand, type $B_{2m}$ NG modes are described by two real scalar fields (or one complex scalar) forming a canonically conjugated pair with dispersion relation $\omega\propto \boldsymbol{p}^{2m}$. 
The possible existence of type $A_m$ and type $B_{2m}$ NG bosons in a given number of spatial dimensions $d$ and at zero temperature is constrained by the nonrelativistic version of the celebrated Coleman-Hohenberg-Mermin-Wagner (CHMW) theorem~\cite{Griffin2013a,Griffin2015b,Watanabe2014a}. In short, a NG boson of type $A_m$ may only exist if $m<d$. This both constrains the dispersion relation of type $A$ NG bosons for fixed $d$, and places a lower bound on the dimension $d$ below which type $A_m$ NG modes with fixed $m$ cannot exist.  On the other hand, type $B_{2m}$ NG modes are not constrained beyond the simple requirement that $m$ be positive.

The construction and properties of EFTs for NG bosons are by now well understood both in relativistic~\cite{Coleman1969a,Callan1969a,Leutwyler1994b} and nonrelativistic~\cite{Leutwyler1994a,Watanabe2014a,Andersen:2014} systems with spontaneously broken internal, i.e.~coordinate-independent, symmetry. Spontaneously broken coordinate-dependent symmetries are more subtle in that the number of NG modes they produce may be lower than the naive count of spontaneously broken generators. This happens as a rule when different global symmetries generate locally indistinguishable fluctuations of the order parameter~\cite{Low2002a,Watanabe2013a,Brauner2014a,Brauner2020a}. While such \emph{redundant symmetries} impose nonlinear constraints on the low-energy effective action similar to any other spontaneously broken symmetry, it is not obvious what they imply for actual physical observables. This was clarified by Cheung et al.~\cite{Cheung2015a}, who showed that redundant symmetry manifests itself in the scattering amplitudes of NG bosons through soft theorems for the $S$-matrix.

Soft theorems describe universal behavior of scattering amplitudes when the momentum of one or more particles is taken to zero. In the present paper, we will address the \emph{single soft limit} of scattering amplitudes of NG bosons. In this case, the momentum of one selected particle is rescaled with a soft factor, $\vek p\to\eps\vek p$. The momenta of all the other particles in the scattering process have to be adjusted in order to maintain overall energy and momentum conservation; it is assumed that none of these other momenta vanish in the limit $\eps\to0$. The asymptotic behavior of an $n$-particle scattering amplitude $A_n$ in the limit $\eps\to0$ is then characterized by a \emph{soft scaling parameter} $\sigma$ such that
\begin{equation}
A_n\propto\eps^\sigma,\qquad\eps\to0.
\end{equation}
As a rule, spontaneous breaking of a global symmetry implies that the scattering amplitudes of the associated NG boson satisfy $\sigma\geq1$; this is known as the \emph{Adler zero} property. Exceptions to this rule where $\sigma=0$ have been known for a long time, although their origin has only been clarified recently~\cite{Kampf2020a,Cheung2021}. On the other hand, theories with $\sigma>1$ are said to possess \emph{enhanced soft limits}.  

All currently known examples of EFTs for NG bosons with $\sigma>1$ feature redundant symmetry, even though the existence of such a symmetry has not been proven to be a necessary condition for enhanced soft limits. It is therefore sensible to search for new possible theories with $\sigma>1$ by focusing on the symmetry. Direct classification of possible Lie algebra structures admitting redundant symmetries has been carried out to construct catalogs of Lorentz-invariant theories with enhanced soft limits~\cite{Bogers2018a,Bogers2018b,Roest:2019}. An advantage of this approach is that it offers a unified treatment of the landscape of EFTs regardless of the number of NG degrees of freedom (flavors). An obvious disadvantage is that the presence of a redundant symmetry is an unproven assumption.

In fact, the historically first exploration of possible relativistic EFTs with enhanced soft limits was carried out using \emph{on-shell} scattering amplitude methods~\cite{Cheung2015a,Cheung2016a,Cheung2017a,Elvang:2018s}. The basic idea is roughly as follows. Within a generic EFT for a scalar field, imposing the condition that the scattering amplitudes in the soft limit have the Adler zero property ($\sigma=1$) will constrain the Wilson coefficients of the many operators one can add to the effective Lagrangian. The role of such constraints is to ensure cancellations between various contributions to the $S$-matrix, which ultimately leads to the Adler zero in the soft limit. Imposing subsequently enhanced scaling of scattering amplitudes, with given fixed $\sigma>1$, is expected to impose even stricter conditions on the Wilson coefficients at the leading order of the low-energy expansion. In extreme cases, the constraints from soft theorems are so stringent that the leading order of the EFT boils down to a one-parameter family of Lagrangians. Examples of such \emph{exceptional} theories are the nonlinear sigma model (NLSM), the Dirac-Born-Infeld (DBI) theory, and the special Galileon~\cite{Cheung2015a}.

It turns out that the theories singled out by requiring enhanced soft limits possess $S$-matrices that are recursively constructable through novel on-shell recursion relations~\cite{Cheung2016a}. More generally, it was shown that soft theorems implied by redundant symmetries can be used to recursively construct higher point on-shell amplitudes~\cite{Luo2016a}. Apart from analyzing the properties of already known theories, the novel on-shell recursion relations for EFTs have been utilized to further explore the landscape of EFTs with enhanced symmetries and soft limits~\cite{Cheung2017a,Elvang:2018s,Cheung2018c,Low2019a}. Very recently, the same philosophy was also applied to EFTs with universal albeit not necessarily vanishing soft behavior of the $S$-matrix~\cite{Kampf:2021multi}. The combination of on-shell recursion relations and soft theorems has developed into a full-fledged \emph{soft bootstrap} program~\cite{Elvang:2018s}. This is an algorithmic procedure for searching for local EFTs with given infrared properties, based on on-shell soft data and consistency conditions for the $S$-matrix. Common examples of soft data include the spectrum of massless particles, unbroken symmetries, and soft theorems. Further details can be found e.g.~in refs.~\cite{Elvang:2018s,Low2019a} and references therein.

While considerable effort has been devoted to exploring the properties of relativistic scalar EFTs using state-of-the-art techniques of quantum field theory~\cite{Cachazo2016a,Padilla2016,Cheung2017a,Low2018a,Low2018b,Elvang:2018s,Gonzalez2019,Bellazzini:2020, Arkani-Hamed2020b, Rodina2021}, nonrelativistic scalar EFTs have received considerably less attention. This has left nonrelativistic EFTs with enhanced symmetries and soft limits a terra incognita. The present paper should be understood as an attempt to shed light on this new territory. We will investigate the landscape of EFTs satisfying soft theorems of the form $\sigma>1$. The fact that EFTs with enhanced soft limits have an on-shell reconstructable $S$-matrix will allow us to combine conventional quantum field theory techniques with modern on-shell methods adapted for nonrelativistic EFTs~\cite{us} in order to carve out the landscape. 

This paper is rather lengthy and utilizes a range of different approaches to study nonrelativistic EFTs. In order to help the reader orient in the text, we now give a brief overview of the contents and main results of the individual sections.

%%%%%%%%%%%%%%%%%%%%%%%%%%%%%%%%%%%%%%%%%%%%%%%

\subsection{Outline and results}

\paragraph{Section~\ref{sec:2}}
We review the results of a previous classification of nonrelativistic EFTs with enhanced symmetries~\cite{Brauner2021a}, which provides a basis for the discussion of scattering amplitudes in the rest of the paper. The section covers mainly type $A_1$ and type $B_2$ theories of a single NG mode. A novel type $B_4$ theory, whose quartic dispersion relation is protected by symmetry, is also presented.

\paragraph{Section~\ref{sec:sym}} 
We review how redundant symmetries constrain the soft behavior of scattering amplitudes of NG bosons. It is pointed out that enhanced \emph{spatial} symmetry alone is not sufficient to ensure enhanced soft limits of scattering amplitudes. Instead, we find that the infrared dynamics can depend on what types of NG bosons are present in the system. We derive a new sufficient condition for enhanced soft limits, which requires information about the dispersion relation of the NG bosons. This is one of the main results of the present paper. Finally, we discuss concrete consequences for type $A$ and type $B$ NG bosons, using the theories cataloged in section~\ref{sec:2} for illustration.

\paragraph{Section~\ref{sec:recursion}} 
Here we present the on-shell technology used in our soft bootstrap and infrared classification procedures. This includes soft momentum shifts and on-shell recursion relations, adapted to nonrelativistic EFTs~\cite{us}. The bottom-up exploration of nonrelativistic EFTs in the following two sections relies heavily on the content of this section.  

\paragraph{Section~\ref{sec:bootstrap}}
We use soft bootstrap to carve out the landscape of on-shell constructable type $A_1$ and type $B_2$ theories with enhanced soft limits. The exceptional theories uncovered in the type $A_1$ sector are well-known relativistic theories. This observation leads to the notion of \emph{emergent Lorentz invariance} from the infrared. In the type $B_2$ sector, we recover the theories obtained using the Lie-algebraic classification in section~\ref{sec:2} without any additional surprises. Some details about the numerical setup for calculation of scattering amplitudes can be found in appendix~\ref{sec:numerics}. 

\paragraph{Section~\ref{sec:bounds}}
The observations made in the previous section raise the natural question whether \emph{all} exceptional type $A_1$ theories are necessarily relativistic, and whether there are any exceptional type $B_2$ theories that still remain undiscovered. To prepare the ground for attacking these questions, we first introduce a few parameters that furnish a simple classification scheme for scalar EFTs with enhanced soft limits. We then derive general bounds on the classification parameters, narrowing down the landscape of possible EFTs. Using analytical bootstrap methods, we next prove that all exceptional type $A_1$ theories in fact are relativistic, which allows us to lift existing results for relativistic EFTs to the whole type $A_1$ subfamily. On the type $B_2$ side, we give a proof that there are no exceptional type $B_2$ theories beyond those already discovered through Lie-algebraic methods. Some technical details are relegated to appendix~\ref{sec:4ptansatz}. 

\paragraph{Section~\ref{sec:comparison}}
Here we summarize the main findings of the paper and discuss how the top-down symmetry approach and the bottom-up bootstrap approach provide complementary insights into the landscape of nonrelativistic EFTs.

%%%%%%%%%%%%%%%%%%%%%%%%%%%%%%%%%%%%%%%%%%%%%%%

\section{Nonrelativistic EFTs with enhanced symmetry}
\label{sec:2}

In all currently known examples of EFTs featuring scattering amplitudes with enhanced soft limits, the latter are a consequence of an underlying redundant symmetry. The program to map the landscape of candidate EFTs based on a Lie-algebraic classification was initiated in refs.~\cite{Bogers2018a,Bogers2018b} for Lorentz-invariant theories. In ref.~\cite{Brauner2021a}, the analysis was extended to rotationally invariant EFTs with other than Lorentz boosts or without any boost symmetry whatsoever. In this section, we briefly review the main results of these works, and give some concrete examples of candidate EFTs where scattering amplitudes may be expected to possess an enhanced soft limit. This sets the basis for the follow-up discussion. In some sense, the following sections revolve around two immediate questions: (i) whether all the candidate theories identified here indeed possess enhanced amplitudes in the soft limit and (ii) whether there are any other EFTs with enhanced scattering amplitudes than those singled out here.

Throughout this paper, we will use the notation in which spatial indices are labeled with lowercase Latin letters $r,s,t,u,\dotsc$.\footnote{Temporal components of spacetime vectors will always be indicated with a $0$, not to be confused with~$t$ which, according to our convention, is a spatial index.} Wherever needed, we will use the standard Euclidean metric, $g_{rs}=\delta_{rs}$. The Lie algebra of infinitesimal symmetries of any translationally and rotationally invariant EFT necessarily includes the generators $J_{rs}$ of spatial rotations and $P_r$ of spatial translations. To these we add a set of scalar generators $Q_i$, some of which may be spontaneously broken, thus giving rise to the NG boson content of the EFT. Finally, we allow for an a priori undetermined set of redundant vector generators, $K_{rA}$, where the index $A$ distinguishes different redundant symmetries. In $d>2$ spatial dimensions, rotational invariance fixes some of the commutation relations among these generators,
\begin{align}
[J_{rs},J_{tu}]&=\im(g_{ru}J_{st}+g_{st}J_{ru}-g_{rt}J_{su}-g_{su}J_{rt}),\\
[J_{rs},P_t]&=\im(g_{st}P_r-g_{rt}P_s),\\
[P_r,P_s]&=0,\\
[J_{rs},K_{tA}]&=\im(g_{st}K_{rA}-g_{rt}K_{sA}),\\
[J_{rs},Q_i]&=0,\\
[Q_i,Q_j]&=\im f^k_{ij}Q_k,
\end{align}
where $f^k_{ij}$ are the structure constants of the Lie algebra of the scalar generators. In addition, we assume that the scalars $Q_i$ are translationally invariant, $[P_r,Q_i]=0$. This seems to be necessary in order have the usual Adler zero~\cite{Watanabe:2014,Rothstein:2017}, let alone further enhanced soft limits.

The main result of the Lie-algebraic classification of EFTs carried out in refs.~\cite{Bogers2018a,Bogers2018b,Brauner2021a} can now be summarized as follows. First, there are two particular linear combinations of the scalars $Q_i$ that play a distinguished role, $Q_A$ and $Q_{AB}$. These are defined by the right-hand side of the commutators
\begin{align}
[P_r,K_{sA}]&=\im g_{rs}Q_A,\\
[K_{rA},K_{sB}]&=\im(g_{AB}J_{rs}+g_{rs}Q_{AB}).
\end{align}
Here $g_{AB}$ is a symmetric matrix of coefficients which completely fixes the remaining commutators among $P_r$, $K_{rA}$, $Q_A$ and $Q_{AB}$,
\begin{align}
[K_{rA},Q_B]&=-\im g_{AB}P_r,\\
[K_{rC},Q_{AB}]&=\im(g_{AC}K_{rB}-g_{BC}K_{rA}),\\
[Q_A,Q_B]&=0,\\
[Q_{AB},Q_C]&=\im(g_{BC}Q_A-g_{AC}Q_B),\\
[Q_{AB},Q_{CD}]&=\im(g_{AD}Q_{BC}+g_{BC}Q_{AD}-g_{AC}Q_{BD}-g_{BD}Q_{AC}).
\end{align}
All that remains to have the complete Lie algebra is to find the commutators of the remaining scalars $Q_i$ with $K_{rA}$, $Q_A$ and $Q_{AB}$. To that end, we first introduce another set of coefficients, $d_{Ai}$, defined by the right-hand side of the commutator
\begin{equation}
[Q_i,K_{rA}]=(t_i)^B_{\phantom BA}K_{rB}-\im d_{Ai}P_r.
\end{equation}
Finally, we define two block $(n+1)\times(n+1)$ matrices, where $n$ denotes the range over which the index $A$ on $K_{rA}$ runs,
\begin{equation}
\arraycolsep=1ex
\renewcommand{\arraystretch}{1.5}
(T_i)^A_{\phantom AB}\equiv
\left(\begin{array}{c|c}
(t_i)^A_{\phantom AB} & 0\\
\hline
d_{Bi} & 0
\end{array}\right),\qquad
L_{AB}\equiv
\left(\begin{array}{c|c}
Q_{AB} & \im Q_A\\
\hline
-\im Q_B & 0
\end{array}\right).
\label{block}
\end{equation}
The matrices $T_i$ are constrained by the requirement that they span an affine representation of the Lie algebra of $Q_i$, that is, $[T_i,T_j]=\im f^k_{ij}T_k$. By the same token, the $n\times n$ matrices $t_i$ must span an ordinary (linear) representation of the Lie algebra of $Q_i$. The set of coefficients $g_{AB}$ is required to form a symmetric invariant tensor under the action of $t_i$. Moreover, $g_{AB}$ is related to the coefficients $d_{Ai}$ through
\begin{equation}
g_{AB}=-a^i_Ad_{Bi}=-a^i_Bd_{Ai},
\label{gAB}
\end{equation}
where $a^i_A$ is defined by $Q_A\equiv a^i_AQ_i$. The last remaining commutator can now be expressed compactly as
\begin{equation}
[Q_i,L_{AB}]=(T^T_iL+LT_i)_{AB}.
\end{equation}
Altogether, the commutation relations among the generators $J_{rs}$, $P_r$, $K_{rA}$ and $Q_i$ are completely fixed by the structure constants $f^k_{ij}$ of the subalgebra of scalars $Q_i$, the affine representation $T_i$ of this subalgebra, and the symmetric invariant tensor $g_{AB}$ of this subalgebra. In order to take account of time translation invariance, we also have to add a Hamiltonian. This can be implemented as one of the $Q_i$ scalars.

The Lie-algebraic structure reviewed above has an intriguing geometric interpretation. Namely, the subalgebra spanned on the generators $J_{rs}$, $P_r$, $K_{rA}$, $Q_A$ and $Q_{AB}$ is identical to the algebra of isometries of a $(d+n)$-dimensional pseudo-Euclidean space, endowed with the metric $g_{rs}\oplus g_{AB}$. Here $Q_A$ play the role of translations in the $n$ extra dimensions, $K_{rA}$ generate rotations between the $d$ physical dimensions and the extra dimensions, and $Q_{AB}$ generate rotations that operate exclusively in the extra dimensions. This makes it possible to interpret the resulting EFTs in terms of fluctuations of a $d$-dimensional brane embedded in a higher-dimensional space.

Finding all EFTs with the above Lie-algebraic structure is an open problem. In ref.~\cite{Brauner2021a}, two infinite classes of EFTs with multiple flavors of NG bosons were presented, generalizing the standard (relativistic) DBI and Galileon theories. In the present paper, we restrict the discussion to theories of a single NG boson. In the nonrelativistic domain, this may correspond either to a single real scalar field or to a single complex scalar field. In these special cases, the classification of possible EFTs can be carried out completely, with some further simplifying assumptions on the symmetry content of the EFT. This is the subject of the following two subsections. The classification of theories of a single real scalar in section~\ref{subsec:classificationreal} is taken over from ref.~\cite{Brauner2021a}. The classification of theories of a single complex scalar in section~\ref{subsec:classificationcomplex} is however new. Finally, in section~\ref{subsec:catalog}, we put together the discovered EFTs into a simple catalog. This will serve as a reference for the following sections.

%%%%%%%%%%%%%%%%%%%%%%%%%%%%%%%%%%%%%%%%%%%%%%%

\subsection{Theories of a single real scalar}
\label{subsec:classificationreal}

In ref.~\cite{Brauner2021a} a classification of EFTs for a single real scalar was carried out, which we reproduce here. The classification is based on a minimal extension of the algebra of spacetime symmetries that describes a single species of NG boson and possesses redundant symmetry. Specifically, the index $A$ on $K_{rA}$ is restricted to a single value and thus dropped. Likewise, we only include two scalar generators: the single spontaneously broken scalar $Q_A\equiv Q$, responsible for the existence of the NG boson, and the Hamiltonian $H$. The nontrivial part of the Lie algebra generated by $J_{rs}$, $P_r$, $K_r$, $Q$ and $H$ then takes the form
\begin{equation}
\begin{gathered}
[P_r,K_s]=\im g_{rs}Q,\qquad
[K_r,K_s]=-\im vJ_{rs},\\
[K_r,Q]=\im vP_r,\qquad
[K_r,H]=-\im wK_r+\im uP_r,\qquad
[Q,H]=-\im wQ,
\end{gathered}
\end{equation}
where the parameters $u,v,w$ are only constrained by the requirement that $vw=0$, which means that $v$ and $w$ are mutually exclusive. All other commutators among the generators are either zero or fixed by rotational invariance. Whenever some of the parameters $u,v,w$ are nonzero, they can be removed by a redefinition of the generators. In the end, we only find four distinct EFTs without any tunable parameters. We will review them one by one.

%%%%%%%%%%%%%%%%%%%%%%%%%%%%%%%%%%%%%%%%%%%%%%%

\subsubsection{Spatial Galileon theory}
\label{subsubsec:spatialGalileon}

This corresponds to the special case $u=v=w=0$. The symmetry generated by the broken scalar $Q$ acts on the corresponding NG field $\theta$ as a mere shift, $\theta\to\theta+\eps$. (The same is true for the single-flavor theories reviewed in the following two subsections.) The redundant symmetry generated by $K_r$ acts on $\theta$ as $\theta\to\theta+\beta_rx^r$, where $\beta_r$ is a vector of symmetry parameters. Invariant Lagrangians are constructed out of $\de_0\theta$ and $\de_r\de_s\theta$ (and their derivatives) in a way that respects rotational invariance, that is by contracting all spatial indices. In addition, there is a finite set of quasi-invariant Lagrangians, also known as Wess-Zumino (WZ) terms. These take a form that closely parallels the standard (Lorentz-invariant) Galileon operators,
\begin{equation}
\La_k=\frac1{(d-k)!}\eps^{r_1\dotsb r_kt_{k+1}\dotsb t_d}\eps^{s_1\dotsb s_k}_{\phantom{s_1\dotsb s_k}t_{k+1}\dotsb t_d}\theta(\de_{r_1}\de_{s_1}\theta)\dotsb(\de_{r_k}\de_{s_k}\theta),
\label{WZtermsGalileon}
\end{equation}
where $0\leq k\leq d$. The first of these ($k=0$) is a tadpole that has to be discarded for the EFT to be perturbatively well-defined. The second ($k=1$) is equivalent up to integration by parts to the spatial part of the usual kinetic term. We are then left with $d-1$ possible WZ interaction terms in $d$ spatial dimensions. Upon adding the temporal part of the kinetic term, the full Lagrangian of the spatial Galileon theory becomes
\begin{equation}
\La=\frac12(\de_0\theta)^2-\frac12(\vek\nabla\theta)^2+\sum_{k=2}^dc_k\La_k+\dotsb,
\label{spatialgalileon}
\end{equation}
where $c_k$ are a priori undetermined effective couplings, and the ellipsis stands for invariant operators built out of $\de_0\theta$ and $\de_r\de_s\theta$.

%%%%%%%%%%%%%%%%%%%%%%%%%%%%%%%%%%%%%%%%%%%%%%%

\subsubsection{Spatial DBI theory}
\label{subsubsec:spatialDBI}

There are two mutations of this theory, corresponding to $u=w=0$ and $v\equiv s=\pm 1$. Unlike in the Galileon case, the ``metric'' $g_{AB}$ is nonzero here. Hence we can invoke the geometric interpretation of the theory in terms of fluctuations of a $d$-dimensional brane in a $(d+1)$-dimensional pseudo-Euclidean space. In this interpretation, the scalar $Q$ corresponds to spontaneously broken translations in the extra dimension. For $s=1$, the Lie algebra of spatial symmetries of this theory is isomorphic to $\gr{SO}(d+1)\ltimes\R^{d+1}\simeq\gr{ISO}(d+1)$. For $s=-1$, it is isomorphic to $\gr{SO}(d,1)\ltimes\R^{d+1}$. The time translations generated by the Hamiltonian add an additional factor of $\R$ to this algebra.

Invariance under the shift symmetry generated by $Q$ requires that every field $\t$ in the Lagrangian carries at least one derivative. Unlike in the case of the Galileon, it is now possible to construct nontrivial interactions that contain only one derivative per field, yet are invariant under the redundant higher-dimensional symmetry. These interactions dominate the low-energy expansion of the EFT. The most general effective action containing exactly one derivative per field takes the form
\begin{equation}
S=\int\dd t\,\dd^d\vek x\sqrt{1+s(\vek\nabla\t)^2}\sum_{k=0}^\infty c_k(\nabla_0\t)^k,
\label{DBIaction}
\end{equation}
where $\nabla_0\t$ is the temporal covariant derivative of the NG field $\t$,
\begin{equation}
\label{DBInabla}
\nabla_0\t\equiv\frac{\de_0\t}{\sqrt{1+s(\vek\nabla\t)^2}}.
\end{equation}
The $k=1$ term of the sum in eq.~\eqref{DBIaction} is a total time derivative and can be dropped. Setting $c_0=-s/2$ and $c_2=1/2$ ensures correct normalization of the kinetic term. The construction of subleading interaction terms, containing more than one derivative per field, is somewhat involved; see ref.~\cite{Brauner2021a} for details.

%%%%%%%%%%%%%%%%%%%%%%%%%%%%%%%%%%%%%%%%%%%%%%%

\subsubsection{Galilei-invariant superfluid}
\label{subsubsec:superfluid}

This corresponds to the special case $v=w=0$ and $u=1$. The resulting Lie algebra is the Bargmann algebra, wherein $K_r$ is the generator of Galilei boosts and $Q$ the central charge. Accordingly, the symmetry transformation generated by $K_r$ acts on the spatial coordinate and NG boson field respectively as $x^r\to x^r-\beta^rt$ and $\t\to\t+\beta_rx^r-\frac12\vek\beta^2t$. Spontaneous breakdown of $Q$ (and the associated spontaneous breaking of boost invariance) describes nonrelativistic superfluids.

Just like in the case of the spatial DBI theory, it is possible to construct actions invariant under the Bargmann algebra, containing just one derivative per field. These will dominate the low-energy expansion of the EFT. The most general effective Lagrangian containing exactly one derivative per field reads
\begin{equation}
\La=\sum_{k=1}^\infty c_k(\nabla_0\t)^k,\qquad
\nabla_0\t\equiv\de_0\t-\frac12(\vek\nabla\t)^2.
\label{Lsuperfluid}
\end{equation}
In order to ensure a properly normalized kinetic term, we have to set $c_1=1$ and $c_2=1/2$. Subleading contributions to the effective Lagrangian are constructed out of $\nabla_0\t$ and $\de_r\de_s\t$ and their covariant derivatives. The latter are defined as $\nabla_0\equiv\de_0-(\de^r\t)\de_r$ and $\nabla_r\equiv\de_r$. Remaining free spatial indices on such Galilei-invariant operators are to be contracted in a way that preserves rotational invariance.

Interestingly, the symmetry under the Bargmann algebra admits a set of WZ terms, identical to those of the spatial Galileon theory, eq.~\eqref{WZtermsGalileon}. This is easy to understand as a consequence of the fact that the latter only contain spatial derivatives. An infinitesimal Galilei boost only differs from the infinitesimal spatial Galileon transformation by a time-dependent shift of the coordinate, $x^r\to x^r-\beta^rt$. This does not affect operators that do not contain time derivatives and do not explicitly depend on the coordinates.

%%%%%%%%%%%%%%%%%%%%%%%%%%%%%%%%%%%%%%%%%%%%%%%

\subsubsection{Deformed Galileon theory}
\label{subsubsec:deformedGalileon}

This corresponds to the special case $u=v=0$ and $w=1$. The name of this theory stems from the fact that the corresponding Lie algebra is just the spatial Galileon algebra augmented with nontrivial temporal scaling of $K_r$ and $Q$, that is, with $[H,K_r]=\im K_r$ and $[H,Q]=\im Q$. Accordingly, the transformation rules for $\t$ under the symmetries generated by $Q$ and $K_r$ are twisted, $\theta\to \theta+e^t(\eps+\beta_rx^r)$. The invariant building blocks for the construction of effective Lagrangians are $\nabla_0\t\equiv\de_0\t-\t$ and $\de_r\de_s\t$, and their derivatives.

It is not clear how to even set up a perturbatively well-defined EFT based on the deformed Galileon algebra. Namely, it does not seem possible to construct a kinetic term out of the above basic building blocks. We therefore disregard this theory from further consideration.

%%%%%%%%%%%%%%%%%%%%%%%%%%%%%%%%%%%%%%%%%%%%%%%

\subsection{Theories of a single complex scalar}
\label{subsec:classificationcomplex}

Let us now switch gears and see how we can construct shift-invariant EFTs for a single \emph{complex} scalar. This means that we restrict the general Lie algebra reviewed at the beginning of this section to two possible values of the capital Latin indices, $A=1,2$. This will give us two spontaneously broken scalars $Q_A$ and two real NG fields $\t^A$. In order that the latter can be meaningfully merged into a single complex (Schr\"odinger) field $\psi$, we require the existence of an additional scalar generator, $Q$. This will generate a $\gr{U}(1)$ internal symmetry under which $\psi$ is charged, which is not to be spontaneously broken. In other words, the three generators $Q_A$ and $Q$ together should span the Lie algebra $\gr{ISO}(2)$,
\begin{equation}
[Q,Q_A]=-\im\eps_A^{\phantom AB}Q_B,\qquad
[Q_A,Q_B]=0.
\end{equation}
For simplicity, we assume that the Hamiltonian $H$ commutes with all these other scalars. This is the minimal scalar sector needed for an EFT of a Schr\"odinger scalar. Even in such a restricted setup, the generator $Q_{AB}$ may be present. Namely, it can be incorporated by making $Q_{AB}$ a linear combination of $\eps_{AB}Q$ and $\eps_{AB}H$. Upon carefully imposing the Jacobi identity on the double commutators of all possible combinations of generators, we end up with a two-parameter family of Lie algebras. Those commutators among $J_{rs}$, $P_r$, $K_{rA}$, $Q$, $Q_A$ and $H$ that are neither zero nor completely fixed by rotational invariance, assume the form
\begin{equation}
\begin{gathered}
[P_r,K_{sA}]=\im g_{rs}Q_A,\qquad
[K_{rA},K_{sB}]=\im[\alpha\delta_{AB}J_{rs}+\eps_{AB}g_{rs}(\alpha Q+\beta H)],\\
[K_{rA},Q_B]=-\im\alpha\delta_{AB}P_r,\qquad
[Q,K_{rA}]=-\im\eps_A^{\phantom AB}K_{rB},\qquad
[Q,Q_A]=-\im\eps_A^{\phantom AB}Q_B,
\end{gathered}
\end{equation}
where $\alpha,\beta$ are a priori undetermined parameters. As in the case of EFTs of a single real scalar, this is not really a continuous family of Lie algebras, since those of the parameters which are nonzero can be removed by a redefinition of the generators. At the end of the day, we only find three distinct types of Lie algebras, which we discuss one by one.

%%%%%%%%%%%%%%%%%%%%%%%%%%%%%%%%%%%%%%%%%%%%%%%

\subsubsection{Schr\"odinger-Galileon theory}
\label{subsubsec:SchrodingerGalileon}

This corresponds to the special case $\alpha=\beta=0$. This belongs to the class of multiflavor Galileon-like theories, analyzed in section~3 of ref.~\cite{Brauner2021a}. We can therefore take over all the results thereof. First, the spontaneously broken generators $Q_A$ act upon the two real NG fields $\t^A$ via constant shifts, $\t^A\to\t^A+\eps^A$. Likewise, the redundant vector generators $K_{rA}$ transform the NG fields as $\t^A\to\t^A+\beta_r^Ax^r$. Finally, the unbroken scalar generator $Q$ acts on $\t^A$ linearly through the vector representation of $\gr{SO}(2)$.

Invariant actions can be built out of $\de_0\t^A$ and $\de_r\de_s\t^A$ and their derivatives in a way that preserves spatial rotational and internal $\gr{SO}(2)$ invariance by properly contracting all spatial and internal indices. In addition, there are several WZ terms. Two of them are straightforward multiflavor generalizations of eq.~\eqref{WZtermsGalileon},
\begin{align}
\La_1\to{}&\delta_{AB}\t^A\de_r\de^r\t^B,\\
\notag
\La_3\to{}&\frac1{(d-3)!}(\delta_{AB}\delta_{CD}+\delta_{AC}\delta_{BD}+\delta_{AD}\delta_{BC})\eps^{r_1r_2r_3t_4\dotsb t_d}\eps^{s_1s_2s_3}_{\phantom{s_1s_2s_3}t_4\dotsb t_d}\\
&\times\t^A(\de_{r_1}\de_{s_1}\t^B)(\de_{r_2}\de_{s_2}\t^C)(\de_{r_3}\de_{s_3}\t^D).
\label{L3SchrodingerGalileon}
\end{align}
Finally, there is now a new, genuinely nonrelativistic and multiflavor WZ term with a single derivative, proportional to $\eps_{AB}\t^A\de_0\t^B$. This makes the two fields $\t^A$ canonically conjugate to each other and turns them into a single Schr\"odinger-like degree of freedom with energy proportional to squared momentum. The modified dispersion relation affects power counting in the EFT. It is now necessary to count each temporal derivative as two spatial derivatives. The low-energy expansion of the EFT is then dominated by operators with less than two (equivalent spatial) derivatives per field, which are supplied precisely by the three WZ terms. In terms of the complex field
\begin{equation}
\psi\equiv\frac1{\sqrt2}(\t^1+\im\t^2),
\label{schrodingerfield}
\end{equation}
the leading part of the effective Lagrangian can be written as
\begin{equation}
\La=\he\psi(\im\de_0+\vek\nabla^2)\psi+\La_{\text{int}},
\end{equation}
where the sole interaction term coming from $\La_3$ equals, up to a tunable effective coupling ($\vek\nabla^2$ is the usual shorthand notation for $\de_r\de^r$),
\begin{equation}
\begin{split}
\La_{\text{int}}\propto{}&\de_t\he\psi\de^t\psi(\vek\nabla^2\he\psi\vek\nabla^2\psi-\de_r\de_s\he\psi\de^r\de^s\psi)+\de_r\he\psi\de^r\de^s\he\psi\de_s\de_t\psi\de^t\psi\\
&+\de_r\he\psi\de^r\de^s\psi\de_s\de_t\he\psi\de^t\psi-\vek\nabla^2\he\psi\de_r\he\psi\de^r\de^s\psi\de_s\psi-\vek\nabla^2\psi\de_r\psi\de^r\de^s\he\psi\de_s\he\psi.
\end{split}
\end{equation}

%%%%%%%%%%%%%%%%%%%%%%%%%%%%%%%%%%%%%%%%%%%%%%%

\subsubsection{Schr\"odinger-DBI theory}
\label{subsubsec:SchrodingerDBI}

This corresponds to the special case $\alpha\equiv s=\pm1$ and $\beta=0$. This is a DBI-like theory describing the fluctuations of a $d$-dimensional brane embedded in a $(d+2)$-dimensional pseudo-Euclidean space. For $s=1$, the Lie algebra of spatial symmetries of this theory is isomorphic to $\gr{SO}(d+2)\ltimes\R^{d+2}\simeq\gr{ISO}(d+2)$. For $s=-1$, it is isomorphic to $\gr{SO}(d,2)\ltimes\R^{d+2}$. The $s=1$ mutation of the theory was dubbed ``$\gr{ISO}(2)$ theory'' and detailed in section 4.3 of ref.~\cite{Brauner2021a}. We therefore merely summarize the results with the modifications necessary to account for the two possible values of $s$.

Just like for the spatial DBI theory of a single real scalar, discussed in section~\ref{subsubsec:spatialDBI}, it is possible to construct an action that contains exactly one derivative per field; this dominates the low-energy expansion of the Schr\"odinger-DBI theory. Unlike in the action~\eqref{DBIaction} of the spatial DBI theory, however, terms containing temporal derivatives will be suppressed as a consequence of the modified power counting in type $B_2$ theories. The dominant interactions should therefore be constructed out of operators containing one \emph{spatial} derivative per field. Such interactions are entirely controlled by the induced metric on the brane,
\begin{equation}
G_{rs}\equiv g_{rs}-s\delta_{AB}\de_r\t^A\de_s\t^B,
\end{equation}
or the associated ``metric'' in the NG field space, $\tilde G_{AB}\equiv\delta_{AB}-s\de_r\t^A\de^r\t^B$. Including the WZ term with a single time derivative, which is consistent with the symmetry of the Schr\"odinger-DBI theory, the leading-order action for the theory reads
\begin{equation}
S=\int\dd t\,\dd^d\vek x\,\bigl(\he\psi\im\de_0\psi+c\sqrt{|G|}\bigr),
\label{SchroedingerDBIaction}
\end{equation}
where $c$ is an effective coupling and
\begin{equation}
\lvert G\rvert=\lvert\tilde G\rvert=1-2s\vek\nabla\he\psi\cdot\vek\nabla\psi+(\vek\nabla\he\psi\cdot\vek\nabla\psi)^2-|\vek\nabla\psi\cdot\vek\nabla\psi|^2.
\label{SchrodingerDBIG}
\end{equation}
If desired, subleading terms containing one spatial or temporal derivative per field may be added at will. Unlike in eq.~\eqref{DBIaction}, temporal covariant derivatives of $\psi$ may no longer appear in the Lagrangian on their own. Invariance under the internal $\gr{SO}(2)$ symmetry generated by $Q$ demands that they appear in pairs through the combination
\begin{equation}
\begin{split}
\delta_{AB}\nabla_0\t^A\nabla_0\t^B=\frac1{|G|}\bigl\{&2\de_0\he\psi\de_0\psi(1-s\vek\nabla\he\psi\cdot\vek\nabla\psi)\\
&+s\bigl[(\de_0\he\psi)^2\vek\nabla\psi\cdot\vek\nabla\psi+(\de_0\psi)^2\vek\nabla\he\psi\cdot\vek\nabla\he\psi\bigr]\bigr\}.
\end{split}
\label{SchrodingerDBID0D0}
\end{equation}

%%%%%%%%%%%%%%%%%%%%%%%%%%%%%%%%%%%%%%%%%%%%%%%

\subsubsection{Quarton theory}
\label{subsubsec:typeB4}

This special case with $\alpha=0$ and $\beta=1$ is a novel theory whose existence has, to the best of our knowledge, not been noticed before. We dub it ``quarton'' for reasons that will soon become clear. In order to outline even briefly its features, some technical details are necessary. We relegate those to appendix~\ref{app:B4}. Here we at least mention two distinctive features of the theory.

First, it turns out that the symmetry of the \exotic{} theory does not allow the usual spatial kinetic term in the effective Lagrangian. The gradient expansion of the bilinear part of the Lagrangian starts at order four in spatial derivatives. However, the Schr\"odinger-type term with a single time derivative can still be consistently included. Hence this is an example of a type $B_4$ theory where the dispersion relation $\om\propto\vek p^4$ is natural, i.e.~protected by symmetry. This feature definitely makes the \exotic{} theory worth further study.

Second, it turns out that the low-energy expansion of the effective action of the \exotic{} theory is dominated by interaction terms with exactly two spatial derivatives per field. No interactions with less than two derivatives per field are allowed by symmetry. This means that the scattering amplitudes of NG bosons in this theory will naturally scale with second power of momentum in the single soft limit. From the point of view of classification of EFTs with enhanced soft limits, the \exotic{} theory is therefore to be seen as trivial.

%%%%%%%%%%%%%%%%%%%%%%%%%%%%%%%%%%%%%%%%%%%%%%%

\subsection{Catalog of candidate EFTs for a single NG boson}
\label{subsec:catalog}

The classification of EFTs for a single NG mode, carried out in this section so far, was based entirely on the presence of redundant symmetry. Ultimately, we would however like to check whether or not these EFTs actually feature enhanced soft limits. We will leave out the pathological deformed Galileon theory. Likewise, we will drop the \exotic{} theory, which is of type $B_4$ and where, as explained above, the enhanced soft limit with $\sigma=2$ is expected to be realized trivially. All the other discussed EFTs belong to types $A_1$ and $B_2$, which we focus on in this paper. For the reader's convenience, we list them in table~\ref{tab:catalog}, along with their relativistic counterparts. All these theories have been constructed with a single layer of redundant generators, and we therefore expect a priori the soft scaling parameter $\sigma=2$ for all of them.

\begin{table}[t]
\centering
\begin{tabular}{c|c|c}
Theory & $\sigma$ (expected) & $\sigma$ (actual)\\
\hline\hline
Galileon & $2$ & $2$ or $3$\\
DBI & $2$ & $2$\\
\hline
spatial Galileon & $2$ & $0$ or $1$ or $2$\\
spatial DBI & $2$ & $0$ or $1$ or $2$\\
Galilei-invariant superfluid & $2$ & $0$ or $1$ or $2$\\
Schr\"odinger-Galileon & $2$ & $2$\\
Schr\"odinger-DBI & $2$ & $2$
\end{tabular}
\caption{Overview of type $A_1$ and type $B_2$ EFTs of a single NG boson along with the soft scaling parameter $\sigma$ of their scattering amplitudes. The first two lines correspond to the usual Lorentz-invariant EFTs of a single real massless scalar. The rest of the table lists the EFTs reviewed in this section. The second and third column displays respectively the prediction for $\sigma$ based on the presence of redundant symmetry, and the actual value of $\sigma$ extracted from numerical study of tree-level scattering amplitudes. Appendix~\ref{sec:numerics} details the setup of the numerical analysis. See the text for an explanation of the alternatives in the third column.}
\label{tab:catalog}
\end{table}

In case of the relativistic Galileon, it was found in ref.~\cite{Cheung2015a} that a special choice of couplings, later dubbed special Galileon, makes the amplitudes further enhanced with $\sigma=3$. Soon afterwards, it was clarified that this is a consequence of an additional, hidden symmetry generated by a set of operators that transform as a traceless symmetric Lorentz tensor~\cite{Hinterbichler2015a}. The conclusion that $\sigma$ may be larger than naively expected in case additional ``hidden'' symmetry is present should be uncontroversial.

What is more problematic are the alternative values of $\sigma$ we find for the spatial Galileon and spatial DBI theories and the Galilei-invariant superfluid, as shown in the third column of table~\ref{tab:catalog}. All of these are based on numerical inspection of \emph{tree-level} amplitudes with different choices of effective couplings of the operators allowed by symmetry. We find that, as a rule of thumb, $\sigma=0$ in case the interaction Lagrangian contains cubic vertices. If it does not, then $\sigma=1$ for a generic choice of couplings. Achieving $\sigma=2$ requires a fine-tuning of the couplings. In case of the spatial Galileon theory, we find $\sigma=2$ in case all interaction terms contain, just like those in eq.~\eqref{spatialgalileon}, only spatial derivatives. For the spatial DBI theory, imposing $\sigma=2$ on the class of Lagrangians~\eqref{DBIaction} gives two solutions. One of them is just the relativistic DBI theory. The other turns out to be equivalent to the relativistic DBI theory upon a field redefinition; see section~\ref{subsec:LagscanA1} for more details. Finally, in the Galilei-invariant superfluid the presence of cubic interaction vertices is inevitable. It is however still possible to achieve $\sigma=2$ with a very special choice of couplings, which turns out to give a theory equivalent to the relativistic DBI; see also section~\ref{subsec:LagscanA1}.

Note that the above problem does not appear for the Schr\"odinger-type theories, where we find $\sigma=2$ without further constraints on the effective couplings. We therefore conclude that specifically for type $A_1$ nonrelativistic EFTs, $\sigma=2$ is not guaranteed by spatial redundant symmetry alone. This makes it clear that in case we want to make general statements about the scaling of scattering amplitudes in the soft limit, we need additional physical input. One of the primary goals of the following sections is to understand how to refine the criterion for the enhancement of scattering amplitudes in a way that covers all the cases listed in table~\ref{tab:catalog}.

%%%%%%%%%%%%%%%%%%%%%%%%%%%%%%%%%%%%%%%%%%%%%%%

\section{Enhanced scattering amplitudes from symmetry}
\label{sec:sym}

In this section we set out to understand how the presence of redundant symmetry affects the scaling of scattering amplitudes in the soft limit, and what additional input might be required to ensure that $\sigma>1$. We do so by carefully adapting the nonperturbative argument given in ref.~\cite{Cheung2017a} to nonrelativistic EFTs.

Our strategy is as follows. In section~\ref{subsec:adler}, we review the argument for the existence of Adler zero in a form suitable for nonrelativistic EFTs. To understand how redundant symmetry might imply $\sigma>1$ requires one new technical ingredient. This is the fact that locally indistinguishable (that is redundant) symmetries imply conservation laws that are related by certain linear identities~\cite{Brauner2014a,Brauner2020a}. In order to make the paper self-contained, we present a detailed derivation of these identities in appendix~\ref{sec:Noetherrelations}. Section~\ref{subsec:enhsofbeh} employs them to study the soft limit of scattering amplitudes in nonrelativistic EFTs with redundant symmetry. Finally, in section~\ref{subsec:revisit} we revisit the sample EFTs listed in table~\ref{tab:catalog} to check our understanding of the values of $\sigma$ reported therein.

%%%%%%%%%%%%%%%%%%%%%%%%%%%%%%%%%%%%%%%%%%%%%%%

\subsection{Adler zero}
\label{subsec:adler}

In order to keep the discussion as simple as possible, we will focus on a single NG field $\theta$ and the corresponding Noether current $J^\mu$;\footnote{We use the standard relativistic notation whereby the Greek indices $\mu,\nu,\lambda,\dotsc$ indicate Lorentz vectors. Minkowski inner product of spacetime vectors is denoted with a dot.} the generalization to several flavors of NG bosons is straightforward. The coupling of a NG state with given momentum $\vek p$, $\ket{\theta(\vek p)}$, to the broken current is generally described by the matrix element $\bra{\Omega}J^{\mu}(x)\ket{\theta(\vek p)}$. Since we assume spacetime translation invariance but only spatial rotation invariance, the matrix element is constrained to the following form,
\begin{align}
    \bra{\Omega}J^{\mu}(x)\ket{\theta(\boldsymbol{p})}=e^{-\im p\cdot x}\left[\im p^{\mu}F_1(\abs{\boldsymbol{p}})+\im\delta^{\mu 0}F_2(\abs{\boldsymbol{p}})\right].
\end{align}
Current conservation then fixes the dispersion relation of the NG boson, that is its frequency $\omega$ as a function of the momentum $\vek p$, through
\begin{align}
    \label{onshellcurrent}
    \omega^2(\abs{\boldsymbol{p}})F_1(\abs{\boldsymbol{p}})+\omega(\abs{\boldsymbol{p}}) F_2(\abs{\boldsymbol{p}})-\boldsymbol{p}^2F_1(\abs{\boldsymbol{p}})=0.
\end{align}

Let us now consider the matrix element $\bra{\beta}J^{\mu}(0)\ket{\alpha}$ of the current between some in-state $\ket\alpha$ and out-state $\ket\beta$. These states may include an arbitrary number of NG bosons or other types of particles. By considering the limit where the momentum carried away by the current goes on-shell, we can access the scattering amplitude $\bra{\beta+\theta(\boldsymbol{p})}\ket{\alpha}$ for a process with an additional NG boson inserted in the out-state. Namely, the matrix element of the current features a NG pole around which it factorizes as
\begin{align}
    \label{polecurrent}
    \bra{\beta}J^{\mu}(0)\ket{\alpha}=\frac{\im}{p^0-\omega(\abs{\boldsymbol{p}})}\bra{\Omega}J^{\mu}(0)\ket{\theta(\boldsymbol{p})}\bra{\beta+\theta(\boldsymbol{p})}\ket{\alpha}+R^{\mu}(p),
\end{align}
where $p\equiv p_\alpha-p_\beta$ and the off-shell energy variable $p^0$ should be distinguished from the on-shell energy of the NG boson, $\omega(\abs{\boldsymbol{p}})$. Moreover, $R^\mu(p)$ is a remainder function which is by construction non-singular on-shell, that is in the limit $p^0\to\omega(\vek p)$. Combining eq.~(\ref{polecurrent}) with current conservation yields, 
\begin{align}
    \label{1}
    \frac{p^0\omega(\abs{\boldsymbol{p}}) F_1(\abs{\boldsymbol{p}})+p^0F_2(\abs{\boldsymbol{p}})-\boldsymbol{p}^2F_1(\abs{\boldsymbol{p}})}{p^0-\omega(\abs{\boldsymbol{p}})}\bra{\beta+\theta(\boldsymbol{p})}\ket{\alpha}=p^0R_0(p)+p^rR_r(p).
\end{align}
The pole on the left-hand side is exactly cancelled by using eq.~(\ref{onshellcurrent}), upon which we find
\begin{align}
    \label{NRAdler}
    \bra{\beta+\theta(\boldsymbol{p})}\ket{\alpha}=\frac{p^0R_0(p)+p^rR_r(p)}{\omega(\abs{\boldsymbol{p}})F_1(\abs{\boldsymbol{p}})+F_2(\abs{\boldsymbol{p}})}\biggr\rvert_{p^0=\omega(\abs{\boldsymbol{p}})}.
\end{align}
Note that the denominator on the right-hand side is nonzero as a direct consequence of Goldstone's theorem, which requires  $\bra{\Omega}J^0(0)\ket{\theta}=\im(\omega F_1+F_2)\neq0$. With the additional assumption that $R^\mu(p)$ is non-singular when $p^\mu$ is on-shell \emph{and} the limit $\vek p\to\vek0$ is taken, which does not automatically follow from standard polology rules, we obtain the Adler zero,
\begin{align}
    \label{adler2}
    \lim_{\vek p\rightarrow \vek 0}\bra{\beta+\theta(\boldsymbol{p})}\ket{\alpha}=0.
\end{align}

The Adler zero may be avoided only if the regularity assumption on $R^\mu(p)$ is violated, for instance if the current matrix element~\eqref{polecurrent} has additional singularities for $p^\mu\to0$ besides the NG pole. This may happen if the Noether current can be inserted into the external legs of the amplitude $\bra{\beta}\ket{\alpha}$. This is in turn possible if the expansion of the current in powers of elementary fields contains bilinear terms. Such terms can arise from cubic vertices in the interaction Lagrangian, or from terms in the transformation of $\theta$ under the broken symmetry linear in fields. In section \ref{subsec:revisit}, we provide an explicit illustration of how bilinear terms in the Noether current may spoil the Adler zero property. 

%%%%%%%%%%%%%%%%%%%%%%%%%%%%%%%%%%%%%%%%%%%%%%%

\subsection{Enhanced soft behavior}
\label{subsec:enhsofbeh}

Understanding the soft properties of scattering amplitudes of NG bosons beyond Adler zero requires more detailed knowledge about the remainder function $R^{\mu}(p)$. This may be extracted from the identities among the various Noether currents in case redundant symmetry is present. The authors of ref.~\cite{Cheung2017a} have considered a broad class of generalized shift symmetries of the form
\begin{align}
    \label{genshifttext}
    \theta(x)\rightarrow \theta(x)+\eps_j\bigl[\alpha^j(x)+\alpha_B^j(x)\mathcal{O}^B(x)\bigr],
\end{align}
where $\eps_j$ denotes infinitesimal parameters, $\alpha^j(x)$ and $\alpha_B^j(x)$ are fixed polynomials, and $\mathcal{O}^B(x)$ are local composite operators constructed from $\theta$ and its derivatives. Under some mild regularity assumptions, theories invariant under the shift symmetry~\eqref{genshifttext} feature remainder functions $R^{\mu}({p})$ satisfying the relation
\begin{align}
    \label{Rconstrainttext}
    \Tilde{\alpha}^j(p)p_{\mu}R^{\mu}(p)=0,
\end{align}
valid in the sense of distributions, where the tilde denotes Fourier transform. While a proof of this constraint was already given in ref.~\cite{Cheung2017a}, we reproduce it in appendix~\ref{sec:Noetherrelations} for the sake of completeness. In Lorentz-invariant theories, eq.~(\ref{Rconstrainttext}) implies without further assumptions soft theorems controlling the soft scaling parameter $\sigma$. As we will demonstrate below, this is generally no longer true for nonrelativistic theories.

To see how the relation~\eqref{Rconstrainttext} affects the low-energy dynamics in EFTs with redundant symmetry, let us consider a generalized shift symmetry of the type~\eqref{genshifttext} with the leading, $\theta$-independent polynomial given by
\begin{equation}
\alpha(x)=\eps_{\n_1\dotsb\n_n}x^{\n_1}\dotsb x^{\n_n},
\label{phi}
\end{equation}
where $\eps_{\n_1\dotsb\n_n}$ is a symmetric-tensor infinitesimal parameter and all $\n_i$ are spacetime indices. By means of eq.~\eqref{Rconstrainttext}, this imposes the following constraint on the remainder function $R^\mu(p)$ regardless of the specific form of the polynomials $\alpha^j_B$ or operators $\mathcal O^B$ in eq.~\eqref{genshifttext},
\begin{align}
    \label{soft}
    \lim_{p\rightarrow 0}\partial_{\n_1}\dotsb\partial_{\n_k}[p_{\mu}R^{\mu}(p)]=0,
\end{align}
for any $k=0,\dotsc,n$, where we used the shorthand notation $\de_\n\equiv\de/\de p^\n$. In Lorentz-invariant theories, the set of constraints~\eqref{soft} implies that all Taylor coefficients of $R^\mu(p)$ up to order $n-1$ in momenta vanish. Equation~\eqref{NRAdler} then guarantees that scattering amplitudes of NG bosons have $\sigma\geq n+1$ in the soft limit. For instance, the relativistic Galileon and DBI theories both possess a redundant symmetry, belonging to the class of generalized shift symmetries with $\alpha(x)=\eps_\n x^\n$. This corresponds to $n=1$ and hence $\sigma\geq2$. This is a nontrivial realization of the enhanced soft limit, since the Lagrangian representations of the Galileon and DBI theories contain less than two derivatives per field.

In nonrelativistic theories, rotational invariance requires that the constraints~\eqref{soft} come in multiplets which are symmetric tensors of $\gr{SO}(d)$. It is however not given a priori that all combinations of spatial and temporal indices $\n_i$ appear among the full set of constraints. This depends on the spatial or temporal nature of the polynomial~\eqref{phi}. For instance, when all the $\n_i$ are temporal, then eq.~\eqref{soft} gives no constraints for the spatial part of the remainder function, $R^r(p)$. The temporal part $R^0(p)$, on the other hand, is constrained to have a Taylor series in $p^\mu$ whose coefficients of $(p^0)^k$ vanish for all $k\leq n-1$.

In this paper, we are mainly interested in \emph{spatial} redundant symmetries, for which $\alpha(x)$ is a polynomial in spatial coordinates. In this case, eq.~\eqref{soft} imposes no constraints on $R^0(p)$. On the other hand, $R^r(p)$ is required to have a Taylor series in $p^\mu$ whose purely spatial part starts at order $n$. Given the relation~\eqref{NRAdler} between the remainder function and the scattering amplitude, we see that invariance under generalized spatial shift symmetries is not sufficient to constrain the soft scaling parameter $\sigma$ beyond ordinary Adler zero. 

For an illustration, let us consider the class of theories invariant under generalized spatial shift symmetries of degree $n=1$. This includes the spatial Galileon and spatial DBI theories reviewed in table~\ref{tab:catalog}. In this case, eq.~\eqref{soft} constrains the Taylor expansion of $R^r(p)$ to the form
\begin{equation}
R^r(p)=c^rp^0+c^r_sp^s+\mathcal O(p^2),
\label{RIP}
\end{equation}
where $c^r$ and $c^r_s$ are the respective Taylor coefficients. The temporal part $R^0(p)$ remains unconstrained and will in general have a nonzero limit for $p\to0$. It follows from eq.~\eqref{NRAdler} that if the NG boson has a linear dispersion relation, then its scattering amplitudes will vanish in the soft limit with $\sigma=1$. Enhanced scaling with $\sigma>1$ can only be guaranteed by additional constraints on $R^0(p)$, or alternatively if the energy of the NG boson is proportional to a higher power of momentum.

%%%%%%%%%%%%%%%%%%%%%%%%%%%%%%%%%%%%%%%%%%%%%%%

\paragraph{Type A versus type B NG bosons}

The last observation opens the possibility to achieve enhanced soft limits of scattering amplitudes by combining spatial redundant symmetry with a higher-order dispersion relation of the NG boson. Let us therefore consider generally EFTs of NG bosons in $d$ spatial dimensions, enjoying a generalized spatial shift symmetry of degree $n$. At this point, we must distinguish NG bosons of type $A$ and type $B$, since type $A_m$ modes are forbidden by the generalized CHMW theorem if $m\geq d$. Combining eqs.~\eqref{NRAdler} and~\eqref{soft}, we then get the following bounds for the respective types of NG bosons,\footnote{The reader is reminded that these bounds rely on certain mild technical assumptions behind the derivation of eq.~\eqref{Rconstrainttext}; see appendix~\ref{sec:Noetherrelations} for details.}
\begin{align}
    \label{A}
    \sigma&\geq \min(m,n+1)\quad \text{where} \quad d\geq m+1 &&  \text{for type } A_m,  \\
    \label{B}
    \sigma&\geq \min(2m,n+1)  &&\text{for type } B_{2m}.
\end{align}
This is one of our main results. Note how the scaling of the scattering amplitudes in the soft limit crucially depends on the type of NG boson that is being emitted. This underlines the fact that the soft properties of scattering amplitudes depend not only on the symmetry present in the system, but also on the details of the NG boson spectrum.

In the physically interesting case of $d=3$, the generalized CHMW theorem forbids existence of type $A_m$ NG modes with $m>2$. As a consequence, the most stringent bound that can be imposed on the soft scaling parameter $\sigma$ in type $A$ theories where only $n$ and $m$ are known is $\sigma\geq 2$. There is no corresponding obstruction in type $B_{2m}$ theories though. Based on the bound~\eqref{B} alone, it appears that one can in principle achieve arbitrarily high $\sigma$ by choosing large enough values for $n$ and $2m$.

Another possibility how to guarantee enhanced soft limits is to constrain the temporal part of the remainder function, $R^0(p)$. This can be done on general grounds if the system in question possesses a time-dependent redundant symmetry. Sometimes it may however also be possible to restrict the form of $R^0(p)$ based on the explicit knowledge of the interaction Lagrangian. In extreme cases, the function $R^0(p)$ may be vanishing altogether. A sufficient condition for this to happen, at least at tree level, is the existence of a Lagrangian representation of the theory where the interaction vertices do not contain any temporal derivatives. An example of such a theory is the spatial Galileon~\eqref{spatialgalileon}, restricted to the spatial WZ terms~\eqref{WZtermsGalileon}.

Suppose now that $R^0(p)$ indeed vanishes identically. Then the soft behavior of the scattering amplitudes is completely fixed by the spatial part of the remainder function, $R^r(p)$. If the theory possesses a generalized spatial shift symmetry of degree $n$, then the constraint~\eqref{soft} restricts the form of $R^r(p)$ to
\begin{equation}
R^r(p)=p^0\bar R^r(p)+c^r_{s_1\dotsb s_n}p^{s_1}\dotsb p^{s_n}+\dotsb,
\end{equation}
where $\bar R^r(p)$ is an arbitrary non-singular function of $p^0$ and $\vek p$, and the ellipsis stands for terms of order $n+1$ or higher in spatial momentum. This generalizes the $n=1$ case displayed in eq.~\eqref{RIP}. From eq.~\eqref{NRAdler} we now infer that in type $A_m$ theories, $\sigma\geq\min(m+1,n+1)$, with the same constraint $d\geq m+1$ as in eq.~\eqref{A}. For type $B_{2m}$ theories, on the other hand, $\sigma\geq\min(2m+1,n+1)$ without any constraints on $m$.

%%%%%%%%%%%%%%%%%%%%%%%%%%%%%%%%%%%%%%%%%%%%%%%

\subsection{Revisiting nonrelativistic EFTs with enhanced symmetry}
\label{subsec:revisit}

With an improved understanding of the relationship between enhanced symmetry and enhanced soft limits, we now revisit the catalog of theories presented in section~\ref{subsec:catalog}. In table~\ref{tab:catalog} we summarized possible values of $\sigma$ for tree-level scattering amplitudes in various theories with enhanced symmetries. We are now ready to explain the origin of the discrepancy between the ``expected'' and the ``actual'' values of $\sigma$ displayed in the table.

%%%%%%%%%%%%%%%%%%%%%%%%%%%%%%%%%%%%%%%%%%%%%%%

\paragraph{Relativistic Galileon and DBI theories}

The two Lorentz-invariant theories listed at the top of table~\ref{tab:catalog} were included just for reference. Their properties are by now well-studied, and we refer the reader for instance to appendix C of ref.~\cite{Cheung2017a} for more details.\footnote{Curved space generalizations of these theories have also been considered in the literature, see e.g. ref.~\cite{Bonifacio2021dec} for a recent study on aspects of DBI and special Galileon in de Sitter space.} Here we will just briefly repeat that both of these theories belong to the class of theories where the shift polynomial~\eqref{phi} is linear in all coordinates of the Minkowski spacetime, that is $\alpha(x)=\eps_\n x^\n$. This automatically implies $\sigma\geq2$. Within the parameter space of the Galileon theory, it is however possible to choose couplings in such a way that the action possesses an additional hidden symmetry~\cite{Hinterbichler2015a},
\begin{align}
    \theta\rightarrow\theta +\eps^{\mu\nu}(c^2x_{\mu}x_{\nu}-\partial_{\mu}\theta\partial_{\nu}\theta),
\end{align}
where $\eps^{\mu\nu}$ is a constant traceless symmetric tensor and $c$ is a constant. This generalized shift symmetry of degree $n=2$ is sufficient to ensure $\sigma=3$; the resulting theory is known as the special Galileon.

%%%%%%%%%%%%%%%%%%%%%%%%%%%%%%%%%%%%%%%%%%%%%%%

\paragraph{Schr\"odinger-Galileon and Schr\"odinger-DBI theories}

In section~\ref{subsec:classificationcomplex}, we identified two theories of a single complex (Schr\"odinger) scalar of type $B_2$. Both of these possess one layer of redundant generators, generating a linear spatial generalized shift symmetry with $\alpha(x)=\eps_r x^r$. According to eq.~(\ref{B}) these theories must satisfy, 
\begin{align}
\sigma\geq\min(2m,n+1)=2.
\end{align}
This agrees with the values for $\sigma$ reported in table~\ref{tab:catalog}. 

%%%%%%%%%%%%%%%%%%%%%%%%%%%%%%%%%%%%%%%%%%%%%%%

\paragraph{Spatial DBI theory and the Galilei-invariant superfluid}

In section~\ref{subsec:classificationreal}, we identified three theories of a single real scalar of type $A_1$: the spatial DBI and spatial Galileon theory, and the Galilei-invariant superfluid. All of these theories possess a linear spatial generalized shift symmetry with $\alpha(x)=\eps_rx^r$. Based on eq.~\eqref{A}, we would therefore expect that in all of these theories, $\sigma\geq\min(1,2)=1$. This is certainly an improvement over the naive hope raised in ref.~\cite{Brauner2021a} that these theories should feature enhanced soft limits. We still have a job to do though, namely to account for the empirically found values of $\sigma$, displayed in the last column of table~\ref{tab:catalog}. We will treat jointly the spatial DBI theory and the Galilei-invariant superfluid here, and return to the spatial Galileon theory below.

Let us start with the possibility that $\sigma=0$. This violates the bound~\eqref{A}, and in fact the very Adler zero property. As mentioned briefly at the end of section~\ref{subsec:adler}, such an exception may occur when the Noether current contains bilinear contributions. Let us check this explicitly. We shall for the moment put aside the redundant symmetry, linear in spatial coordinates, which is specific to a given theory. Instead, we will focus on a generic EFT, invariant under the constant shift $\theta\to\theta+\eps$. The leading contributions to the effective Lagrangian of such a theory carry one derivative per each factor of $\theta$. The most general rotationally invariant cubic interaction Lagrangian with one derivative per field can be parameterized as 
\begin{align}
    \label{onecubic}
    &\La_3=c_1(\partial_0\theta)^3+c_2\partial_0\theta(\partial_r\theta)^2.
\end{align}
The corresponding bilinear contributions to the Noether current of the constant shift symmetry are obtained by taking a derivative with respect to $\de_\mu\theta$,
\begin{align}
    \label{currentj}
    J^0_2=3c_1(\partial_0\theta)^2+c_2(\partial_{r}\theta)^2, \qquad J^{r}_2=-2c_2\partial_0\theta\partial^{r}\theta.
\end{align}
Then the remainder function $R^\mu(p)$ defined by eq.~\eqref{polecurrent} receives a contribution, coming from insertion of the bilinear current in all possible external legs of the scattering process,
\begin{align}
    \label{R2}
    R_2^{\mu}(p)=\sum_{i} \frac{\im}{(p+q_i)^2}\bra{q_i,-(p+q_i)}J^{\mu}_2(0)\ket{0}\bra{\beta}\ket{\alpha}.
\end{align}
Here $q_i$ denotes the external momenta of the process $\ket\alpha\to\ket\beta$, for simplicity oriented all outwards so that energy--momentum conservation reads $\sum_{i}q_i=0$. Multiplying both sides of eq.~(\ref{R2}) with $p_{\mu}$ and using eq.~(\ref{currentj}) we find
\begin{equation}
\label{6}
p_{\mu}R_2^{\mu}(p)=\im\bra{\beta}\ket{\alpha}\sum_i(p^0+q_i^0)\frac{(3c_1+c_2)p^0q_i^0+2c_2\vek p\cdot\vek q_i}{p\cdot q_i}.
\end{equation}
This does not respect Adler zero unless some cancellations occur. One possibility is that all the fractions on the right-hand side of eq.~\eqref{6} vanish individually. This however requires $c_1=c_2=0$, that is absence of cubic vertices in the Lagrangian. Another, more interesting possibility is that all the fractions are constant, in which case the part of the right-hand side of eq.~\eqref{6} proportional to $q^0_i$ vanishes upon summation over $i$ thanks to energy conservation. The part proportional to $p^0$ will remain nonzero, but will respect the Adler zero property. This possibility can be realized if $c_1=-c_2\equiv c$, leading to
\begin{equation}
p_\mu R^\mu_2(p)=2\im cp^0\bra{\alpha}\ket{\beta}\times(\text{number of particles in $\ket\alpha$ and $\ket\beta$}).
\end{equation}
Hence the only cubic interaction Lagrangian with one derivative per field that yields Adler zero is
\begin{align}
\La_3=c\partial_0\theta(\partial_{\mu}\theta)^2.
\label{cubicAdler}
\end{align}

It is obvious from eqs.~\eqref{DBIaction} and~\eqref{DBInabla} that the spatial DBI theory admits cubic couplings that are not of the form~\eqref{cubicAdler}. For such choice of couplings, the Adler zero is violated and we find $\sigma=0$. Setting $c_3=0$ in eq.~\eqref{DBIaction} however eliminates the unwanted cubic vertex. In such a restricted parameter space, the spatial DBI theory has generically $\sigma=1$ in accord with eq.~\eqref{A}. Furthermore, the relativistic DBI theory is a special case of the spatial DBI theory~\eqref{DBIaction} in which the couplings $c_k$ are tuned in such a way so as to recover Lorentz invariance. Upon such fine tuning, we find $\sigma=2$. This explains all the possible values of $\sigma$ in the spatial DBI theory, shown in table~\ref{tab:catalog}.

A very similar argument applies to the Galilei-invariant superfluid, cf.~eq.~\eqref{Lsuperfluid}. Here $c_1=1$ and $c_2=1/2$ are fixed by the normalization of the kinetic term. For a generic value of $c_3$, we find a cubic term that is not of the form~\eqref{cubicAdler} and hence $\sigma=0$. However, the Adler zero can be saved by setting $c_3=1/2$, for which we then generally find $\sigma=1$ in accord with eq.~\eqref{A}. Remarkably, even with the inevitable cubic interaction vertex, there is still a very particular tuning of the couplings of the Galilei-invariant superfluid which yields $\sigma=2$. We will explain the origin of this special case in section~\ref{subsec:LagscanA1}.

%%%%%%%%%%%%%%%%%%%%%%%%%%%%%%%%%%%%%%%%%%%%%%%

\paragraph{Spatial Galileon theory}

Let us finally have a look at the spatial Galileon theory. Here the spatial linear shift symmetry, $\theta\to\theta+\eps_rx^r$, admits an interaction term proportional to $(\de_0\theta)^3$. In accord with the above discussion, this will necessarily violate the Adler zero and thus give $\sigma=0$. There is however another cubic interaction vertex, corresponding to the $k=2$ WZ term in eq.~\eqref{WZtermsGalileon}. Let us look at the properties of this interaction more closely.

As the first step, it is convenient to rewrite the $k=2$ WZ term in a way that makes invariance under the constant shift symmetry, $\theta\to\theta+\eps$, manifest. Using integration by parts, we can bring the cubic interaction Lagrangian to the form\footnote{Here the subscript $3$ on $\La$ indicates a cubic interaction vertex, not the $k=3$ WZ term~\eqref{WZtermsGalileon}.}
\begin{equation}
\La_3=c\de_r\theta\de_s\theta\de^r\de^s\theta,
\end{equation}
where $c$ is a generic coupling constant. Owing to the fact that this operator carries no time derivatives, it only gives a bilinear contribution to the spatial part of the Noether current,
\begin{equation}
J_2^r=c(\de_s\theta\de^r\de^s\theta-\de^r\theta\de_s\de^s\theta).
\end{equation}
Inserting the current in all external legs of the scattering process $\ket\alpha\to\ket\beta$ again gives a contribution to the remainder function $R^\mu(p)$, given by eq.~\eqref{R2}. Working out the details, we get the spatial Galileon equivalent of eq.~\eqref{6},
\begin{equation}
p_\mu R^\mu_2(p)=\im c\bra{\beta}\ket{\alpha}\sum_i\frac{(\vek p\cdot\vek q_i)^2-\vek p^2\vek q_i^2}{p\cdot q_i}.
\end{equation}
In the soft limit, this gives a contribution to the scattering amplitudes of the NG boson that scales with $\sigma=1$.

Finally, there is a special tuning of the couplings of the spatial Galileon theory which leads to $\sigma=2$. This can be achieved by dropping the cubic WZ term and only including the $k\geq3$ WZ terms in the Lagrangian~\eqref{spatialgalileon}. In this special case, the temporal part of the remainder function, $R^0(p)$, vanishes simply because $J^0=\de_0\theta$ is a noninteracting current that only gives the pole contribution in eq.~\eqref{polecurrent}. Then, in accord with the discussion at the end of section~\ref{subsec:enhsofbeh}, we find enhanced scaling with $\sigma=2$ in spite of the fact that the spatial Galileon is a type $A_1$ theory. We should however note that such a fine tuning of the couplings is not protected by symmetry. We therefore expect the enhanced soft behavior to be destroyed by radiative corrections. More generally, the argument presented at the end of section~\ref{subsec:enhsofbeh}, based on the assumption $R^0(p)=0$, is expected to only hold at tree level.

%%%%%%%%%%%%%%%%%%%%%%%%%%%%%%%%%%%%%%%%%%%%%%%

\section{Soft recursion}
\label{sec:recursion}

Employing different classification and organizing principles for EFTs can sometimes help illuminate nontrivial relations among different theories. Having considered scalar nonrelativistic EFTs with enhanced soft limits from a top-down symmetry perspective, we now shift gears and initiate their bottom-up amplitude study. Bottom-up approaches to scattering amplitudes in relativistic EFTs have provided novel perspectives on important theories appearing in different areas of the modern $S$-matrix program. For instance, the special Galileon theory~\cite{Cheung2015a,Hinterbichler2015a,Novotny2017a} was first discovered in the context of soft limits of scattering amplitudes. Moreover, the exceptional scalar EFTs obtained from soft bootstrap studies~\cite{Cheung2017a} coincide with the EFTs constructed from the Cachazo-He-Yuan (CHY) representation~\cite{Cachazo2014a,Cachazo2014c} and are precisely the scalar EFTs that are known to satisfy Bern-Carrasco-Johansson (BCJ) duality~\cite{Bern2010,Bern2019}. These results suggest a rich interplay between CHY representation, BCJ duality, and soft limits~\cite{Carrasco2016,Cheung2018a,Gonzalez2019,Low2019b,Low2020a}. In addition, new insights into the soft structure of the $S$-matrix have been obtained in the recently established program of asymptotic symmetries~\cite{Strominger2013,Cachazo2014b,Cheung2016b,Pasterski2016,Strominger2017,Arkani-Hamed2020} and from studying the geometry of field space~\cite{Volkov1973a,Cheung2021}.

Establishing a bottom-up approach to nonrelativistic scalar EFTs with enhanced soft limits pursues multiple goals. First of all, it will be interesting to see how the bottom-up picture complements the top-down approach developed in sections~\ref{sec:2} and~\ref{sec:sym}, and allows us to tie up the loose ends the latter has left. Second, the bottom-up approach will also shed some light on the role that Lorentz invariance plays in EFTs with exceptional soft behavior. More ambitiously, in light of the rich results from relativistic bootstrapping one may hope that nonrelativistic bootstrap will provide a foundation for extending the $S$-matrix program to theories without Lorentz invariance.

In the following two sections, we will explore the landscape of nonrelativistic EFTs through various bootstrapping techniques. These rely heavily on the on-shell recursion technology, recently extended to nonrelativistic EFTs by the present authors~\cite{us}. In this prequel to the present paper, we showed in particular that it is possible to recursively reconstruct the tree-level $S$-matrix of nonrelativistic EFTs with enhanced soft limits. In the rest of this section, we will briefly review and slightly extend the nonrelativistic on-shell recursion machinery established in ref.~\cite{us}. This will prepare the ground for the bootstrap analysis in the following two sections. For the sake of simplicity, the discussion is restricted to EFTs for NG bosons with linear or quadratic dispersion relation, although the general formalism is applicable to all type $A$ and type $B$ theories.  

%%%%%%%%%%%%%%%%%%%%%%%%%%%%%%%%%%%%%%%%%%%%%%%

\subsection{Soft shifts: linear dispersion relation}
\label{sec:softshift1}

Soft momentum shifts are complex-valued deformations of external momenta in a scattering process which maintain on-shell conditions and conservation of total energy and momentum while probing the soft limit for the external momenta. Apart from playing an integral role in recursion relations, the soft momentum shifts also provide an invaluable tool for studying the landscape of EFTs on its own.

We start the discussion with theories of type $A_1$ NG bosons. Apart from trivial rescaling of coordinates, these have kinematics that is identical to Lorentz-invariant theories; possible breaking of Lorentz invariance only shows in the particle interactions.\footnote{This is true for theories of a single flavor of NG boson, or for multi-flavor theories where all NG boson species have the same speed of propagation.} What follows below is therefore essentially just a brief review of soft momentum shifts as developed in ref.~\cite{Cheung2017a}.

In the rest of this section, we will use a convention distinguishing particles entering and leaving the scattering process by a sign. We will choose $e_i=-1$ for particles in the initial state and $e_i=+1$ for particles in the final state. The laws of conservation of energy and momentum for type $A_1$ NG bosons are thus subsumed into the four-vector condition
\begin{equation}
\sum_{i=1}^ne_ip_i=0,
\label{EMconservation}
\end{equation}
where $n$ is the number of particles participating in the scattering process and $p_i$ are their momenta. Finally, we will use the symbol $D\equiv d+1$ to denote the dimension of spacetime.

%%%%%%%%%%%%%%%%%%%%%%%%%%%%%%%%%%%%%%%%%%%%%%%

\paragraph{All-line shift}

The \emph{all-line soft shift} of external momenta is defined by
\begin{align}
    &\hat{p}_i\equiv p_i(1-za_i), \qquad 1\leq i\leq n,
\end{align}
where $z\in\C$ is a parameter. Imposing energy and momentum conservation on the shifted variables $\hat p_i$ implies the following constraints on the coefficients $a_i$, 
\begin{align}
    \label{all-line-2}
    \sum_{i=1}^ne_ia_ip_i=0.
\end{align}
Simple linear algebra guarantees that the set of equations~(\ref{all-line-2}) for the variables $a_i$ (with $p_i$ being fixed) has a solution space of dimension $n-D$. However, a one-dimensional subspace of solutions corresponds to all the $a_i$s being equal; this is guaranteed by the energy and momentum conservation~\eqref{EMconservation}. Since we would eventually like to use the soft momentum shift to probe the single soft limit for the individual particles, we need all the $a_i$s to be different. We are thus forced to mod out the subspace of trivial solutions. The existence of nontrivial all-line shifts for a generic kinematical configuration therefore requires that $n\geq D+2$.

%%%%%%%%%%%%%%%%%%%%%%%%%%%%%%%%%%%%%%%%%%%%%%%

\paragraph{All-but-one-line shift}

As the name suggests, the \emph{all-but-one-line soft shift} treats one of the $n$ momenta differently from the others. This shift allows one to probe the single soft limit of $n-1$ particles in the scattering process. It is defined by
\begin{align}
    \hat{p}_i&\equiv p_i(1-za_i), \qquad 1\leq i\leq n-1, \\
    \hat{p}_n&\equiv p_n+zq_n,
\end{align}
where momentum conservation and on-shell conditions imply the following constraints,
\begin{align}
    e_nq_n=\sum_{i=1}^{n-1}e_ia_ip_i, \qquad q_n^2=p_n\cdot q_n=0.
    \label{qndef}
\end{align}
We can view the first relation in eq.~\eqref{qndef} as a definition of $q_n$. The second relation therein then constitutes two homogeneous constraints (one linear and one quadratic) on the $n-1$ unknown variables $a_i$. Possible solutions for $a_i$ therefore carve out an $(n-3)$-dimensional surface in $\R^{n-1}$. This necessarily includes a one-dimensional subspace where all the $a_i$s are equal. The existence of other solutions, where the $a_i$s are different, can therefore only be guaranteed if $n-3>1$. We conclude that the all-but-one-line shift is only useful if $n\geq5$, regardless of the spacetime dimension $D$.

%%%%%%%%%%%%%%%%%%%%%%%%%%%%%%%%%%%%%%%%%%%%%%%

\paragraph{All-but-two-line shift}

Finally, the \emph{all-but-two-line soft shift} allows one to access the single soft limit of $n-2$ particles in the scattering process. It is defined by
\begin{align}
    \hat{p}_i&\equiv p_i(1-za_i), \qquad 1\leq i\leq n-2, \\
    \hat{p}_{n-1}&\equiv p_{n-1}+zq_{n-1},\\
    \hat{p}_n&\equiv p_n+zq_n,
\end{align}
where momentum conservation and on-shell conditions imply the constraints
\begin{align}
    e_{n-1}q_{n-1}+e_nq_n=\sum_{i=1}^{n-2}e_ia_ip_i, \qquad q_{n-1}^2=q_n^2=p_{n-1}\cdot q_{n-1}=p_n\cdot q_n=0.
    \label{qndef2}
\end{align}
Here we can choose the $a_i$s arbitrarily and thus ensure that they are all different as desired. Equation~\eqref{qndef2} then constitutes $D+4$ constraints on the $2D$ unknown components of $q_{n-1}$ and $q_n$. Hence for $D\geq4$ we can always find an all-but-two-line shift that allows us to probe the single soft limit of the first $n-2$ particles. For $D=3$ we have to be a bit more careful, since we can no longer pick all the $a_i$s arbitrarily. We can however still fix $a_i$ with $1\leq i\leq n-3$. Then $a_{n-2}$ together with $q_{n-1}$ and $q_n$ constitute $2D+1=7$ variables, constrained by the $D+4=7$ relations in eq.~\eqref{qndef2}. However, for $n=4$ (where we only choose $a_1$ beforehand) we cannot exclude the possibility that the solution we find for $a_2$ coincides with $a_1$. In $D=3$, the existence of an all-but-two-line shift with different $a_i$s is therefore only guaranteed for $n\geq5$.

%%%%%%%%%%%%%%%%%%%%%%%%%%%%%%%%%%%%%%%%%%%%%%%

\subsection{Soft shifts: quadratic dispersion relation}
\label{sec:softshift2}

Let us now see how to define soft momentum shifts for NG bosons with a quadratic dispersion relation, that is either type $A_2$ or type $B_2$. The line of reasoning follows closely the steps we took for NG bosons with a linear dispersion relation. We can therefore afford to be more concise.

%%%%%%%%%%%%%%%%%%%%%%%%%%%%%%%%%%%%%%%%%%%%%%%

\paragraph{All-line shift}

The all-line soft shift for theories with a quadratic dispersion reads
\begin{align}
    \boldsymbol{\hat{p}}_i&\equiv \boldsymbol{p}_i(1-a_iz),\\
    \hat{p}^0_i&\equiv\boldsymbol{\hat{p}}_i^{2}=\boldsymbol{p}_i^{2}(1-a_iz)^{2}.
\end{align}
Momentum and energy conservation impose the following constraints,
\begin{align}
    \label{B2m11}
    \sum_{i=1}^ne_ia_i\boldsymbol{p}_i&=0, \\
    \label{B2m22}
    \sum_{i=1}^ne_i(1-za_i)^{2}\boldsymbol{p}_i^{2}&=0.
\end{align}
Similar to the all-line soft shift in type $A_1$ theories, the existence of nontrivial solutions to eq.~(\ref{B2m11}) requires $n\geq d+2=D+1$. The two additional constraints imposed by eq.~(\ref{B2m22}) imply that nontrivial all-line soft shifts for NG bosons with quadratic dispersion relation require $n\geq d+4=D+3$. 

%%%%%%%%%%%%%%%%%%%%%%%%%%%%%%%%%%%%%%%%%%%%%%%

\paragraph{All-but-one-line shift}

The all-but-one-line soft shift takes the following form, 
\begin{align}
    \label{all-but-one-1}
    \boldsymbol{\hat{p}}_i&\equiv \boldsymbol{p}_i(1-a_iz), \qquad &&1\leq i\leq n-1,\\
    \label{all-but-one-2}
    \hat{p}^0_i&\equiv\boldsymbol{\hat{p}}_i^{2}=\boldsymbol{p}_i^{2}(1-a_iz)^{2}, \qquad &&1\leq i\leq n-1,\\
    \label{all-but-one-3}
    \hat{\boldsymbol{p}}_n&\equiv \boldsymbol{p}_n+z\boldsymbol{q}_n, \\
    \label{all-but-one-4}
    \hat{p}^0_n&\equiv\hat{\vek p}_n^2= p_n^0+2z\boldsymbol{p}_n\cdot\boldsymbol{q}_n+z^2\boldsymbol{q}_n^2. 
\end{align}
Momentum and energy conservation require the following conditions to be satisfied,
\begin{align}
    \label{B2m1}
    \sum_{i=1}^{n-1}e_ia_i\boldsymbol{p}_i&=e_n\boldsymbol{q}_n, \\
    \label{B2m2}
    \sum_{i=1}^{n-1}e_ia_i\boldsymbol{p}_i^2&=e_n\boldsymbol{p}_n\cdot\boldsymbol{q}_n, \\
    \label{B2m3}
    \sum_{i=1}^{n-1}e_ia_i^2\boldsymbol{p}_i^2+e_n\boldsymbol{q}_n^2&=0.
\end{align}
The existence of a nontrivial all-but-one-line soft shift can be asserted using the same argument as for type $A_1$ theories. We treat eq.~\eqref{B2m1} as a definition of $\vek q_n$. The remaining equations~\eqref{B2m2} and~\eqref{B2m3} then constitute two homogeneous constraints on the $n-1$ variables $a_i$. Given that setting all the $a_i$s equal to each other still generates a one-dimensional subspace of solutions, we end up with the lower bound $n\geq5$ required for a nontrivial solution with different $a_i$s to exist.

%%%%%%%%%%%%%%%%%%%%%%%%%%%%%%%%%%%%%%%%%%%%%%%

\paragraph{All-but-two-line shift}
Finally, the all-but-two-line soft shift is defined by
\begin{align}
    \boldsymbol{\hat{p}}_i&\equiv \boldsymbol{p}_i(1-a_iz), \qquad &&1\leq i\leq n-2,\\
    \hat{p}^0_i&\equiv\boldsymbol{\hat{p}}_i^{2}=\boldsymbol{p}_i^{2}(1-a_iz)^{2}, \qquad &&1\leq i\leq n-2,\\
    \hat{\boldsymbol{p}}_{n-1}&\equiv \boldsymbol{p}_{n-1}+z\boldsymbol{q}_{n-1}, \\
    \hat{p}^0_{n-1}&\equiv\hat{\boldsymbol{p}}_{n-1}^2= p_{n-1}^0+2z\boldsymbol{p}_{n-1}\cdot\boldsymbol{q}_{n-1}+z^2\boldsymbol{q}_{n-1}^2, \\
    \hat{\boldsymbol{p}}_n&\equiv \boldsymbol{p}_n+z\boldsymbol{q}_n, \\
    \hat{p}^0_n&\equiv\hat{\boldsymbol{p}}_n^2= p_n^0+2z\boldsymbol{p}_n\cdot\boldsymbol{q}_n+z^2\boldsymbol{q}_n^2.
\end{align}
Momentum and energy conservation impose the following constraints,
\begin{align}
    \sum_{i=1}^{n-2}e_ia_i\boldsymbol{p}_i&=e_{n-1}\boldsymbol{q}_{n-1}+e_n\boldsymbol{q}_{n}, \\
    \sum_{i=1}^{n-2}e_ia_i\boldsymbol{p}_i^2&=e_{n-1}\boldsymbol{p}_{n-1}\cdot\boldsymbol{q}_{n-1}+e_n\boldsymbol{p}_n\cdot\boldsymbol{q}_n, \\
    \sum_{i=1}^{n-2}e_ia_i^2\boldsymbol{p}_i^2+e_{n-1}\boldsymbol{q}_{n-1}^2+e_n\boldsymbol{q}_n^2&=0.
\end{align}
Here we can assign the $a_i$s arbitrary (different) values. The vectors $\vek q_{n-1}$ and $\vek q_n$ then constitute $2d$ unknown variables, constrained by $d+2$ relations implied by energy and momentum conservation. This ensures the existence of a solution for any $d\geq2$. In order that the solution does not degenerate to the all-line shift with equal $a_i$s, we require that we can choose at least two different $a_i$s a priori, that is $n\geq4$.

%%%%%%%%%%%%%%%%%%%%%%%%%%%%%%%%%%%%%%%%%%%%%%%

\subsection{Recursion relations}
\label{sec:4.3}

We now have all the tools we need to derive recursion relations for type $A_1$ and type $B_2$ EFTs with enhanced soft limit, following closely ref.~\cite{us}. As the first step towards deriving a soft recursion relation for an $n$-particle tree-level scattering amplitude $A_n$, we promote it to a complex function $\hat{A}_n(z)$ of the shifted momenta $\hat p_i$. The original amplitude $A_n=\hat{A}_n(0)$ can be recovered by means of a contour integral, 
\begin{align}
    \label{contour}
    A_n=\frac{1}{2\pi\im}\oint\dd z \frac{\hat{A}_n(z)}{zF_{n_s}(z)},
\end{align}
where the contour is an infinitesimal circle enclosing the origin of the complex plane. The denominator factor $F_{n_s}(z)$ is defined as
\begin{align}
    F_{n_s}(z)=\prod_{i=1}^{n_s}(1-a_iz)^{\sigma},
\end{align}
where $n_s$ denotes the number of external legs whose single soft limit is accessible by the momentum shift employed.\footnote{This means explicitly that $n_s=n$ for the all-line shift, $n_s=n-1$ for the all-but-one-line shift, and $n_s=n-2$ for the all-but-two-line shift.} The integrand $\hat{A}_n(z)/F_{n_s}(z)$ is designed so that the poles from $F_{n_s}(z)$ are cancelled by the soft behavior of $\hat{A}_n(z)$, since the latter by assumption scales like $(1-a_iz)^{\sigma}$ in the soft limit of the $i$-th particle, $z\rightarrow 1/a_i$. Thus, the sole singularities in $\hat{A}_n(z)/F_{n_s}(z)$ are those stemming from the unitarity poles corresponding to the factorization channels of $\hat{A}_n(z)$, plus possibly the pole at infinity. If the contribution of the latter vanishes, then $A_n$ is simply equal to the sum of residues at the factorization channel poles,
\begin{align}
    \label{NRrec}
    A_n=-\sum_I\sum_{i=1}^{2}\Res_{z=z_I^{i}}\frac{\hat{A}_n(z)}{zF_{n_s}(z)}.
\end{align}
Here $I$ labels different factorization channels and $z_I^i$, $i=1,2$ are the solutions to the on-shell condition, which is a quadratic polynomial in $z$,\footnote{There is a mathematical possibility that the leading, $z^2$ term of the polynomial vanishes. It is normally possible to avoid this by shifting the momentum shift parameters by a suitable constant, $a_i\to a_i+c$. This will work as long as the intermediate momentum $P_I$ is not identically zero, which is a kinematical singularity that we will not include in our considerations. }
\begin{align}
    \bigl(\hat{P}_I^0\bigr)^2-\hat{\boldsymbol{P}}_I^{2}&=0 && \text{for type }A_1,  \\
    \hat{P}_I^0-\hat{\boldsymbol{P}}_I^{2}&=0 && \text{for type }B_{2},
\end{align}
where the intermediate momentum $P_I$ is defined by
\begin{align}
    P_I\equiv \sum_{i\in I}e_ip_i.
\end{align}
Factorization dictates that on the pole at $z=z^i_I$, the amplitude $\hat A_n(z)$ in eq.~(\ref{NRrec}) can be expressed in terms of on-shell lower-point amplitudes $A_L^{(I)}$ and $A_R^{(I)}$ as $\hat{A}_L^{(I)}(z)\hat{A}_R^{(I)}(z)/D^{(I)}(z)$, where $D^{(I)}$ denotes the inverse propagator associated with the factorization channel $I$,
\begin{align}
    D^{(I)}(z)&\equiv\bigl(\hat{P}_I^0\bigr)^2-\hat{\boldsymbol{P}}_I^{2} && \text{for type } A_1,  \\
    D^{(I)}(z)&\equiv\hat{P}_I^0-\hat{\boldsymbol{P}}_I^{2} && \text{for type } B_{2}.
\end{align}
It follows that
\begin{align}
    \label{Angeneral}
    A_n=-\sum_I\sum_{i=1}^{2}\Res_{z=z_I^{i}}\frac{\hat{A}_L^{(I)}(z)\hat{A}_R^{(I)}(z)}{zF_{n_s}(z)D^{(I)}(z)}.
\end{align}

The recursion relation~\eqref{Angeneral} is suitable for reconstruction of higher-point amplitudes from lower-point ones using symbolic computation. For manual calculation of scattering amplitudes via recursion, a different way of evaluating $A_n$ is often more convenient. To that end, we observe that the contribution to eq.~\eqref{Angeneral} from the factorization channel $I$ matches the sum of residues at $z=z_I^i$, $i=1,2$ of the meromorphic function
\begin{align}
    \label{analyticf}
    \frac{\hat{A}_L^{(I)}(z)\hat{A}_R^{(I)}(z)}{zF_{n_s}(z)D^{(I)}(z)}.
\end{align}
This function may also have nonvanishing residues at $z=1/a_i$ and $z=0$ due to the fact that $\hat{A}_L^{(I)}(z)$ and $\hat{A}_R^{(I)}(z)$ are off-shell for $z\neq z_I^i$, and therefore no longer necessarily cancel the zeros of $F_{n_s}(z)$. In the special case where $\hat{A}_L^{(I)}(z)$ and $\hat{A}_R^{(I)}(z)$ are both local functions of momenta, i.e.~have no poles, one can apply Cauchy's theorem to the function~(\ref{analyticf}) to recast the amplitude $A_n$ as a sum over residues at $z=0$ and $z=1/a_i$ alone,
\begin{align}
    \label{master}
    A_n=&\sum_I\frac{\hat{A}_L^{(I)}(0)\hat{A}_R^{(I)}(0)}{D^{(I)}(0)}+\sum_I\sum_{i=1}^{n_{s}}\Res_{z=1/a_i}\frac{\hat{A}_L^{(I)}(z)\hat{A}_R^{(I)}(z)}{zF_{n_s}(z)D^{(I)}(z)}.
\end{align}
The first term on the right-hand side of eq.~(\ref{master}) corresponds to the sum over Feynman diagrams with an internal propagator line, whereas the second (double) sum corresponds to contributions from $n$-point contact operators. 

The above-derived expressions for $A_n$ rely on the absence of a pole at infinity. A simple way to check whether the contribution at infinity really vanishes is by performing a uniform rescaling of momenta $\{\vek p_i\}\rightarrow  \{\lambda\vek p_i\}$. If the $n$-point amplitude scales as $A_n\rightarrow \lambda^mA_{n}$, then it is guaranteed that $\hat{A}_n(z)/F_{n_s}(z)$ \textit{at worst} goes like $z^{m-n_{s}\sigma}$ at large $z$. Thus, the contribution to $A_n$ from the contour at infinity vanishes if
\begin{align}
\label{valid}
m<n_s\sigma.
\end{align}
A more precise sufficient criterion for eq.~(\ref{valid}) to be satisfied was put forward in ref.~\cite{us}.

It is worth pointing out that soft recursion relations are also applicable to theories with universal albeit not necessarily vanishing soft behavior, see refs.~\cite{Luo2016a,Rodina_2019,Cheung2021} and references therein for details. The principal idea then is to redefine $F_{n_s}(z)$ so that the contribution from the pole at infinity still vanishes, and use subleading terms from soft theorems to determine additional poles from $F_{n_s}(z)$ in the contour integral~(\ref{contour}). Soft theorems encoding information beyond the soft scaling of scattering amplitudes can be used as input in soft bootstrap techniques to extend the exploration of the EFT landscape~\cite{Kampf:2020gal,Kampf:2021multi}, but this is beyond the scope of the present work.

%%%%%%%%%%%%%%%%%%%%%%%%%%%%%%%%%%%%%%%%%%%%%%%

\section{Soft bootstrap}
\label{sec:bootstrap}

In this and the next section we use the soft recursion relations developed in section~\ref{sec:recursion} to examine the existence of single-flavor EFTs with nontrivial soft behavior from a bottom-up perspective. This represents a first extension of the soft bootstrap program to nonrelativistic (effective) field theories. The soft bootstrap program is already well established in the literature, and detailed treatments of it with various applications can be found e.g.~in refs.~\cite{Elvang2018a, Elvang:2018s, Low2019a, Elvang:2020lue} and references therein. The following two paragraphs give a brief working introduction to soft bootstrap techniques. Applications to the landscape of type $A_1$ and type $B_2$ theories follow respectively in sections~\ref{subsec:bootstrapA1} and~\ref{subsec:bootstrapB2}.

The standard approach to exploring the landscape of EFTs through soft bootstrap can be understood as a two-step procedure. The first step involves writing down all possible amplitudes $A_m$ with $m$ smaller than some fixed value $n$, which are consistent with the imposed spectrum of particles, symmetries, and soft scaling properties. We will refer to these amplitudes as \textit{seed amplitudes}. Normally $n$ is so small that all seed amplitudes are polynomials in the particle momenta and energies. The problem of classifying all possible seed amplitudes with the given properties is then equivalent to classifying all contact operators in the Lagrangian with $m$ fields, modulo equations of motion and integration by parts. In the present work the problem of classifying seed amplitudes never exceeds a level of difficulty that is tractable by brute force. However, we point out that there are powerful mathematical tools that can be used to identify all independent kinematical polynomials with given symmetries and kinematical constraints in a systematic manner~\cite{Henning:2015,Henning:2017}. 

In the second step, the seed amplitudes are used as input in the soft recursion relations. The result is supposed to be a physical amplitude, and must therefore be independent of the unphysical momentum shift parameters $a_i$. Any dependence on $a_i$ rules out the existence of a consistent theory with the given seed amplitude. If, on the other hand, the result of recursion is $a_i$-independent, then there \textit{might} be an underlying theory with the assumed properties. One cannot use soft bootstrap as a tool to rigorously assert the existence of a consistent theory, as that would require extending the recursion inductively to all $n$. This already hints that soft bootstrap is best suited for narrowing down the landscape of candidate theories with given soft behavior.

%%%%%%%%%%%%%%%%%%%%%%%%%%%%%%%%%%%%%%%%%%%%%%%

\subsection{Type $A_1$ bootstrap}
\label{subsec:bootstrapA1}

The discussion in sections~\ref{sec:softshift1} and~\ref{sec:4.3} shows that in $D\leq4$ dimensions, which is the case of most physical interest, all amplitudes with $n\geq6$ can be reconstructed from lower-point seeds using recursion based on the all-line soft shift. We do not need to take into account possible seed three-point amplitudes. First, in Lorentz-invariant theories, there are no such amplitudes due to kinematical constraints. Second, we argue below in section~\ref{subsec:3pt} that even in nonrelativistic type $A_1$ theories, existence of any three-point amplitude necessarily leads to trivial soft behavior, whereby the soft scaling parameter $\sigma$ cannot exceed the value implied by mere counting of derivatives in the contact operator. It thus remains to classify all possible seed four-point and five-point amplitudes. We shall now elaborate the $n=4$ case in detail, and return to the $n=5$ case below.

%%%%%%%%%%%%%%%%%%%%%%%%%%%%%%%%%%%%%%%%%%%%%%%

\subsubsection{Seed four-point amplitudes}
\label{subsec:seedA1}

For the sake of simplicity, we will replace the notation $p^0_i$ for the particle energies with $\om_i$. We will return to the convention whereby all the four-momenta are treated by default as outgoing; incoming particles then carry negative energy. This will simplify the implementation of the permutation (Bose) symmetry of the seed amplitudes. Rotational invariance requires that the seed amplitude $A_4$ is a polynomial in the energies $\om_i$ and the dot products $\vek p_i\cdot\vek p_j$. The latter can always be expressed in terms of the relativistic Mandelstam variables,
\begin{equation}
s\equiv(p_1+p_2)^2=2p_1\cdot p_2,\quad
t\equiv(p_1+p_3)^2=2p_1\cdot p_3,\quad
u\equiv(p_1+p_4)^2=2p_1\cdot p_4,
\end{equation}
and the energies. Altogether, the seed amplitude $A_4$ can therefore be sought as a polynomial in $s,t,u$ and $\om_i$ that is invariant under the action of the permutation group $S_4$, modulo the energy--momentum conservation constraints
\begin{equation}
s+t+u=0,\qquad
\om_1+\om_2+\om_3+\om_4=0.
\label{ideal}
\end{equation}

The classification of seed four-point amplitudes now in principle reduces to an exercise in group theory. The set of monomials $\{s,t,u\}$ carries a representation, $R_1$, of $S_4$; the set of monomials $\{\om_1,\om_2,\om_3,\om_4\}$ carries a representation $R_2$ thereof. In order to ensure a nontrivial realization of an enhanced soft limit of the four-point amplitude, that is $\sigma=2$, we can restrict to polynomials of total degree in momenta less than $8$. This amounts to inspecting the tensor products $R_1^{\otimes m}\otimes R_2^{\otimes n}$ with $2m+n<8$, and finding all singlets of $S_4$ in their decomposition into irreducible representations.

In practice, the problem can be solved even more straightforwardly using symbolic computation, without the theory of group representations. It is convenient to start with the monomial bases $\{\bar s,\bar t,\bar u\}$ and $\{\bar\om_1,\bar\om_2,\bar\om_3,\bar\om_4\}$, defined by subtracting respectively $(s+t+u)/3$ and $(\om_1+\om_2+\om_3+\om_4)/4$ from each element. This ensures that we do not carry trivial components vanishing by energy and momentum conservation throughout the calculation. In the next step, we form a Kronecker product of $m$ factors of the Mandelstam basis and $n$ factors of the energy basis, and symmetrize the result with respect to permutations from $S_4$. The resulting set of symmetric polynomials is typically largely redundant and only contains a handful of linearly independent elements. In the last step, the resulting candidate seed amplitudes have to be inspected one by one for the value of $\sigma$ they imply.

\begin{table}[t]
\centering
\begin{tabular}{c|c|c}
$(m,n)$ & Label & Amplitude\\
\hline
\hline
$(2,0)$ & $A_4^{(1)}$ & $s^2+t^2+u^2$\\
$(2,2)$ & $A_4^{(2)}$ & $(s^2+t^2+u^2)(\om_1^2+\om_2^2+\om_3^2+\om_4^2)$\\
$(2,2)$ & $A_4^{(3)}$ & $s^2(\om_1\om_2+\om_3\om_4)+t^2(\om_1\om_3+\om_2\om_4)+u^2(\om_1\om_4+\om_2\om_3)$\\
$(2,3)$ & $A_4^{(4)}$ & $(s^2+t^2+u^2)(\om_1\om_2\om_3+\om_1\om_2\om_4+\om_1\om_3\om_4+\om_2\om_3\om_4)$\\
$(3,0)$ & $A_4^{(5)}$ & $s^3+t^3+u^3$
\end{tabular}
\caption{Seed four-point amplitudes in type $A_1$ theories with $2m+n<8$ and the soft scaling parameter $\sigma\geq2$. Here $s,t,u$ are the usual relativistic Mandelstam variables, whereas $\omega_i$ are the energies of the participating particles.}
\label{tab:seedA1}
\end{table}

We find altogether five independent amplitudes with the required properties, as displayed in table~\ref{tab:seedA1}. Some of them are easy to identify. The only two seed four-point amplitudes that are manifestly Lorentz-invariant are $A_4^{(1)}$ and $A_4^{(5)}$. These satisfy respectively $\sigma=2$ and $\sigma=3$ and correspond to the relativistic DBI and special Galileon theories~\cite{Cheung2015a}. Note that both of these theories are exceptional in the sense that they correspond to the lowest possible number of derivatives per field for given $\sigma$. Furthermore, it is easy to check that the quartic ($k=3$) WZ term~\eqref{WZtermsGalileon} of the spatial Galileon theory generates an amplitude that is a linear combination of $A_4^{(3)}$ and $A_4^{(5)}$. It remains to clarify what, if any, consistent EFTs the other amplitudes in table~\ref{tab:seedA1} correspond to.

Before doing so, let us remark that rotationally invariant four-point amplitudes of a single real massless scalar of type $A_1$ were classified recently in ref.~\cite{Bonifacio:2021} using the Hilbert series techniques~\cite{Henning:2015,Henning:2017}. Our result is consistent with theirs in that (i) there is a single four-derivative amplitude ($E_2$) with $\sigma=2$, (ii) there are two six-derivative amplitudes ($E_2e_2$, $S_2$) with $\sigma=2$ and one ($E_3$) with $\sigma=3$, (iii) there is one additional seven-derivative amplitude ($E_2e_3$) with $\sigma=2$. Finding the exact mapping between our result and that of ref.~\cite{Bonifacio:2021} is impeded by the fact that they represent the amplitudes by polynomials in $\om_i$ and $(\vek p_i+\vek p_j)^2$. Choosing the basis monomials as we do here makes it straightforward to identify the subset of Lorentz-invariant seed amplitudes.

%%%%%%%%%%%%%%%%%%%%%%%%%%%%%%%%%%%%%%%%%%%%%%%

\subsubsection{Consistency constraints from soft bootstrap}

Let us now inspect the full set of candidate seed amplitudes as displayed in table~\ref{tab:seedA1}. We employ the recursion relation~\eqref{Angeneral} to generate the six-point amplitude from the seed four-point amplitude. In order not to exclude a priori any possibility, we have to allow the latter to be an arbitrary linear combination of the basis amplitudes in table~\ref{tab:seedA1},
\begin{equation}
A_4=\sum_{i=1}^5c_iA_4^{(i)}.
\end{equation}
Based on the symmetry approach laid out in sections~\ref{sec:2} and~\ref{sec:sym}, we do not expect it to be possible, for instance, to combine the seed amplitudes of the DBI and Galileon theories. However, it is one of the goals of our soft bootstrap analysis to check whether there might possibly be other theories featuring nontrivial enhanced soft limits than those predicted by the Lie-algebraic classification of section~\ref{sec:2}.

The result is simple to state. We find that independence of the six-point amplitude on the momentum shift parameters $a_i$ requires $c_2=c_4=0$. The amplitudes $A_4^{(2)}$ and $A_4^{(4)}$ therefore do not correspond to any physically consistent theory. Moreover, $c_1c_3=0$, which means that the amplitudes $A_4^{(1)}$ and $A_4^{(3)}$ are mutually exclusive; no nontrivial linear combination of them is consistent. The contribution of $A_4^{(5)}$ turns out to be unconstrained at the six-point level. Further constraints might be obtained by extending the soft bootstrap to the eight-point amplitude. This however cannot be done without first considering possible seed five-point amplitudes.

Altogether, our soft bootstrap program for type $A_1$ EFTs has turned out partially successful. We have been able to very efficiently isolate mere three seed four-point amplitudes, $A_4^{(1)}$, $A_4^{(3)}$ and $A_4^{(5)}$, two of which are Lorentz-invariant. We even have three specific EFTs that produce such amplitudes: the relativistic DBI and special Galileon theories, and the nonrelativistic spatial Galileon theory. At this stage, we however cannot exclude possible ``hybrid'' theories whose four-point amplitudes would be linear combinations of the amplitude of either the relativistic DBI or the spatial Galileon and the amplitude of the special Galileon.

%%%%%%%%%%%%%%%%%%%%%%%%%%%%%%%%%%%%%%%%%%%%%%%

\subsubsection{Effective Lagrangian scan}
\label{subsec:LagscanA1}

To complement the above soft bootstrap analysis, we have approached the problem of finding EFTs with enhanced soft limits from the brute-force Lagrangian perspective. Our goal was to gain additional insight into the Lagrangian representation of the seed four-point amplitudes listed in table~\ref{tab:seedA1}, and to extend the analysis to five-point amplitudes without having to deal with the nonlinear kinematical constraints on the on-shell four-momenta in a five-particle scattering process.

As the first step, we used symbolic computation to generate all contact operators for a single real scalar field $\theta$ with up to five factors of $\theta$ under the constraint that each factor of $\theta$ carries one or two (temporal or spatial) derivatives.\footnote{Requiring at least one derivative on each $\theta$ ensures manifest invariance under the shift symmetry that makes $\theta$ a NG boson.} Moreover, we only included in the analysis operators where not all factors of $\theta$ carry two derivatives, that is the average number of derivatives per field is less than two. This is the Lagrangian representation of the requirement that the enhanced scaling with $\sigma=2$ in the soft limit be nontrivial.

Importantly, we discarded a priori all cubic interaction vertices. It is known that in derivatively coupled Lorentz-invariant theories of a single scalar, cubic vertices can always be removed by a nonlinear field redefinition~\cite{Cheung2017a}. This is no longer the case in theories lacking Lorentz invariance. However, as already remarked above and as shown in detail in section~\ref{subsec:3pt}, there are no type $A_1$ theories with a nontrivial three-point on-shell amplitude where the soft limit would be enhanced beyond naive counting of derivatives. Whatever cubic operator present in the Lagrangian must therefore give a vanishing three-point amplitude. Still, the role of cubic interaction vertices in nonrelativistic EFTs is less trivial than in their Lorentz-invariant counterparts. We will get back to this point below.

With the collection of interaction operators at hand, we computed the four-point and five-point amplitudes following the algorithm detailed in appendix~\ref{sec:numerics}. Imposing the scaling of these amplitudes in the soft limit with $\sigma=2$, we then obtained constraints on the effective couplings. While the resulting set of candidate interaction operators is highly redundant, there is only a small number of corresponding four-point amplitudes. These turn out to exactly correspond to the amplitudes listed in table~\ref{tab:seedA1}. The corresponding result for the five-point amplitude is quite surprising. In spite of the large basis of candidate operators, there turns out to be only one seed amplitude that realizes nontrivially $\sigma=2$ scaling. This is Lorentz-invariant, and corresponds to the (Lorentz-invariant version of the) $k=4$ Galileon WZ term~\eqref{WZtermsGalileon}. There are no genuinely nonrelativistic enhanced five-point amplitudes, at least in $D=4$ spacetime dimensions. 

The Lagrangian scan also offers additional insight beyond mere verification of the results of section~\ref{subsec:seedA1}. We were thus able to inspect directly an interesting subclass of theories where each factor of $\theta$ carries exactly one derivative. These theories are defined by the class of Lagrangian densities
\begin{equation}
\La=\frac12(\de_0\theta)^2-\frac12(\vek\nabla\theta)^2+\sum_{n=3}^\infty\sum_{k=0}^{\lfloor n/2\rfloor}c_{n,2k}[(\vek\nabla\theta)^2]^k(\de_0\theta)^{n-2k},
\label{LDBIlike}
\end{equation}
where $n$ labels the valency of the interaction vertex and $2k$ the number of spatial derivatives. Requiring Adler zero ($\sigma=1$) restricts the cubic vertex to a single parameter $c_3$, in terms of which $c_{3,0}=c_3$ and $c_{3,2}=-c_3$. This constraint follows from the inspection of the four-point amplitude, but turns out to guarantee the Adler zero property also for all higher-point amplitudes. This agrees with the general argument given in section~\ref{subsec:revisit}.

Imposing furthermore enhanced scaling with $\sigma=2$ similarly reduces the three couplings $c_{n,2k}$ at $n=4$ to a single free parameter, $c_4\equiv c_{4,0}$. It turns out that all the other couplings in the Lagrangian~\eqref{LDBIlike} are then uniquely determined by $c_3$ and $c_4$; we have checked this numerically for amplitudes up to $n=8$. We thus end up with a two-parameter family of EFTs featuring enhanced scaling with $\sigma=2$, represented by
\begin{equation}
\La=\frac1{8\tilde c_4}\biggl\{1-2c_3\de_0\theta-\sqrt{(1-2c_3\de_0\theta)^2-8\tilde c_4[(\de_0\theta)^2-(\vek\nabla\theta)^2]}\biggr\},
\label{LpseudoDBI}
\end{equation}
where $\tilde c_4\equiv c_4-2c_3^2$. Note that for $c_3=0$, this recovers the relativistic DBI theory. For $c_3\neq0$, this however appears to be a genuinely nonrelativistic theory. In fact, it is possible to tune the couplings to make eq.~\eqref{LpseudoDBI} a special case of an EFT with the symmetries of a Galilei-invariant superfluid. Namely, by setting $c_3=1/2$ and $\tilde c_4=1/8$ (that is $c_4=5/8$), the Lagrangian~\eqref{LpseudoDBI} becomes
\begin{equation}
\La=-\sqrt{1-2\de_0\theta+(\vek\nabla\theta)^2}
\end{equation}
up to a total time derivative, which is indeed a special case of eq.~\eqref{Lsuperfluid}.

Obviously, the class~\eqref{LpseudoDBI} includes theories with different symmetries. It may therefore come as a surprise that, upon a closer look, the on-shell amplitudes generated by eq.~\eqref{LpseudoDBI} are independent of $c_3$. The amplitudes of the entire class of theories are identical to those of the relativistic DBI theory. This point is worth stressing. We started with spacetime translation and spatial rotation invariance, yet by imposing enhanced soft limit with $\sigma=2$, we ended up with the amplitudes of a Lorentz-invariant theory. In this sense, Lorentz invariance has \emph{emerged} as a consequence of our assumptions on the soft behavior of scattering amplitudes.

It may still appear puzzling that the entire class of Lagrangians~\eqref{LpseudoDBI} should map to a single relativistic theory. The resolution of this paradox is that the parameter $c_3$ can be removed from the theory by a time-dependent shift of $\theta$, followed by a rescaling of the time coordinate. Note that this is not the type of field redefinition one usually considers in EFT. The common lore is to perform a nonlinear redefinition that preserves the kinetic term. The corrections generated by the redefinition within the kinetic term may then serve e.g.~to remove the cubic coupling. Here, the cubic vertex is removed by a correction to the \emph{quartic} vertex, generated by the shift of $\theta$. We are not aware of a general argument that would guarantee a priori that such field redefinitions leave the $S$-matix unchanged. Should we however take such a generalized notion of reparameterization invariance of the $S$-matrix for granted, then we can discard cubic interaction vertices from the outset, as we after all did in our scan of effective Lagrangians with less than two derivatives per field.

%%%%%%%%%%%%%%%%%%%%%%%%%%%%%%%%%%%%%%%%%%%%%%%

\subsection{Type $B_2$ bootstrap}
\label{subsec:bootstrapB2}

According to sections~\ref{sec:softshift2} and~\ref{sec:4.3}, the soft recursion based on the all-line shift can be used to reconstruct amplitudes in type $B_2$ theories that satisfy $n\geq d+4$. This means that in the most interesting case of $d=3$ spatial dimensions, we would need seed amplitudes up to $n=6$. This would impose on us the necessity to deal with the nontrivial analytic structure of the six-point amplitude, which cannot be captured by contact operators in the effective Lagrangian. In order to circumvent this problem, we switch temporarily to $d=2$ spatial dimensions and restrict to theories conserving particle number, i.e.~theories of a Schr\"odinger scalar. In such theories, only amplitudes $A_n$ with even $n$ exist. Cubic vertices are not an issue and all we need to do is to classify seed four-point amplitudes.

Before we can do that, we have to deal with the type $B_2$ kinematics though. The seed amplitudes are going to be polynomials in the energies $\om_i=\vek p_i^2$ and the rotationally invariant dot products
\begin{equation}
s_{ij}\equiv\vek p_i\cdot\vek p_j.
\end{equation}
But these are not all independent due to energy and momentum conservation. We need to find the type $B_2$ equivalent of the constraints~\eqref{ideal}. The energy and momentum conservation conditions take the form
\begin{equation}
\sum_je_j\om_j=0,\qquad
\sum_je_j\vek p_j=\vek0,
\end{equation}
where the signs $e_j$ distinguish particles in the initial and final state. Dotting now the momentum conservation condition into $\vek p_i$ gives an explicit expression for the energies in terms of $s_{ij}$,
\begin{equation}
\om_i=-e_i\Sum{j}{j\neq i}e_js_{ij}.
\label{energy}
\end{equation}
The notation used to indicate the summation range is such that the first line indicates the summation variable(s), whereas the second line indicates in parentheses possible constraints on these variables.

Equation~\eqref{energy} exhausts all possible rotationally invariant constraints implied by momentum conservation. Using now energy conservation in combination with eq.~\eqref{energy} gives
\begin{equation}
0=\sum_i\Sum{j}{j\neq i}e_js_{ij}=\Sum{i,j}{i\neq j}e_js_{ij}=\frac12\Sum{i,j}{i\neq j}(e_i+e_j)s_{ij}.
\end{equation}
Obviously, only such pairs $i,j$ that both the $i$-th and the $j$-th particle belongs to the initial or the final state contribute to the sum. This implies one additional constraint,
\begin{equation}
\Sum{i,j\in\text{IN}}{i\neq j}s_{ij}=\Sum{k,l\in\text{OUT}}{k\neq l}s_{kl},
\label{inout}
\end{equation}
where the ``IN'' and ``OUT'' in the subscripts indicate that only incoming and outgoing particles are to be included in the sum.

%%%%%%%%%%%%%%%%%%%%%%%%%%%%%%%%%%%%%%%%%%%%%%%

\subsubsection{Seed four-point amplitudes}

So far we have not used anywhere the assumption that the given type $B_2$ theory conserves particle number. This implies that the numbers of particles in the initial and the final state of the scattering process must match. For $n=4$, we label the incoming particles by convention with the indices $1,3$, and the outgoing particles with $2,4$. The constraint~\eqref{inout} then reduces to
\begin{equation}
s_{13}=s_{24}.
\end{equation}
Altogether, the four-particle scattering process is characterized by five kinematical parameters, which may be chosen as
\begin{equation}
\frac{s_{13}+s_{24}}2,s_{12},s_{14},s_{23},s_{34}.
\label{B2variables}
\end{equation}
Note that this is the same number of independent kinematical variables as for type $A_1$ kinematics, where the seven variables $s,t,u$ and $\omega_{1,2,3,4}$ are constrained by the two linear relations in eq.~\eqref{ideal}.

\begin{table}[t]
\centering
\begin{tabular}{c|c}
Label & Amplitude\\
\hline
\hline
$A_4^{(1)}$ & $(s_{13}+s_{24})^2$\\
$A_4^{(2)}$ & $\frac12(s_{13}+s_{24})(s_{12}+s_{14}+s_{23}+s_{34})-(s_{12}+s_{14})(s_{23}+s_{34})$\\
$A_4^{(3)}$ & $(s_{13}+s_{24})^2(s_{12}+s_{14}+s_{23}+s_{34})$\\
$A_4^{(4)}$ & $(s_{13}+s_{24})(s_{12}s_{23}+s_{14}s_{34})$\\
$A_4^{(5)}$ & $(s_{13}+s_{24})(s_{12}s_{34}+s_{14}s_{23})$\\
$A_4^{(6)}$ & $\frac12(s_{13}+s_{24})(s_{12}s_{14}+s_{23}s_{34})-(s_{12}s_{14}s_{23}+s_{12}s_{14}s_{34}+s_{12}s_{23}s_{34}+s_{14}s_{23}s_{34})$\\[1ex]
$A_4^{(7)}$ & 
$\begin{aligned}
&{\textstyle\frac12}(s_{13}+s_{24})(s_{12}^2+s_{14}^2+s_{23}^2+s_{34}^2)\\
&-(s_{12}^2s_{34}+s_{12}s_{34}^2+s_{14}^2s_{23}+s_{14}s_{23}^2)-(s_{12}^2s_{23}+s_{12}s_{23}^2+s_{14}^2s_{34}+s_{14}s_{34}^2)
\end{aligned}$
\\[2ex]
$A_4^{(8)}$ & $(s_{13}+s_{24})^3$
\end{tabular}
\caption{Seed four-point amplitudes in type $B_2$ theories with soft scaling exponent $\sigma\geq2$. The nonrelativistic ``Mandelstam variables'' $s_{ij}$ are defined by $s_{ij}\equiv\vek p_i\cdot\vek p_j$.}
\label{tab:seedB2}
\end{table}

The four-point amplitude has to be invariant under the permutation group $S_2\times S_2$, which allows to swap independently the two particles in the initial and final state. As in the type $A_1$ case, we are looking for seed amplitudes that realize nontrivially enhanced scaling in the soft limit, $\sigma=2$. This amounts to restricting to permutation-invariant polynomials of degree less than four in the kinematical variables~\eqref{B2variables}. There turn out to be altogether eight candidate amplitudes that admit an enhanced soft limit as the momentum of the first particle is taken to zero, as displayed in table~\ref{tab:seedB2}.

Note that only for $A_4^{(i)}$ with $i\in\{1,3,4,5,8\}$ the enhanced scaling in the soft limit is manifest. For $i\in\{2,6,7\}$ the scaling only becomes visible once we use the relation $s_{12}-s_{13}+s_{14}=s_{11}=\om_1$. Nevertheless, not all of these eight amplitudes are physical. The amplitude $A_4^{(2)}$ is not invariant under the exchange of incoming and outgoing momenta. The same applies to $A_4^{(4)}$ and $A_4^{(6)}$, although their linear combination $\frac12A_4^{(4)}+A_4^{(6)}$ does not have this problem. Altogether, we therefore end up with six physically sensible candidate four-point amplitudes. Based on our previous discussions, we are able to identify a priori two of these amplitudes. The sole amplitude corresponding to one derivative per field, $A_4^{(1)}$, is generated by the Schr\"odinger-DBI theory. In addition, a detailed calculation shows that the Schr\"odinger-Galileon theory possesses a four-point amplitude that is proportional to $\frac14A_4^{(3)}-\frac12A_4^{(4)}-A_4^{(5)}-A_4^{(6)}$.

%%%%%%%%%%%%%%%%%%%%%%%%%%%%%%%%%%%%%%%%%%%%%%%

\subsubsection{Consistency constraints from soft bootstrap}

It remains to be clarified whether there are other consistent EFTs than the Schr\"odinger-DBI and Schr\"odinger-Galileon theories that give rise to some of the amplitudes in table~\ref{tab:seedB2} or their linear combinations. To that end, we again use the recursion relation~\eqref{Angeneral}. With the seed four-point amplitude at hand, we want to check whether or not the recursively constructed six-point amplitude is independent of the momentum shift parameters $a_i$ used in the recursion. It is here that we use the lower spatial dimension, $d=2$; this guarantees the existence of nontrivial solutions for $a_i$ that allow us to take the soft limit for one particle at a time. Note that this constraint on the dimension of space rules out a priori the Schr\"odinger-Galileon theory, since the WZ term~\eqref{L3SchrodingerGalileon} only exists for $d\geq3$.

What we have done in practice was to take the generic seed
\begin{equation}
A_4=c_1A_4^{(1)}+c_3A_4^{(3)}+c_4\Bigl({\textstyle\frac12}A_4^{(4)}+A_4^{(6)}\Bigr)+c_5A_4^{(5)}+c_7A_4^{(7)}+c_8A_4^{(8)}
\end{equation}
and evaluate the six-point amplitude using eq.~\eqref{Angeneral}. The outcome of the recursion is that the six-point amplitude is only consistent if all the coefficients $c_i$ but $c_1$ are zero. Thus, the only consistent EFT left is the Schr\"odinger-DBI theory. To this we have to add the Schr\"odinger-Galileon theory which we know to have amplitudes with $\sigma=2$, but which cannot be captured by the soft bootstrap in $d=2$ spatial dimensions.

Altogether, our findings for the type $A_1$ and type $B_2$ theories confirm our expectation that soft bootstrap is very efficient in narrowing down the landscape of physically consistent EFTs with enhanced soft limits. The detailed analysis has however not revealed any new theory that was not already known based on the symmetry-based Lie-algebraic classification carried out in section~\ref{sec:2}.

%%%%%%%%%%%%%%%%%%%%%%%%%%%%%%%%%%%%%%%%%%%%%%%

\subsubsection{Effective Lagrangian scan}
\label{subsec:LagscanB2}

Let us conclude the discussion of soft bootstrap with a brief report on a complementary, brute-force Lagrangian scan of EFTs for a Schr\"odinger scalar, similar to the scan of type $A_1$ theories discussed in section~\ref{subsec:LagscanA1}. We have considered a class of effective Lagrangians with interactions containing exactly one derivative per field,
\begin{equation}
\La=\he\psi(\im\de_0+\vek\nabla^2)\psi+\sum_{n=2}^\infty\La_{\text{int}}^{(2n)},
\end{equation}
where $\La_{\text{int}}^{(2n)}$ is a charge-conjugation-invariant interaction Lagrangian with valency $2n$, built out of $\de_0\psi,\vek\nabla\psi$ and their complex conjugates. For instance, the most general quartic interaction Lagrangian of this type reads
\begin{equation}
\begin{split}
\La_{\text{int}}^{(4)}={}&c_{4,1}(\de_0\he\psi\de_0\psi)^2+c_{4,2}(\de_0\he\psi\de_0\psi)(\vek\nabla\he\psi\cdot\vek\nabla\psi)+c_{4,3}(\vek\nabla\he\psi\cdot\vek\nabla\psi)^2\\
&+c_{4,4}|\vek\nabla\psi\cdot\vek\nabla\psi|^2+\Bigl[c_{4,5}(\de_0\he\psi)^2\vek\nabla\psi\cdot\vek\nabla\psi+\bar c_{4,5}(\de_0\psi)^2\vek\nabla\he\psi\cdot\vek\nabla\he\psi\Bigr],
\end{split}
\end{equation}
where all $c_{4,i}$ with $1\leq i\leq4$ are real but $c_{4,5}$ may be complex.

The ordinary Adler zero is automatically guaranteed for this class of effective Lagrangians as a consequence of the shift symmetry and the absence of cubic vertices. Imposing enhanced soft limit ($\sigma=2$) restricts the parameter space to one free parameter at each order $2n$. The most general allowed interaction Lagrangian with one derivative per field can then be folded into the form
\begin{equation}
\La_{\text{int}}=-1+s\vek\nabla\he\psi\cdot\vek\nabla\psi+\sqrt{|G|}\left[1+\sum_{n=2}^\infty c_{2n}(\delta_{AB}\nabla_0\theta^A\nabla_0\theta^B)^n\right],
\end{equation}
where $|G|$ and $\delta_{AB}\nabla_0\theta^A\nabla_0\theta^B$ are given respectively by eqs.~\eqref{SchrodingerDBIG} and~\eqref{SchrodingerDBID0D0}. We have checked the validity of this result numerically up to and including the eight-point amplitude. This confirms that the Schr\"odinger-DBI theory is the sole type $B_2$ theory conserving particle number that has interactions with a single derivative per field and features enhanced scaling of scattering amplitudes in the soft limit.

%%%%%%%%%%%%%%%%%%%%%%%%%%%%%%%%%%%%%%%%%%%%%%%

\section{Bounds on the EFT landscape}
\label{sec:bounds}

The numerical bootstrap in section~\ref{sec:bootstrap} did not lead to the discovery of any novel theories not already known. However, the scope of the analysis was restricted to seed amplitudes with less than two derivatives per field, leaving out possible higher-derivative theories with enhanced soft limits. In this section, we will apply a series of \textit{analytical} consistency checks to further narrow down the landscape of type $A_1$ and type $B_2$ EFTs. The result is a collection of constraints on how enhanced soft limits different types of EFTs can possess. This represents an extension of bounds on relativistic EFT space obtained by Cheung et al.~\cite{Cheung2017a}. 

We will start by reconsidering type $A_1$ theories to review and extend important ideas and results from the relativistic literature. This leads to new insights into the emergence of Lorentz invariance, observed in section~\ref{subsec:bootstrapA1}, and thus provides a clearer understanding of the role played by Lorentz invariance in the type $A_1$ landscape. It will also serve as a warm-up for the novel challenge of bounding the type $B_2$ landscape, which culminates with a derivation of a no-go theorem for exceptional type $B_2$ theories with $\sigma>2$.  

%%%%%%%%%%%%%%%%%%%%%%%%%%%%%%%%%%%%%%%%%%%%%%%

\subsection{Type $A_1$ theories}
\label{subsec:boundsA1}

In ref.~\cite{Cheung2017a}, the authors derived bounds on the soft scaling parameter $\sigma$ as a function of the average number of derivatives per field in the Lagrangian. We will refer to these bounds as \emph{leading interaction bounds}. To obtain similar results for the whole class of nonrelativistic type $A_1$ theories, we will adopt a new classification scheme applicable to EFTs with enhanced soft limits.\footnote{The classification scheme employed in refs.~\cite{Cheung2015a,Cheung2016a,Cheung2017a} applies to EFTs both with and without enhanced soft limits. The scheme put forward here is restricted to EFTs with enhanced soft limits. However, it makes it possible to include fundamental operators with different average numbers of derivatives per field.} Since EFTs with enhanced soft limits are on-shell constructable, their $S$-matrices are completely determined by fundamental operators. The latter are defined as the lowest-dimension operators whose on-shell matrix elements are needed to recursively construct any tree-level amplitude of the theory at the leading order of its derivative expansion. It therefore seems sensible to classify EFTs with enhanced soft limits in terms of the properties of their fundamental operators. 

Our classification scheme consists of four parameters, $D$, $\sigma$, $v$ and $\tau$. The $D$ and $\sigma$ have already been used extensively throughout this paper. Furthermore, $v$ is the highest number of external legs such that the amplitude $A_v$ is nonvanishing and local in momenta. Finally, $\tau\equiv\max\{\tau_n\}_{n\leq v}$, where $\tau_n$ denotes the average number of derivatives per field in the fundamental operator with $n$ fields. In type $A_1$ theories, the number of derivatives refers to the sum of spatial and temporal derivatives. Intuitively, $\tau/\sigma$ measures \textit{how} enhanced soft limits a given theory has. As shown below, leading interaction bounds are simply inequalities relating the parameters $\tau$, $\sigma$ and $v$.
  
%%%%%%%%%%%%%%%%%%%%%%%%%%%%%%%%%%%%%%%%%%%%%%%

\subsubsection{Soft limit of leading interaction}
\label{sec:6.1.1}

An EFT with enhanced soft limits parameterized by $(\tau, \sigma, D, v)$ has a nonvanishing local amplitude $A_v$, which is a polynomial function in spatial momenta and frequencies (hereafter collectively denoted as momenta). A soft momentum shift lifts $A_v$ to a complex polynomial of degree $\tau_v v$ in $z$, $A_v\rightarrow \hat{A}_v(z)$. Vanishing of the amplitude in the single soft limit for a particular external leg corresponds to a zero of this polynomial. Denoting as $v_s$ the number of external legs whose single soft limit is accessible using the chosen momentum shift, the total number of zeros, counting multiplicity, must be at least $v_s\sigma$. At the same time, this cannot be higher than the degree of the polynomial, hence
\begin{align}
    \label{degreevszeros}
    \tau\geq\tau_v\geq\frac{v_s\sigma}{v}.
\end{align}
The most stringent bound on $\tau$ requires maximal $v_s$. However, the applicability of soft shifts also depends on $v$ and $D$. Below we summarize bounds obtained using different types of soft shifts, along with the values of $v$ and $D$ for which these shifts are applicable.

%%%%%%%%%%%%%%%%%%%%%%%%%%%%%%%%%%%%%%%%%%%%%%%

\paragraph{All-but-two-line shift bounds}

The most general bounds arise from the all-but-two-line shift as this allows the smallest value of $v_s$ but can be used for any $v\geq4$ as long as $D\geq4$,
\begin{align}
    \label{twoconstraint1}
    \tau\geq \frac{v-2}{v}\sigma, \qquad v\geq 4, \quad D\geq 4.
\end{align}
Since $(v-2)/v$ is a monotonously increasing function of $v$, we get the least stringent but most universal bound by evaluating eq.~\eqref{twoconstraint1} at $v=4$,
\begin{align}
    \label{sickbound}
    \tau\geq \frac{\sigma}{2}, \qquad v\geq 4, \quad D\geq 4.
\end{align}
This inequality is saturated by the relativistic exceptional theories: the NLSM with $(\tau,\sigma)=(1/2,1)$, the DBI theory with $(\tau,\sigma)=(1,2)$, and the special Galileon with $(\tau,\sigma)=(3/2,3)$.

%%%%%%%%%%%%%%%%%%%%%%%%%%%%%%%%%%%%%%%%%%%%%%%

\paragraph{All-line shift bounds}

The most stringent bounds arise from the all-line soft shift, which is applicable when $v\geq D+2$, 
\begin{align}
    \label{all-line}
    \tau\geq \sigma, \qquad v\geq D+2.
\end{align}
However, for operators with at least $\sigma$ derivatives per field, the enhanced soft limit with the soft scaling parameter $\sigma$ will be realized trivially. The bound~(\ref{all-line}) therefore forbids the existence of EFTs with nontrivially enhanced soft limits and $v\geq D+2$. Notice how this is consistent with the properties of known EFTs such as the Galileon. 
%%%%%%%%%%%%%%%%%%%%%%%%%%%%%%%%%%%%%%%%%%%%%%%

\paragraph{All-but-one-line shift bounds}

In line with the discussion in section~\ref{sec:softshift1}, the bound obtained from the all-but-one-line soft shift is valid when $v\geq 5$,
\begin{align}
    \label{all-but-one-0}
    \tau\geq \frac{v-1}{v}\sigma, \qquad v\geq 5.
\end{align}
This provides a more stringent bound than eq.~(\ref{twoconstraint1}) for theories where $v\geq 5$. Equation~(\ref{all-but-one-0}) is saturated by the quintic ($k=4$) Galileon WZ term~\eqref{WZtermsGalileon}, both the spatial Galileon and its relativistic version, where $(\tau,\sigma)=(8/5,2)$.

%%%%%%%%%%%%%%%%%%%%%%%%%%%%%%%%%%%%%%%%%%%%%%%

\paragraph{Summary}

Theories with $v=4$ in $D\geq 4$ are subject to the universal bound $\tau\geq{\sigma}/{2}$. EFTs living on the line $\tau={\sigma}/{2}$ are called \emph{exceptional}. This terminology indicates that such theories have maximally enhanced scattering amplitudes for given average number of derivatives per field $\tau$. Theories with nontrivial five-point contact amplitudes are moreover subject to the more stringent bound $\tau\geq 4\sigma/5$. If the six-point amplitude is also local, then the theory cannot have enhanced soft limits in $D=4$. 

The case of $D=3$ spacetime dimensions requires separate treatment. Here the all-line-shift bound implies that theories with a contact five-point amplitude cannot have enhanced soft limits. Hence, all three-dimensional theories with enhanced soft limits have $v=4$.

%%%%%%%%%%%%%%%%%%%%%%%%%%%%%%%%%%%%%%%%%%%%%%%

\subsubsection{Three-point amplitudes}
\label{subsec:3pt}

The bounds on the EFT space discussed above revolved largely around local four-point and five-point amplitudes. We have ignored possible three-point amplitudes. Let us now explain why this is justified.

First of all, in (derivatively coupled) Lorentz-invariant theories, there are no three-point amplitudes due to the lack of nonzero relativistic invariants in three-particle kinematics. Hence, any nonvanishing three-point amplitude in a nonrelativistic type $A_1$ theory must necessarily be a function of the energies $\omega_{1,2,3}$ alone. By permutation invariance, the three-point amplitude must then be a function of $\omega_1\omega_2+\omega_1\omega_3+\omega_2\omega_3$ and $\omega_1\omega_2\omega_3$. (The third independent elementary symmetric polynomial, $\omega_1+\omega_2+\omega_3$, vanishes by energy conservation.) For a given average number of derivatives per field $\tau_3$, we may thus express the generic three-point amplitude as
\begin{equation}
    \label{3ptansatz}
    \begin{split}
    \sum_{2a+3b=3\tau_3}&\lambda_b(\omega_1\omega_2+\omega_1\omega_3+\omega_2\omega_3)^a(\omega_1\omega_2\omega_3)^b\\
    &=\sum_{2a+3b=3\tau_3}\lambda_b[\omega_1\omega_2-(\omega_1+\omega_2)^2]^a[-\omega_1\omega_2(\omega_1+\omega_2)]^b,
    \end{split}
\end{equation}
where $\lambda_b$ are arbitrary constants. In the single soft limit $\omega_1\rightarrow 0$, eq.~(\ref{3ptansatz}) scales as $\omega_1^b$. The maximum value that $b$ can take is $\tau_3$. Hence $\sigma\leq\tau_3$, implying that theories with a nonvanishing three-point amplitude cannot have nontrivially enhanced soft limits.

We conclude that even in nonrelativistic type $A_1$ theories, it is justified to discard possible three-point amplitudes, as long as one is interested only in EFTs with nontrivially enhanced soft limits. A direct consequence is that all five-point amplitudes in such theories, consistent with factorization, are necessarily local functions of momenta.

%%%%%%%%%%%%%%%%%%%%%%%%%%%%%%%%%%%%%%%%%%%%%%%

\subsubsection{Bounds on exceptional theories from soft recursion}

Consistency of the $S$-matrix requires that higher-point amplitudes should not depend on the specific way in which recursion is applied, for instance on the choice of the momentum shift parameters $a_i$. In ref.~\cite{Cheung2017a}, Cheung et al.~use this simple albeit powerful statement to obtain further bounds on the EFT parameter space. In particular, they apply soft recursion to a completely general local and Lorentz-invariant four-point amplitude ansatz and require the resulting six-point amplitude to be independent of unphysical parameters. This is shown to put surprisingly powerful constraints on the four-point amplitude ansatz. One of their main results is a no-go theorem for exceptional theories with ``super-enhanced'' soft behavior, $\sigma>3$.

In section~\ref{subsec:bootstrapA1}, we observed emergence of Lorentz invariance in the class of nonrelativistic type $A_1$ theories from the assumed exceptionally soft behavior. It therefore seems natural to ask whether a similar no-go theorem also applies to the whole class of type $A_1$ EFTs. The answer to this question is indeed yes. First, it follows from the bound~\eqref{all-but-one-0} that no exceptional theory can have any five- or higher-point contact amplitude. Moreover, we argued in section~\ref{subsec:3pt} that theories with a nontrivial three-point amplitude cannot have enhanced soft limits at all. To prove that all exceptional type $A_1$ theories are necessarily Lorentz-invariant, it is therefore sufficient to show that all exceptional four-point amplitudes are. We do this in detail in appendix~\ref{sec:A1}. This allows the no-go theorem of ref.~\cite{Cheung2017a} for consistent exceptional theories with $\sigma>3$ to be lifted to a more general no-go theorem valid for all type $A_1$ theories in $D\geq 4$ spacetime dimensions.

%%%%%%%%%%%%%%%%%%%%%%%%%%%%%%%%%%%%%%%%%%%%%%%

\subsection{Theories with quadratic dispersion relation}

Our next objective is to study nonrelativistic EFTs with enhanced soft limits where the NG bosons have a quadratic dispersion relation, i.e.~type $A_2$ and type $B_2$ theories. The first step is to adapt the leading interaction bounds from section~\ref{subsec:boundsA1}. This requires a modification of the definition of the classification parameter $\tau$, which now counts the average number of \emph{spatial} derivatives per field, whereby every temporal derivative is counted as two spatial derivatives. In the second step, we use the leading interaction bounds to define exceptional theories of type $B_2$. We show that locality and factorization imply strong constraints on the soft scaling parameter $\sigma$. These lead to a no-go theorem for the existence of ``super-exceptional'' type $B_2$ theories with $\sigma>2$.  

%%%%%%%%%%%%%%%%%%%%%%%%%%%%%%%%%%%%%%%%%%%%%%%

\paragraph{All-but-two-line shift}

The least stringent bounds on EFTs where the NG boson has a quadratic dispersion relation arise from the all-but-two-line soft shift. This shift is applicable when $v\geq4$ and $D\geq 3$, as shown in section~\ref{sec:softshift2}. Hence
\begin{align}    
\label{twoconstraint}
    \tau\geq \frac{v-2}{v}\sigma, \qquad v\geq4,\quad D\geq 3.
\end{align}
Since $(v-2)/v$ is a monotonously increasing function of $v$, the least stringent but universal bound valid for all $v\geq4$ follows by substituting $v=4$,
\begin{align}
    \label{twocont2}
    \tau\geq \frac{\sigma}{2}, \qquad v\geq 4, \quad D\geq 3.
\end{align}
Similarly to type $A_1$ theories, we define exceptional EFTs with a quadratic dispersion relation as those for which $\tau=\sigma/2$. This bound is saturated for the Schr\"odinger-DBI theory with $(\tau,\sigma)=(1,2)$. In addition, there is a nonrelativistic version of the NLSM which is also exceptional by our definition, although it does not have enhanced soft limits in the usual sense. This NLSM lives on $\C P^1$ and has $(\tau,\sigma)=(1/2,1)$. It describes the low-energy dynamics of ferromagnets~\cite{Volkov1971a,Leutwyler1994a}. We are not aware of any other exceptional EFTs of type $A_2$ or $B_2$.

%%%%%%%%%%%%%%%%%%%%%%%%%%%%%%%%%%%%%%%%%%%%%%%

\paragraph{All-but-one-line shift}

In complete analogy with type $A_1$ theories we obtain the following bound from the all-but-one-line soft shift, 
\begin{align}
    \tau\geq \frac{v-1}{v}\sigma, \qquad v\geq 5.
\end{align}
This again provides a more stringent bound for $v\geq 5$ than the inequality in eq.~(\ref{twoconstraint}).

%%%%%%%%%%%%%%%%%%%%%%%%%%%%%%%%%%%%%%%%%%%%%%%

\paragraph{All-line shift}

The following constraint is obtained from the all-line soft shift and is only valid when $v\geq D+3$, 
\begin{align}
    \tau\geq \sigma, \qquad v\geq D+3.
\end{align}
Thus, there cannot exist EFTs of type $A_2$ or $B_2$ with nontrivially enhanced soft limits that have $v\geq D+3$. 

%%%%%%%%%%%%%%%%%%%%%%%%%%%%%%%%%%%%%%%%%%%%%%%

\paragraph{Summary}

As a consequence of conservation of particle number, there are no interaction vertices with an odd number of fields in Schr\"odinger-type theories. This leads to the following refined bounds for type $B_2$ theories,
\begin{align}
    \label{B2sums}
    \tau&\geq \frac{\sigma}{2}, &v&\geq 4, \quad D\geq 3, \\
    \tau&\geq \frac{v-1}{v}\sigma, &v&\geq 6, \\
    \label{B2sumf}
    \tau&\geq \sigma, &v&\geq D+3.
\end{align}
Hence, any type $B_2$ theory with $\tau<5\sigma/6$ must have a nonvanishing four-point amplitude. 

Type $A_2$ theories can have both odd and even interaction vertices. However, due to the generalized CHMW theorem they may only exist in $D\geq4$ spacetime dimensions. This leads to the following refined bounds,
\begin{align}
    \tau&\geq \frac{\sigma}{2},&v&\geq 4, \quad D\geq 4, \\
    \tau&\geq \frac{v-1}{v}\sigma,&v&\geq 5, \\
    \tau&\geq \sigma,&v&\geq D+3\geq 7.
\end{align}
We conclude that any type $A_2$ EFT with $\tau<4\sigma/5$ must have a nonvanishing four-point amplitude.

%%%%%%%%%%%%%%%%%%%%%%%%%%%%%%%%%%%%%%%%%%%%%%%

\subsection{Bounds on exceptional type $B_2$ theories}

The above-derived bounds show that exceptional theories satisfy $\sigma=\tau/2$ regardless of the dispersion relation of the NG boson. While all exceptional theories of type $A_1$ have already been identified~\cite{Cheung2017a}, finding all exceptional theories where NG bosons have a quadratic dispersion relation remains an open problem. So far, the only known such theories are the $\C P^1$ NLSM and the Schr\"{o}dinger-DBI theory, which are both of type $B_2$. In this subsection, we will prove that there cannot be any ``super-exceptional'' type $B_2$ theories with $\sigma\geq3$, respecting locality and factorization. When combined with the soft bootstrap approach to type $B_2$ theories, developed in section~\ref{subsec:bootstrapB2}, this rules out the existence of any other exceptional Schr\"odinger-type theories than the above two.

The $S$-matrix of type $B_2$ exceptional theories is fully constructable from the four-point seed amplitude via on-shell recursion. In appendix~\ref{sec:B2}, we prove that the four-point amplitude in exceptional type $B_2$ theories takes the following simple form, 
\begin{align}
\label{B2ansatztext}
    A_4=\lambda_{\sigma}s_{13}^{\sigma},
\end{align}
where $\lambda_\sigma$ is a parameter. We use the same kinematical conventions as in section~\ref{subsec:bootstrapB2}, that is, the particles $1$ and $3$ are incoming and the particles $2$ and $4$ outgoing, and $s_{13}\equiv\vek p_1\cdot\vek p_3$. For the time being, we do not assume anything about $\sigma$ except that it is a positive integer.

We shall adopt an analytical bootstrap approach similar to that of ref.~\cite{Cheung2017a}. There, it was shown that consistency of six-point amplitudes recursively constructed from a generic Lorentz-invariant four-point seed only allows very specific EFTs to have enhanced soft limits. In fact, very similar methods have also been used to study the consistency of theories of massless particles of higher spin in four-dimensional Minkowski spacetime~\cite{Benincasa:2007xk,McGady2013} through  Britto-Cachazo-Feng-Witten-type recursion~\cite{Britto2004a,Britto2005a}.

Our strategy to rule out the existence of super-exceptional theories (in $d\geq3$ spatial dimensions) will be as follows. First, we derive a condition that recursively constructed six-point amplitudes in consistent (super-)exceptional type $B_2$ theories must satisfy. This does most of the job and the derivation given in section~\ref{subsec:6.3.1} is rather technical. Since the consistency condition turns out to be purely kinematical, it is subsequently easy to show that it cannot be satisfied for a generic four-particle kinematical configuration in any type $B_2$ theory where $\sigma\geq3$. This we do in section~\ref{subsec:6.3.2}.

%%%%%%%%%%%%%%%%%%%%%%%%%%%%%%%%%%%%%%%%%%%%%%%

\subsubsection{Six-point amplitude}
\label{subsec:6.3.1}

\begin{figure}[t]
    \centering
    \includegraphics[width=0.75\textwidth]{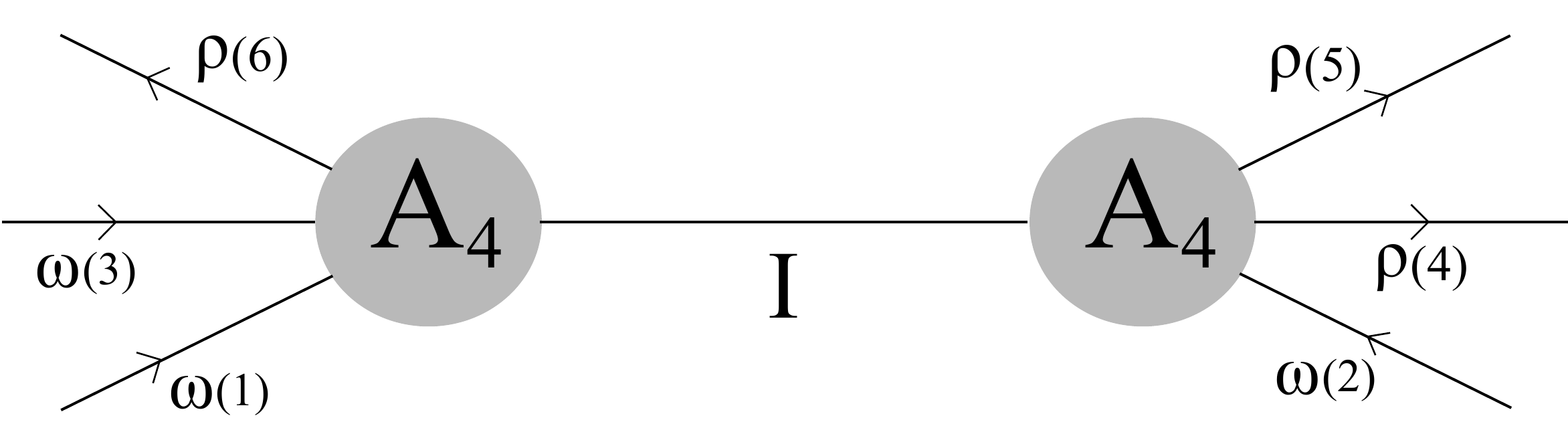}
    \caption{Graphic representation of a generic contact contribution to the six-point amplitude. Here $1,2,3$ label incoming particles and $\omega$ their $S_3$ permutation, whereas $4,5,6$ label outgoing particles and $\rho$ their $S_3$ permutation.}
    \label{fig:33}
\end{figure}

Six-point amplitudes have the special feature that they can be decomposed into factorization and contact terms, cf.~eq.~(\ref{master}). Using eq.~\eqref{master} together with the all-but-one-line shift ($n_s=5$), the second, contact contribution to the six-point amplitude acquires the form
\begin{align}
    \label{6contact}
    A_6^{\rm{contact}}=\sum_I\sum_{i=1}^{5}\Res_{z=1/a_i}\frac{\hat{A}_L^{(I)}(z)\hat{A}_R^{(I)}(z)}{zF(z)D^{(I)}(z)},
\end{align}
where we have defined $F(z)$ as
\begin{align}
    F(z)\equiv \prod_{i=1}^5f^{\sigma}_i(z)\equiv \prod_{i=1}^5(1-a_iz)^{\sigma}. 
\end{align}
The shifted subamplitudes $\hat{A}_L^{(I)}(z)$ and $\hat{A}_R^{(I)}(z)$ are local four-point amplitudes given by eq.~(\ref{B2ansatztext}) with equal values of $\sigma$. In order to distinguish different factorization channels, we will adopt the notation whereby the labels $1,2,3$ indicate particles in the initial state and the labels $4,5,6$ particles in the final state. Then the factorization channel $I$ is uniquely specified by choosing one outgoing particle for the ``left'' subamplitude and one incoming particle for the ``right'' subamplitude; see fig.~\ref{fig:33}. Combining this notation with eq.~(\ref{B2ansatztext}), the contact contribution to the six-point amplitude~\eqref{6contact} can be expressed as follows,
\begin{align}
    \label{a6contact}
    A_6^{\rm{contact}}&=\frac14\sum_{\omega,\rho}\sum_{i=1}^5\Res_{z=1/a_i}\left[\frac{\lambda^2_{\sigma}s^{\sigma}_{\omega(1)\omega(3)}s^{\sigma}_{\omega(2)(\rho(\hat{4})+\rho(\hat{5})-\omega(\hat{2}))}}{2z\hat{X}}\frac{1}{f^{\sigma}_{4}(z)f^{\sigma}_{5}(z)}\right], \\
    \hat{X}&=\hat{\boldsymbol{p}}_{\omega(1)}\cdot\hat{\boldsymbol{p}}_{\rho(6)}+\hat{\boldsymbol{p}}_{\omega(3)}\cdot\hat{\boldsymbol{p}}_{\rho(6)}-\hat{\boldsymbol{p}}_{\omega(1)}\cdot\hat{\boldsymbol{p}}_{\omega(3)}-\hat{\boldsymbol{p}}_{\rho(6)}^2\nonumber \\
    &=\hat{\boldsymbol{p}}_{\omega(2)}\cdot\hat{\boldsymbol{p}}_{\rho(4)}+\hat{\boldsymbol{p}}_{\omega(2)}\cdot\hat{\boldsymbol{p}}_{\rho(5)}-\hat{\boldsymbol{p}}_{\rho(4)}\cdot \hat{\boldsymbol{p}}_{\rho(5)}-\hat{\boldsymbol{p}}_{\omega(2)}^2.
    \label{Xdef}
\end{align}
Recall that the momenta of particles $1$ to $5$ are shifted according to eq.~(\ref{all-but-one-1}), whereas that of particle $6$ is shifted by eq.~(\ref{all-but-one-3}). We have also employed a shorthand notation whereby $s_{i\hat j}=\vek p_i\cdot\hat{\vek p}_j$, $s_{i(\hat j+\hat k)}=\vek p_i\cdot(\hat{\vek p}_j+\hat{\vek p}_k)$ and similar. The overall factor of $1/4$ in eq.~\eqref{a6contact} accounts for the fact there are nine factorization channels; the sum over all permutations $\omega,\rho$ of incoming and outgoing particles counts each channel four times.

In the following, we scrutinize the dependence of eq.~\eqref{a6contact} on the parameters $a_i$. Since this contact contribution to the six-point amplitude is a rational function of the $a_i$s, any unphysical dependence on $a_i$ is likely to manifest itself by spurious poles. Demanding absence of such spurious poles in eq.~\eqref{a6contact} is therefore a simple yet strong necessary condition for the amplitude to be physically consistent.

First, note that the factors $f_i^\sigma(z)$, $i\leq3$ in the denominator have been cancelled by the momentum shift. Hence only the residua at $z=1/a_4$ and $z=1/a_5$ will be nonzero. Since eq.~\eqref{a6contact} is manifestly invariant under the exchange $4\leftrightarrow5$, we can without loss of generality restrict to the residue at $z=1/a_4$. This may give rise to two types of spurious poles in $a_4$. First, the $f_5^\sigma(z)$ factor in the denominator will give a contribution singular as $a_4\to a_5$. Second, the factor of $\hat X$ will give contributions singular as $a_4\to a_i$, $i\leq3$. Below, we will focus on spurious poles of the second type, coming from the intermediate propagator in the diagram in fig.~\ref{fig:33}. Thanks to the invariance under $S_3$ permutations of the oncoming particles, it is sufficient to inspect the case $i=1$, that is singularities of the amplitude of the type $1/(a_4-a_1)$.

It follows from the definition of $\hat X$ in eq.~\eqref{Xdef} that when $\rho(6)=4$, the contribution of the residue at $z=1/a_4$ to eq.~\eqref{a6contact} will be proportional to $1/[(a_4-a_{\omega(1)})(a_4-a_{\omega(3)})]$. On the other hand, if $\rho(4)=4$ or $\rho(5)=4$, we find a spurious pole of the type $1/(a_4-a_{\omega(2)})$. In either case, particles $1$ and $4$ must be on the same side of the factorization channel $I$ to generate a spurious pole contribution of the form $1/(a_4-a_1)$. This leaves us with two possibilities, depending on whether the third particle that is on the same side of the factorization channel as particles $1$ and $4$ is incoming or outgoing. In both cases, there are two different factorization channels that contribute. We consider these contributions below. 

%%%%%%%%%%%%%%%%%%%%%%%%%%%%%%%%%%%%%%%%%%%%%%%

\paragraph{Particles $1$ and $4$ plus another outgoing particle}

\begin{figure}[t]
    \centering
    \includegraphics[width=0.75\textwidth]{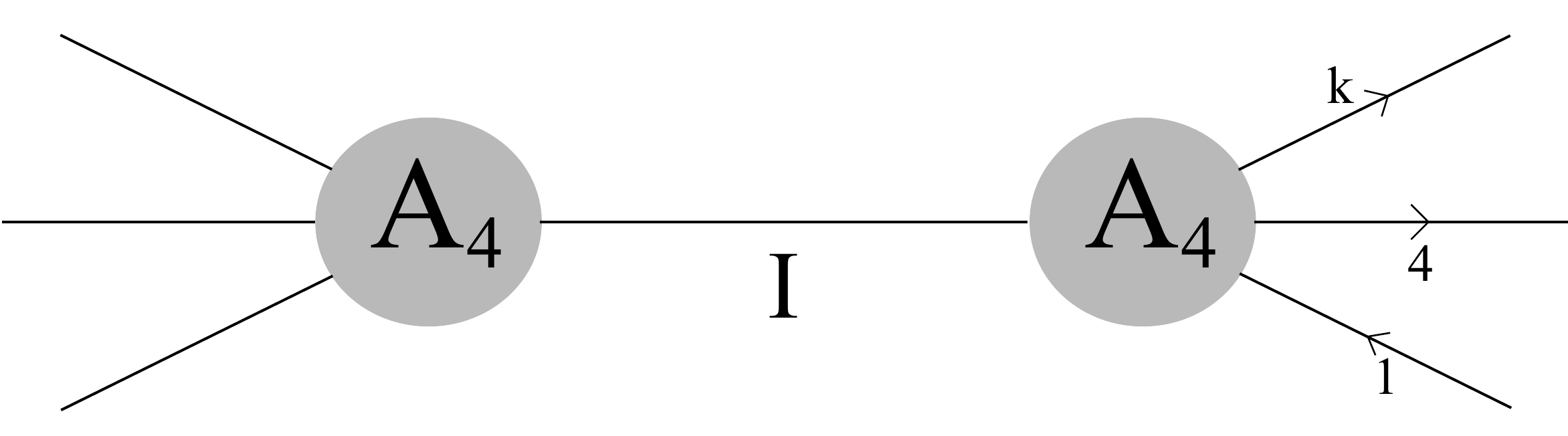}
    \caption{Sketch of a factorization channel with spurious pole $1/(a_4-a_1)$ where particles $1$ and $4$ are on the same side with another outgoing particle $k$.}
    \label{fig:2out1in}
\end{figure}

In this case, particles $1$ and $4$ are attached to the right subamplitude in fig.~\ref{fig:33}. The second outgoing particle may be either $5$ or $6$. Setting without loss of generality $\rho(4)=4$ and accordingly $\omega(2)=1$, we thus have $\rho(5)=k\in\{5,6\}$; see fig.~\ref{fig:2out1in} for a sketch. The inverse propagator of the respective factorization channels becomes
\begin{align}
\label{2in1out}
    \hat{P}_0-\hat{P}_I^2=2(\hat{\boldsymbol{p}}_{1}\cdot\hat{\boldsymbol{p}}_{4}+\hat{\boldsymbol{p}}_{1}\cdot\hat{\boldsymbol{p}}_{k}-\hat{\boldsymbol{p}}_{4}\cdot\hat{\boldsymbol{p}}_{k}-\hat{\boldsymbol{p}}^2_{1})\equiv 2\hat{X}_k,
\end{align}
which defines a new variable $\hat{X}_k$ for future convenience. The contributions of the two factorization channels with $k=5,6$ to the residue at $z=1/a_4$ in eq.~\eqref{a6contact} together read
\begin{align}
    \label{21-1}
    \frac12\lambda_{\sigma}^2s_{23}^{\sigma}\sum_{k=5}^6\Res_{z=1/a_4}\frac{s^{\sigma}_{1(\hat{4}+\hat{k}-\hat{1})}}{zf^{\sigma}_{4}(z)f^{\sigma}_{5}(z)\hat{X}_k}.
\end{align}

Since the pole at $z=1/a_4$ is not simple for any $\sigma\geq2$, the residue requires taking derivatives. To that end, note that the inverse propagator $2\hat X_k$ vanishes in the limit $z\to1/a_4$ and $a_4\to a_1$, whereas its derivative does not,
\begin{align}
    \left.\hat{X}_k\right\rvert_{z=1/a_4}&=\left.f_1(z)s_{1(\hat k-\hat 1)}\right\rvert_{z=1/a_4},\\
    \left.\frac{\dd\hat{X}_k}{\dd z}\right\rvert_{z=1/a_4}&=a_4\hat{\vek p}_k\cdot(\vek p_4-\vek p_1)\Bigr\rvert_{z=1/a_4}+\mathcal O(a_4-a_1).
\end{align}
The leading spurious pole at $a_4\to a_1$ therefore comes from the contribution where all the $\sigma-1$ derivatives involved in the calculation of the residue act on $\hat X_k$. This leading spurious pole then takes the form
\begin{align}
    \label{summand}
    \left.-\frac12\lambda_\sigma^2s_{23}^\sigma\sum_{k=5}^6
    \frac{[\hat{\vek p}_k\cdot(\vek p_4-\vek p_1)]^{\sigma-1}}{[f_1(z)f_5(z)]^\sigma}\right\rvert_{z=1/a_4}.
\end{align}

%%%%%%%%%%%%%%%%%%%%%%%%%%%%%%%%%%%%%%%%%%%%%%%

\paragraph{Particles $1$ and $4$ plus another incoming particle}

\begin{figure}[t]
    \centering
    \includegraphics[width=0.75\textwidth]{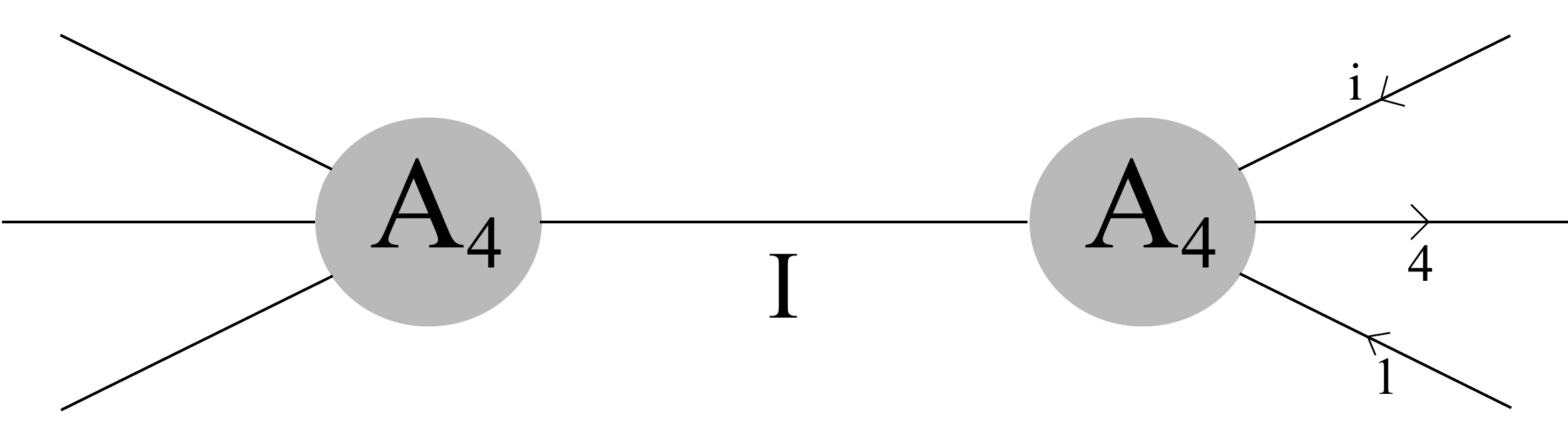}
    \caption{Sketch of a factorization channel with spurious pole $1/(a_4-a_1)$ where particles $1$ and $4$ are on the same side with another incoming particle $i$.}
    \label{fig:2in1out}
\end{figure}

Here the additional incoming particle on the same side of the factorization channel as particles $1$ and $4$ may be either particle $2$ or particle $3$. We can set without loss of generality $\rho(6)=4$, $\omega(1)=1$ and $\omega(3)=i\in\{2,3\}$; see fig.~\ref{fig:2in1out} for a sketch. The inverse propagator associated with the two possible factorization channels then is
\begin{align}
\label{2out1in}
    \hat{P}_0-\hat{P}_I^2=2(\hat{\boldsymbol{p}}_{1}\cdot\hat{\boldsymbol{p}}_{4}+\hat{\boldsymbol{p}}_{i}\cdot\hat{\boldsymbol{p}}_{4}-\hat{\boldsymbol{p}}_{1}\cdot\hat{\boldsymbol{p}}_{i}-\hat{\boldsymbol{p}}^2_{4})\equiv 2\hat{X}_i.
\end{align}
The contributions of these two factorization channels to the residue at $z=1/a_4$ in eq.~\eqref{a6contact} now read
\begin{equation}
    \frac12\lambda_\sigma^2\sum_{i=2}^3\Res_{z=1/a_4}\frac{s_{1i}^\sigma s_{\bar i(\hat 5+\hat 6-\hat{\bar i})}^\sigma}{zf_4^\sigma(z)f_5^\sigma(z)\hat X_i},
\end{equation}
where we defined the label $\bar i$ to denote the incoming particle on the opposite side of the factorization channel than particles $1$ and $4$.

Following the same steps as in the previous case, we observe that the inverse propagator $2\hat X_i$ vanishes in the limit $z\to1/a_4$ and $a_4\to a_1$, whereas its derivative does not,
\begin{align}
        \left.\hat{X}_i\right\rvert_{z=1/a_4}&=-f_1(z)f_i(z)s_{1i}\Bigr\rvert_{z=1/a_4},\\
        \left.\frac{\dd\hat{X}_i}{\dd z}\right\rvert_{z=1/a_4}&=a_4\hat{\vek p}_i\cdot(\vek p_1-\vek p_4)\Bigr\rvert_{z=1/a_4}+\mathcal O(a_4-a_1).
\end{align}
The leading spurious pole at $a_4\to a_1$ then comes from the contribution where all the $\sigma-1$ derivatives act on $\hat X_i$. This leading spurious pole then takes the form
\begin{align}
    \label{12-1}
    \left.-\frac12\lambda_\sigma^2\sum_{i=2}^3\frac{s^{\sigma}_{\bar{i}(\hat{5}+\hat{6}-\hat{\bar{i}})}[\hat{\boldsymbol{p}}_{i}\cdot(\boldsymbol{p}_{1}-\boldsymbol{p}_4)]^{\sigma-1}}{[-f_1(z)f_i(z)f_5(z)]^\sigma}\right\rvert_{z=1/a_4}.
\end{align}
To recast this expression into a more useful form we employ the fact that sending $a_4\rightarrow a_1$ and simultaneously setting $z=1/a_4$ corresponds to a double soft limit where both particles $1$ and $4$ become soft. In this limit $\hat{p}_2,\hat{p}_3,\hat{p}_5,\hat{p}_6$ form a (shifted) four-particle Schr\"{o}dinger kinematics with $\hat{p}_2+\hat{p}_3=\hat{p}_5+\hat{p}_6$, which in turn implies
\begin{align}
    \label{rs}
    s^{\sigma}_{\bar{i}(\hat{5}+\hat{6}-\hat{\bar{i}})}=s^{\sigma}_{\bar{i}\hat{i}}=s^{\sigma}_{23}f^{\sigma}_i(z).
\end{align}
This reduces eq.~(\ref{12-1}) to
\begin{align}
    \label{12-2}
    \left.\frac12\lambda_\sigma^2s^{\sigma}_{23}\sum_{i=2}^3\frac{[\hat{\boldsymbol{p}}_{i}\cdot(\boldsymbol{p}_{4}-\boldsymbol{p}_1)]^{\sigma-1}}{[f_1(z)f_5(z)]^\sigma}\right\rvert_{z=1/a_4}.
\end{align}

Putting this together with eq.~\eqref{summand} gives the leading contribution to the spurious pole at $a_4\to a_1$ in the six-point amplitude,
\begin{equation}
\begin{split}
    \frac{\lambda_\sigma^2s_{23}^\sigma}{2[f_1(z)f_5(z)]^\sigma}\Bigl\{&\left[\hat{\boldsymbol{p}}_2\cdot(\boldsymbol{p}_4-\boldsymbol{p}_1)\right]^{\sigma-1}+\left[\hat{\boldsymbol{p}}_3\cdot(\boldsymbol{p}_4-\boldsymbol{p}_1)\right]^{\sigma-1}-\left[\hat{\boldsymbol{p}}_5\cdot(\boldsymbol{p}_4-\boldsymbol{p}_1)\right]^{\sigma-1}\\
    &-\left(\hat{\boldsymbol{p}}_6\cdot(\boldsymbol{p}_4-\boldsymbol{p}_1)\right)^{\sigma-1}\Bigr\}\Biggr\rvert_{z=1/a_4}.
\end{split}
\label{consistcond}
\end{equation}
Vanishing of this expression is a necessary condition for the recursively constructed exceptional six-point amplitude to be physically consistent. Equivalently, a sufficient condition for an exceptional type $B_2$ theory with soft scaling parameter $\sigma$ to be unphysical is
\begin{equation}
\label{excond}
\begin{split}
    \Bigl\{&\left[\hat{\boldsymbol{p}}_2\cdot(\boldsymbol{p}_4-\boldsymbol{p}_1)\right]^{\sigma-1}+\left[\hat{\boldsymbol{p}}_3\cdot(\boldsymbol{p}_4-\boldsymbol{p}_1)\right]^{\sigma-1}-\left[\hat{\boldsymbol{p}}_5\cdot(\boldsymbol{p}_4-\boldsymbol{p}_1)\right]^{\sigma-1}\\
    &-\left(\hat{\boldsymbol{p}}_6\cdot(\boldsymbol{p}_4-\boldsymbol{p}_1)\right)^{\sigma-1}\Bigr\}\Bigr\rvert_{z=1/a_4}\neq0.
\end{split}
\end{equation}
It is useful to verify that the already known exceptional type $B_2$ theories actually pass the proposed test. The $\C P^1$ NLSM with $\sigma=1$ passes trivially. It is likewise easy to see that the Schr\"{o}dinger-DBI theory, where $\sigma=2$, passes thanks to the four-particle kinematical relation $\hat{\boldsymbol{p}}_2+\hat{\boldsymbol{p}}_3-\hat{\boldsymbol{p}}_5-\hat{\boldsymbol{p}}_6=\vek0$, valid at $z=1/a_4$ in the limit $a_4\to a_1$.

%%%%%%%%%%%%%%%%%%%%%%%%%%%%%%%%%%%%%%%%%%%%%%%

\subsubsection{Beyond $\sigma=2$}
\label{subsec:6.3.2}

It is now easy to see that the leading contribution to the spurious pole, displayed in eq.~\eqref{consistcond}, does not vanish for a generic kinematical configuration for any $\sigma\geq2$. It is again sufficient to consider the limit in which $\hat p_1$ and $\hat p_4$ vanish so that the momenta $\hat p_2,\hat p_3,\hat p_5,\hat p_6$ define on-shell four-particle Schr\"odinger kinematics. Using the shorthand notation $\vek p_4-\vek p_1\equiv\vek v$, we are then in other words asking whether the condition
\begin{equation}
(\hat{\vek p}_2\cdot\vek v)^{\sigma-1}+(\hat{\vek p}_3\cdot\vek v)^{\sigma-1}=(\hat{\vek p}_5\cdot\vek v)^{\sigma-1}+(\hat{\vek p}_6\cdot\vek v)^{\sigma-1}
\label{jabber1}
\end{equation}
with some fixed vector $\vek v$ can be satisfied simultaneously with the energy and momentum conservation conditions
\begin{equation}
\hat{\vek p}_2^2+\hat{\vek p}_3^2=\hat{\vek p}_5^2+\hat{\vek p}_6^2,\qquad
\hat{\vek p}_2+\hat{\vek p}_3=\hat{\vek p}_5+\hat{\vek p}_6.
\label{jabber2}
\end{equation}

It is once again obvious that eq.~\eqref{jabber1} is satisfied for both $\sigma=1$ and $\sigma=2$. To show that it cannot in general hold for $\sigma\geq3$, it is sufficient to find a suitable kinematical configuration that violates it. The menu is vast, but one particularly simple and transparent choice is the limit in which one of the momenta, say $\hat{\vek p}_2$, is very small. In this limit, the remaining momenta satisfy the three-particle kinematical conditions
\begin{equation}
\hat{\vek p}_3^2=\hat{\vek p}_5^2+\hat{\vek p}_6^2,\qquad
\hat{\vek p}_3=\hat{\vek p}_5+\hat{\vek p}_6.
\label{jabber2}
\end{equation}
We can for instance choose $\hat{\vek p}_3$ to point in the direction of $\vek v$, and both $\hat{\vek p}_{5},\hat{\vek p}_{6}$ to make the angle $\pi/4$ with $\vek v$. Then $\abs{\hat{\vek p}_5}=\abs{\hat{\vek p}_6}=\abs{\hat{\vek p}_3}/\sqrt2$, and eq.~\eqref{jabber1} reduces to
\begin{equation}
1=2^{2-\sigma}.
\label{jabber3}
\end{equation}
This is satisfied for $\sigma=2$ but no $\sigma\geq3$, as we wanted to show.\footnote{Note that eq.~\eqref{jabber3} does not contradict the conclusion that $\sigma=1$ passes the consistency test. Upon taking the limit $\hat{\vek p}_2\to\vek0$, we removed the first term from eq.~\eqref{jabber1}, which is of course not the case for $\sigma=1$.}

%%%%%%%%%%%%%%%%%%%%%%%%%%%%%%%%%%%%%%%%%%%%%%%

\section{Summary and comparison of different approaches}
\label{sec:comparison}

In this paper, we have employed an arsenal of different methods to explore nonrelativistic EFTs with enhanced soft limits. Here we briefly summarize our main findings, focusing on the pros and cons of the various approaches. We divide the discussion into two parts, addressing respectively the symmetry-based top-down approach and the bottom-up approach based on on-shell recursion.

As stressed in the introduction, the motivation behind the Lie-algebraic approach to the classification of EFTs with enhanced soft limits is that all currently known examples of such theories do possess redundant symmetry. In fact, the authors of ref.~\cite{Cheung2017a} have proven the one-way implication that invariance under polynomial shifts of NG fields of order $\sigma$ in spacetime coordinates guarantees scaling of the scattering amplitudes in the soft limit with the scaling parameter $\sigma$ or higher. With this in mind, one of us worked out a classification of nonrelativistic EFTs with spatial redundant symmetry~\cite{Brauner2021a}, which roughly speaking corresponds to invariance under shifts polynomial in \emph{spatial} coordinates. This classification is reviewed and slightly extended in section~\ref{sec:2}. The first main observation of section~\ref{sec:sym} is that invariance under spatial polynomial symmetry is no longer sufficient to guarantee particular scaling of scattering amplitudes in the soft limit. The second main result is that it is still possible to derive soft theorems controlling the soft scaling parameter $\sigma$, if one uses as an additional input information about the dispersion relation of the NG bosons. Thus, invariance under polynomial shift symmetry of degree $n$ in spatial coordinates, together with dispersion relation $\omega\propto|\vek p|^m$, implies (under some mild regularity assumptions) that $\sigma\geq\min(m,n+1)$. By combining this result with the generalized CHMW theorem, we showed that the infrared behavior of scattering amplitudes crucially depends on what type of NG bosons are being scattered.

The Lie-algebraic approach has proven very successful in identifying concrete examples of EFTs with enhanced soft limits. One of its advantages is that it can treat on the same footing theories with an in principle arbitrary number of NG flavors. Moreover, it can easily generate subleading contributions to the effective Lagrangian. The main drawback of the Lie-algebraic approach probably is that it is based on an a priori assumption on the soft scaling parameter $\sigma$. The setup used previously in refs.~\cite{Bogers2018a,Brauner2021a} as well as here contains a single layer of (spacetime or spatial) vector redundant generators, which corresponds to $\sigma=2$. The complexity of the classification problem rapidly increases as additional layers of redundant generators are added for higher $\sigma$. While it was still feasible in ref.~\cite{Bogers2018b} to identify the special Galileon theory by adding a set of rank-two tensor generators corresponding to $\sigma=3$, going to even higher $\sigma$ would be very cumbersome. This should be contrasted with the on-shell recursion approach of ref.~\cite{Cheung2017a}, which allowed to rule out all theories with $\sigma>3$ at once.

The bottom-up approach that we employed in sections~\ref{sec:recursion}, \ref{sec:bootstrap} and~\ref{sec:bounds} is based on three main ingredients: seed amplitudes, on-shell recursion relations and soft bootstrap. The seed amplitudes, constructed in section~\ref{sec:bootstrap} for type $A_1$ and type $B_2$ theories, implement the basic properties of locality, energy and momentum conservation and Bose symmetry. They correspond to contact operators in the effective Lagrangian but, unlike the latter, do not suffer from ambiguities due to integration by parts and field redefinitions. Thus, they offer a very efficient way to encode the leading interactions among a given set of NG bosons. The crucial step of the bottom-up approach is the on-shell recursion, allowing one to iteratively reconstruct higher-point amplitudes from the seeds. The nonrelativistic recursion relations, derived previously in ref.~\cite{us} and reviewed in section~\ref{sec:recursion}, rely on the assumption that the scattering amplitudes have an enhanced soft limit, with given value of $\sigma$.

The soft bootstrap is an algorithmic procedure based on on-shell recursion that allows one to check which of the seed amplitudes actually correspond to a consistent EFT, and to rule out a priori theories with certain values of $\sigma$. Thus, in section~\ref{sec:bootstrap} we used numerical soft bootstrap to construct six-point amplitudes out of the seeds found therein. By investigating the consistency of the six-point amplitude, we were able to rule out a range of combinations of the seed amplitudes as unphysical. Consistent combinations of seed amplitudes were mapped to EFTs found using the Lie-algebraic classification in section~\ref{sec:2}. Interestingly, we did not discover any exceptional type $A_1$ theories beyond the well-known relativistic exceptional theories. In this sense, Lorentz symmetry emerged from the numerical bootstrap.

In section~\ref{sec:bounds}, we obtained bounds on the landscapes of type $A_1$, $A_2$ and $B_2$ EFTs by combining analytic soft bootstrap with the generalized CHMW theorem. These bounds constrain \textit{how} enhanced the scattering amplitudes of an EFT may be without violating locality. Motivated by the emergence of  Lorentz symmetry observed in section~\ref{sec:bootstrap}, we went on to prove that all exceptional type $A_1$ theories in fact are Lorentz-invariant. This promotes the relativistic no-go theorem for exceptional EFTs with $\sigma>3$~\cite{Cheung2017a} to the whole type $A_1$ subfamily in $D\geq 4$ dimensions. We also proved a novel no-go theorem for exceptional Schr\"odinger-type theories with $\sigma>2$. This means that the only exceptional theories of a single complex Schr\"odinger scalar in $D\geq 4$ dimensions are the $\C P^1$ NLSM and the Schr\"odinger-DBI theory. There is no type $B_2$ analog of the special Galileon. 

As is clear from the above, the top-down approach based on symmetry and the bottom-up approach based on on-shell recursion are largely complementary to each other. On the one hand, the top-down approach is able to quickly produce concrete examples of EFTs with desired properties. It however becomes extremely inefficient when the task is to show that no \emph{other} EFTs with the prescribed soft properties exist. On the other hand, the bottom-up approach is an invaluable tool to discard the existence of theories with given particle spectrum and soft behavior. However, it cannot be used to prove that a given set of seed amplitudes actually yields a consistent complete tree-level $S$-matrix; that would require infinitely many recursion steps. The complete EFT has to be constructed by other means, for instance the coset construction~\cite{Coleman1969a,Callan1969a}.

%%%%%%%%%%%%%%%%%%%%%%%%%%%%%%%%%%%%%%%%%%%%%%%

\section{Outlook}
\label{sec:outlook}

The modern scattering amplitude program aspires not only to supply practitioners with efficient tools for computation of scattering amplitudes, but also, and perhaps more importantly, to build new foundations of quantum field theory itself. Should this ambition succeed, it is mandatory not to remain limited to the realm of Lorentz-invariant field theory. The present work along with ref.~\cite{us} constitutes a first step in the program of extending results of the study of Lorentz-invariant scattering amplitudes to theories without Lorentz invariance. Here we considered only the properties of amplitudes of a single type of NG boson in the single soft limit. There are however many other aspects of nonrelativistic scattering amplitudes that await being explored.

Apart from extending the present study to theories with multiple NG boson flavors~\cite{Brauner2021a}, some natural directions for further study include other kinematical limits than the single soft limit, theories of particles with nonzero spin, or scattering amplitudes in nonrelativistic string theory. More ambitiously, one could use the bottom-up bootstrap approach to search for hidden structures in nonrelativistic field theories that may or may not resemble known relations and dualities among relativistic theories. The recently proposed Kawai-Lewellen-Tye bootstrap program~\cite{Chi2021} for generalizing the double copy could be an inspiring starting point for work in this direction. Last but not least, the analogy with Lorentz-invariant theories suggests that the exceptional type $B_2$ EFTs found here may provide interesting benchmark models for future studies of nonrelativistic theories with special properties.

%%%%%%%%%%%%%%%%%%%%%%%%%%%%%%%%%%%%%%%%%%%%%%%

\acknowledgments

T.B.~would like to thank Karol Kampf and Riccardo Penco for discussions regarding generation of random on-shell kinematical variables for scattering amplitudes. M.A.M.~acknowledges partial support from the DFG Collaborative Research Centre ``Neutrinos and Dark Matter in Astro- and Particle Physics'' (SFB 1258). M.A.M.~also acknowledges the hospitality of the University of Stavanger where part of the work was done. This work has been supported in part by the grant no.~PR-10614 within the ToppForsk-UiS program of the University of Stavanger and the University Fund.

%%%%%%%%%%%%%%%%%%%%%%%%%%%%%%%%%%%%%%%%%%%%%%%

\appendix

%%%%%%%%%%%%%%%%%%%%%%%%%%%%%%%%%%%%%%%%%%%%%%%

\section{Numerical calculation of scattering amplitudes}
\label{sec:numerics}

Some of the material presented in the main text of this paper relies on numerical evaluation of tree-level scattering amplitudes. This applies in particular to the scans for new theories featuring enhanced soft limits. However, we have also used numerical computation to check the expected behavior of scattering amplitudes predicted using other methods, for instance on-shell recursion. In this appendix, we briefly discuss the bundle of \textsc{Wolfram Mathematica}\textsuperscript{\textregistered} codes that we used for the purpose.

We do not reproduce here the complete code used for our numerical computations. We do, however, describe the main algorithms used. Our approach to evaluation of tree-level scattering amplitudes is modular in nature. The code consists of three fairly independent units, dedicated respectively to the construction of topologies of Feynman diagrams, extraction of Feynman rules from a given Lagrangian, and generation of on-shell kinematical variables consistent with energy and momentum conservation. We discuss these one by one.

%%%%%%%%%%%%%%%%%%%%%%%%%%%%%%%%%%%%%%%%%%%%%%%

\subsection{Feynman diagram topologies}
\label{subsec:topologies}

With an increasing number $n$ of participating particles, the number of Feynman diagrams grows rapidly. It is therefore necessary to automatize the construction of all topologies of Feynman diagrams for given $n$. In the terminology of graph theory, this amounts to finding all connected tree graphs with exactly $n$ vertices of unit valency (``external legs'') and all other vertices of valency greater than or equal to three (``interaction vertices''). Each of the external legs is decorated with a label, representing the momentum of the particle entering the graph through that leg.

\begin{table}[t]
\centering
\begin{tabular}{c|c|c|c|c|c|c}
Diagrams & All & \parbox{2.6cm}{Only cubic and\\ quartic vertices} & \parbox{1.8cm}{Only cubic\\ vertices} & \parbox{1.6cm}{No cubic vertices} & \parbox{1.7cm}{Only even\\ vertices} & \parbox{1.7cm}{Charge\\ conserving}\\
\hline
\hline
4-point & 4 & 4 & 3 & 1 & 1 & 1\\
5-point & 26 & 25 & 15 & 1 & 0 & 0\\
6-point & 236 & 220 & 105 & 11 & 11 & 10\\
7-point & 2\,752 & 2\,485 & 945 & 36 & 0 & 0\\
8-point & 39\,208 & 34\,300 & 10\,395 & 372 & 337 & 265\\
9-point & 660\,032 & 559\,405 & 135\,135 & 2\,311 & 0 & 0\\
10-point & 12\,818\,912 & 10\,525\,900 & 2\,027\,025 & 26\,252 & 20\,267 & 13\,401
\end{tabular}
\caption{Some benchmark values for the numbers of tree-level Feynman diagrams with up to ten participating particles. The second column displays the total number of diagrams without any restrictions on the interaction vertices. The notation for the other columns is largely self-explanatory. ``Even vertices'' are vertices with even valency. ``Charge-conserving'' diagrams are oriented diagrams with only even vertices such that the numbers of legs entering and leaving each vertex are equal.}
\label{tab:graphcount}
\end{table}

The construction was carried out recursively using the built-in graph theory tools of \textsc{Wolfram Mathematica}\textsuperscript{\textregistered}. Starting with a single seed three-point graph, the set of all $(n+1)$-point graphs was obtained from the already available set of $n$-point graphs by attaching a new external leg in all possible ways. The latter include connecting the new leg to all already existing interaction vertices in the graph, and to all edges (``propagators'') of the graph, thereby creating a new cubic vertex. The growth of the number of Feynman diagram topologies with $n$ is well illustrated by table~\ref{tab:graphcount}. Evaluating the tree-level amplitude on a desktop computer quickly becomes infeasible for $n>10$, unless we can make some assumptions on the structure of interaction vertices. As the table shows, the computational workload is for instance much lower in theories with only even interaction vertices. In fact, for $n\leq10$, the total number of diagrams is largely dominated by graphs containing cubic vertices. In practice, it is therefore clearly advantageous to remove cubic interaction terms from the Lagrangian by a field redefinition. This is always possible in derivatively coupled Lorentz-invariant scalar theories~\cite{Cheung2017a}.

Once all the Feynman graph topologies were generated, the individual diagrams were decorated with labels facilitating later application of Feynman rules. Thus, each propagator was labeled with the momentum flowing through it. In tree diagrams, this momentum is most easily determined by cutting the propagator and adding up all external momenta appearing in one of the resulting connected components of the graph. With all the propagators decorated, each interaction vertex was decorated with a list of momenta flowing through the propagators adjacent to the vertex.

The above steps are sufficient for theories of a single real scalar. In multiflavor theories (not considered in this paper), an additional label indicating flavor is needed for each external leg and each propagator. In theories of a complex scalar, the graphs have to be oriented so as to indicate the flow of the $\gr{U}(1)$ charge. One only needs to keep those diagrams where all interaction vertices have even valency and equal numbers of incoming and outgoing propagators. Such graphs are represented by the last column of table~\ref{tab:graphcount}.

The Feynman diagram topologies are independent of the choice of interaction Lagrangian. They were therefore generated once for all, separately for real scalar and complex scalar theories, and stored for later reuse in an output file.

%%%%%%%%%%%%%%%%%%%%%%%%%%%%%%%%%%%%%%%%%%%%%%%

\subsection{Feynman rules}
\label{subsec:FRules}

Generating the set of Feynman rules from a given interaction Lagrangian is an exercise in list manipulation. It is practically convenient to split up the Lagrangian explicitly into parts containing interaction vertices with different valency,
\begin{equation}
\La_{\text{int}}=\sum_{n=3}^{n_{\text{max}}}\La_n.
\end{equation}
Each $\La_n$ is then expanded into a sum of monomials. A given monomial consists of a numerical coefficient multiplying an operator composed solely of the NG field and its derivatives. For the purposes of automatic generation of Feynman rules, the operator can be represented as a nested list. At the top level, each element of the list corresponds to one factor of the NG field $\t$ in the operator. Each such element is itself a list of indices indicating derivatives acting on $\t$; for instance $\{0,r,s,t\}$ stands for $\de_0\de_r\de_s\de_t\t$. This encodes all the information needed to generate the Feynman rules.

There are some details of the procedure one needs to pay particular attention to. For instance, one has to remember to sum over all permutations of legs in each monomial to get the correct Feynman rule. Also, dummy indices appearing in a given monomial need to be labeled with a unique identifier each. In fact, it appears practically convenient to avoid using dummy indices altogether as far as possible. This can be achieved by representing derivatives of $\t$ with tensors. Manipulation and simplification of tensor contractions was done using the \textsc{TensorSimplify} package by Carl Woll. Once generated, the Feynman rules characterizing a given class of Lagrangians were stored for later reuse in an output file, just like the Feynman diagram topologies.

%%%%%%%%%%%%%%%%%%%%%%%%%%%%%%%%%%%%%%%%%%%%%%%

\subsection{Random kinematics}
\label{subsec:kinematics}

Importantly, vanishing of the scattering amplitudes of NG bosons, whether through the ordinary Adler zero or its enhancements, requires that all the particles participating in the scattering process be on-shell. In addition, the total energy and momentum in the scattering process must be conserved. For the textbook problem of scattering of four massless scalars in Lorentz-invariant theories, these requirements boil down to the fact that the invariant amplitude is a function of the Mandelstam variables $s,t,u$ that satisfy the constraint $s+t+u=0$. For a higher number of particles, or in theories lacking Lorentz invariance, the on-shell and conservation constraints may however be highly nontrivial. It is then often more feasible to replace analytic manipulations of amplitudes as functions of external momenta and energies with ``numerical experiments.'' In the latter, one generates randomly a set of momenta satisfying all the kinematical constraints and evaluates the amplitude numerically. Generic features of a given theory can then be tested by simply repeating the computation with different sets of random momenta. In this subsection, we describe in detail how to generate random sets of kinematical data in two classes of theories: type $A_1$ theories of a real scalar, and type $B_2$ theories of a complex scalar.

%%%%%%%%%%%%%%%%%%%%%%%%%%%%%%%%%%%%%%%%%%%%%%%

\subsubsection{Type $A_1$ theories}

Theories of type $A_1$ possess the same kinematics as Lorentz-invariant theories; Lorentz boosts are only broken by the interactions. We can therefore use the insight gained by studying genuinely Lorentz-invariant theories. The discussion below mirrors closely the random kinematics generator contained in the \textsc{BCFW} package introduced in ref.~\cite{Bourjaily2010a}.

The starting point is the observation that in $d=3$ spatial dimensions, any on-shell massless four-momentum $p^\mu$ can be represented by a $2\times2$ Hermitian matrix $P$, defined by
\begin{equation}
P\equiv\begin{pmatrix}
p^0+p^3 & p^1-\im p^2\\
p^1+\im p^2 & p^0-p^3
\end{pmatrix},
\end{equation}
where the on-shell condition is equivalent to $\det P=0$. As a consequence, the matrix $P$ can be represented in terms of the projector to its sole eigenvector with nonzero eigenvalue, or equivalently as $P=\pm\lambda\lambda^\dagger$, where $\lambda\in\C^2$ and the sign corresponds to the sign of the energy $p^0$. This in turn determines whether the four-momentum represents a particle in the initial or final state of the scattering process. We use the convention wherein the positive sign of $p^0$ corresponds to outgoing particles and the negative sign to incoming particles.

A set of $n$ four-momenta satisfying all the kinematical constrains can then be generated as follows. We choose randomly (including random signs) the first $n-2$ four-momenta. The first of these is rescaled by the soft factor $\eps$, which allows one to probe the soft limit by taking $\eps\to0$. Let us now denote
\begin{equation}
\sum_{i=1}^{n-2}P_i\equiv
\begin{pmatrix}
\alpha & \beta+\im\gamma\\
\beta-\im\gamma & \delta
\end{pmatrix}
\label{Psum}
\end{equation}
with $\alpha$, $\beta$, $\gamma$ and $\delta$ being all real. Let us also introduce a notation for the spinor $\lambda_n$ representing the last particle,
\begin{equation}
\lambda_n=\begin{pmatrix}
w\\
x
\end{pmatrix},
\end{equation}
where we assume without loss of generality that $w$ is real. By the conservation of energy and momentum, the four-momentum of the $(n-1)$-th particle is then fixed by
\begin{equation}
P_{n-1}=-\sum_{i=1}^{n-2}P_i-\lambda_n\lambda_n^\dagger=-\begin{pmatrix}
\alpha+w^2 & \beta+\im\gamma+wx^*\\
\beta-\im\gamma+wx & \delta+xx^*
\end{pmatrix}.
\label{Ps}
\end{equation}
The only kinematical constraint that remains to be satisfied is that the four-momentum represented by $P_{n-1}$ actually is on-shell, that is, $\det P_{n-1}=0$. This gives us a single condition for the real variable $w$ and the complex variable $x$. We can use it for instance to solve for $w$ in terms of $x$, which gives two solutions,
\begin{align}
w_\pm&=\frac1{2\delta}\Bigl[x(\beta+\im\gamma)+x^*(\beta-\im\gamma)\pm\sqrt\Delta\Bigr],\\
\Delta&\equiv[x(\beta+\im\gamma)+x^*(\beta-\im\gamma)]^2+4\delta(\beta^2+\gamma^2-\alpha\delta-\alpha xx^*).
\end{align}
A particularly simple setup follows if we choose $x=0$, that is orient the $n$-th particle along the third coordinate axis. Namely, we then find
\begin{equation}
P_{n-1}=-\begin{pmatrix}
\frac{\beta^2+\gamma^2}{\delta} & \beta+\im\gamma\\
 \beta-\im\gamma & \delta
\end{pmatrix},\qquad
P_n=\begin{pmatrix}
\frac{\beta^2+\gamma^2}{\delta}-\alpha & 0\\
0 & 0
\end{pmatrix}.
\label{last2P}
\end{equation}
The great advantage of this choice is that the four-momenta of the $(n-1)$-th and the $n$-th particle are rational functions of the four-momenta of the first $n-2$ particles. It is therefore possible to generate consistent kinematics where all $n$ four-momenta take rational values. This is very useful for the numerical analysis of the scaling of scattering amplitudes in the limit $\eps\to0$, as one can take advantage of infinite-precision arithmetic.

The above-described algorithm to generate random kinematics however has a problem for $n=4$. In this case, the sum in eq.~\eqref{Psum} reduces to a single term in the limit $\eps\to0$ where the first four-momentum vanishes. Accordingly, in this limit, $\beta^2+\gamma^2-\alpha\delta\to0$. But then by eq.~\eqref{last2P} the $n$-th four-momentum also vanishes. In other words, for $n=4$ this generator does not capture a single soft limit, but rather a limit in which two of the four four-momenta vanish simultaneously. This problem can be avoided by not setting $x$ to zero, but rather to some other randomly chosen number. The price to pay is that the four-momenta~\eqref{last2P} are then no longer rational-valued.

%%%%%%%%%%%%%%%%%%%%%%%%%%%%%%%%%%%%%%%%%%%%%%%

\subsubsection{Type $B_2$ theories conserving the number of particles}

For theories of a complex scalar, conserving the number of particles, only scattering processes with equal numbers $n$ of incoming and outgoing particles are possible. We adopt the convention that the first, third etc.~particle is incoming, whereas the second, fourth etc.~particle is outgoing. In terms of spatial momenta $\vek p_i$ and the corresponding energies $\om_i$, the energy and momentum conservation laws then dictate that
\begin{equation}
\sum_{i=1}^{2n}(-1)^i\om_i=0,\qquad
\sum_{i=1}^{2n}(-1)^i\vek p_i=\vek 0.
\end{equation}
The energies are by definition all positive and satisfy the on-shell condition $\om_i=\vek p_i^2$.

The algorithm to generate consistent random kinematics now goes as follows. We choose at will $2n-2$ of the spatial momenta and calculate the corresponding energies. As for type $A_1$ theories, we can again introduce the soft factor $\eps$ for the first particle via the replacement $(\om_1,\vek p_1)\to(\eps^2\om_1,\eps\vek p_1)$. Let us take the staggered sum of these $2n-2$ energies and momenta,
\begin{equation}
\mathcal E\equiv\sum_{i=1}^{2n-2}(-1)^{i+1}\om_i=\sum_{i=1}^{2n-2}(-1)^{i+1}\vek p_i^2,\qquad
\vek\Pi\equiv\sum_{i=1}^{2n-2}(-1)^{i+1}\vek p_i.
\end{equation}
To find the remaining momenta $\vek p_{2n-1}$ and $\vek p_{2n}$, we need to choose the spatial dimension $d$. Here we will work out in detail the case $d=3$, but other choices of $d$ require just a minor adjustment of the argument. Let us now write $\vek p_{2n-1}$ and $\vek p_{2n}$ as
\begin{equation}
\vek p_{2n-1}=(a,b,c),\qquad
\vek p_{2n}=(A,B,C).
\end{equation}
The corresponding energies are defined by the on-shell condition. The only remaining constraints to be satisfied are those implied by energy and momentum conservation. Out of the six components of $\vek p_{2n-1}$ and $\vek p_{2n}$, we can therefore choose two, e.g.~$A$ and $B$, arbitrarily. The remaining components are then uniquely determined,
\begin{equation}
\begin{gathered}
a=A-\Pi^1,\qquad
c=\frac1{2\Pi^3}\bigl[\mathcal E+(\Pi^1)^2+(\Pi^2)^2-(\Pi^3)^2-2A\Pi^1-2B\Pi^2\bigr],\\
b=B-\Pi^2,\qquad
C=\frac1{2\Pi^3}(\mathcal E+\vek\Pi^2-2A\Pi^1-2B\Pi^2).
\end{gathered}
\end{equation}
For type $B_2$ kinematics, it is therefore always possible to randomly generate rational-valued kinematical data. There is no problem to generate data even for $n=2$ (four particles) so that the single soft limit can be taken by setting $\eps\to0$.

%%%%%%%%%%%%%%%%%%%%%%%%%%%%%%%%%%%%%%%%%%%%%%%

\section{Exceptional contact four-point amplitudes}
\label{sec:4ptansatz}

The no-go theorems for the existence of super-exceptional theories, discussed in section~\ref{sec:bounds}, rely on detailed understanding of contact four-point amplitudes in exceptional theories. We work these out in the present appendix. Note that the arguments given below are formally only valid in $D\geq4$ spacetime dimensions. This is because we assume that there are no other constraints on the (relativistic or nonrelativistic) Mandelstam variables than those implied by energy and momentum conservation and the on-shell condition. 

%%%%%%%%%%%%%%%%%%%%%%%%%%%%%%%%%%%%%%%%%%%%%%%

\subsection{Type $A_1$ theories}
\label{sec:A1}

In the case of type $A_1$ theories we will follow the convention introduced in section~\ref{subsec:seedA1} and express the contact four-point amplitude in terms of the relativistic Mandelstam variables $s,t,u$ and the energies $\omega_1,\omega_2,\omega_3,\omega_4$. Any four-point amplitude constructed purely out of $s,t,u$ is automatically exceptional. Below, we show that there are no other exceptional four-point amplitudes. That is, we show that no four-point amplitude explicitly depending on the energies can be exceptional.

We start by writing down the most general manifestly $S_4$-invariant polynomial in our variables,
\begin{align}
    A_4=\sum_{a,b,c,d,e,f,g} \lambda_{a\dotsb g}s^at^bu^c\omega_1^d\omega_2^e\omega_3^f\omega_4^g+\text{permutations},
\end{align}
where $a,\dotsc,g$ are non-negative integers and $\lambda_{a\dotsb g}$ a priori undetermined coefficients. Next, we eliminate the ambiguity in the polynomial form of $A_4$ at the cost of sacrificing manifest $S_4$-invariance. We do so by expressing $s$ and $\omega_4$ in terms of the other variables using energy and momentum conservation,
\begin{equation}
s^a=(-t-u)^a,\qquad
\omega_4^g=(-\omega_1-\omega_2-\omega_3)^g. 
\label{swap12}
\end{equation}
The amplitude $A_4$ now becomes a sum of monomials of the form
\begin{equation}
t^\alpha u^\beta\omega_1^x\omega_2^y\omega_3^z.
\label{A1monomial}
\end{equation}
Since there are by assumption no further relations among the remaining variables $t,u$ and $\omega_1,\omega_2,\omega_3$, the exceptional scaling of the amplitude must now be manifest \emph{term by term}, that is for each monomial contributing to $A_4$.

Demanding exceptional scaling of the amplitude in the single soft limit for the first particle requires that each monomial of the form~\eqref{A1monomial} satisfies the constraint $x\geq y+z$. However, should the amplitude be at the same time exceptional in the soft limit for the second particle, the corresponding inequality $y\geq x+z$ must hold. Finally, requiring exceptional scaling for the third particle leads to $z\geq x+y$. Obviously, all these three inequalities can only be satisfied simultaneously if $x=y=z=0$. This concludes the argument that the contact four-point amplitude in any exceptional theory in $D\geq4$ spacetime dimensions is necessarily a polynomial in the relativistic Mandelstam variables, hence is Lorentz-invariant.

%%%%%%%%%%%%%%%%%%%%%%%%%%%%%%%%%%%%%%%%%%%%%%%

\subsection{Type $B_2$ theories}
\label{sec:B2}

As explained in section~\ref{subsec:bootstrapB2}, the four-point amplitude in type $B_2$ (Schr\"odinger-type) theories can be expressed in terms of the following kinematical variables,
\begin{align}
    \label{B2mandels}
    s_{13}=s_{24},\quad s_{12},\quad s_{14}, \quad s_{23}, \quad s_{34}, \qquad\qquad s_{ij}\equiv \boldsymbol{p}_i\cdot\boldsymbol{p}_j.
\end{align}
The particles in the initial (final) state are labeled with indices $1,3$ ($2,4$). In $d\geq3$ spatial dimensions, which we will henceforth assume, the variables~\eqref{B2mandels} are mutually independent. The contact four-point amplitude is then given by a unique polynomial in these variables that is required to be invariant under the permutation group $S_2\times S_2$, acting separately on the incoming and outgoing particles, as well as under charge conjugation ($C$). Note that thanks to the constraint $s_{13}=s_{24}$, $s_{13}$ itself is invariant under all the required symmetries. It is thus convenient to highlight the dependence of the four-point amplitude on $s_{13}$. A generic exceptional amplitude with soft scaling parameter $\sigma$ and $\tau=\sigma/2$ derivatives per field is then a polynomial of degree $\sigma$ in the variables~\eqref{B2mandels} that can be written in the form
\begin{align}
    \label{sum1}
    A_4=\sum_{n=0}^{\sigma}c_ns_{13}^n(\dotsb)_{\sigma-n},
\end{align}
where $c_n$ are undetermined coefficients and the shorthand notation $(\dotsb)_m$ indicates a polynomial of degree $m$ in the variables $s_{12},s_{14},s_{23},s_{34}$. Since the variables~\eqref{B2mandels} are algebraically independent and the coefficients of $s_{13}^n$ in eq.~\eqref{sum1} thus unique, all the polynomials $(\dotsb)_{\sigma-n}$ therein are necessarily themselves invariant under both $S_2\times S_2$ and $C$.

Unlike in the case of type $A_1$ kinematics, we can \emph{not} impose the assumed exceptional scaling individually on each term in eq.~\eqref{sum1}. The reason for this is that our variables~\eqref{B2mandels}, while independent, satisfy a set of momentum conservation relations of the type
\begin{equation}
s_{12}-s_{13}+s_{14}=s_{11}=\vek p_1^2,
\label{cancellation}
\end{equation}
which may lead to soft scaling with a higher $\sigma$ than naively expected. We will therefore follow a different approach. Namely, we will use induction in $\sigma$ to prove that any exceptional four-point amplitude with soft scaling parameter $\sigma$ is necessarily proportional to $s_{13}^\sigma$. In other words, only the coefficient $c_\sigma$ in eq.~\eqref{sum1} may be nonzero. The exceptional theories with $\sigma=1$ and $\sigma=2$ are already known and can be identified using the approach of section~\ref{subsec:bootstrapB2}. The only such exceptional theories are the $\C P^1$ NLSM and the Schr\"odinger-DBI theory, both of which satisfy our induction hypothesis.

Let us now take the induction step. We write eq.~\eqref{sum1} as
\begin{equation}
A_4=c_0(\dotsb)_\sigma+s_{13}\sum_{n=1}^{\sigma}c_ns_{13}^{n-1}(\dotsb)_{\sigma-n}.
\label{inductionstep}
\end{equation}
We shall focus on the first term, assuming first that this term vanishes. Then the sum in the second term, $\sum_{n=1}^\sigma c_ns_{13}^{n-1}(\dotsb)_{\sigma-n}$, has the soft scaling parameter $\sigma-1$, and moreover has all the symmetries required of a four-point amplitude. By our induction hypothesis, it is therefore proportional to $s_{13}^{\sigma-1}$, hence $A_4\propto s_{13}^\sigma$ as we wanted to prove.

It remains to address the case where the first term on the right-hand side of eq.~\eqref{inductionstep} is nonzero. To that end, let us inspect more closely the properties of the polynomial $(\dotsb)_\sigma$. A general ansatz for $(\dotsb)_{\sigma}$ can be written as
\begin{align}
    \label{b2generell}
    (\dotsb)_{\sigma}=\Sum{a,b,c,d}{a+b+c+d=\sigma}\lambda_{abcd}s_{12}^as_{14}^bs_{23}^cs_{34}^d,
\end{align}
where $\lambda_{abcd}$ are a priori undetermined coefficients, and we employed the summation range notation introduced below eq.~\eqref{energy}. Terms with $c+d>\sigma/2$ are automatically forbidden, as they would inevitably spoil the exceptional scaling properties of $A_4$. $S_2\times S_2$ permutation invariance of eq.~(\ref{b2generell}) then implies that also terms where $c+d<\sigma/2$ are forbidden. This constrains possible contributions to $(\dotsb)_\sigma$ to terms with $a+b=c+d=\sigma/2$,
\begin{align}
    \label{b2generell2}
    (\dotsb)_{\sigma}=\Sum{a,b,c,d}{a+b=c+d=\sigma/2}\lambda_{abcd}s_{12}^as_{14}^bs_{23}^cs_{34}^d,
\end{align}
which among others implies that $(\dotsb)_\sigma$ can only be nonzero if $\sigma$ is even. Imposing moreover $C$-invariance, eq.~(\ref{b2generell2}) can be further reduced to a polynomial of the following form, 
\begin{align}
    \label{ansatzfinalb2}
    (\dotsb)_{\sigma}=\Sum{a,b}{a+b=\sigma/2}\lambda_{a}s_{12}^as_{14}^bs_{23}^bs_{34}^a. 
\end{align}

Suppose now that we take the single soft limit for the first particle. The only way how the soft scaling parameter could be enhanced beyond naive counting of powers of $\vek p_1$ in the amplitude is if cancellations of the type~\eqref{cancellation} occur. This means that should terms of the type $s_{23}^bs_{34}^a$ with $a+b=\sigma/2$ appear in the amplitude at all, then $A_4$ must necessarily take the following form to preserve its exceptional scaling, 
\begin{align}
    A_4=(s_{12}-s_{13}+s_{14})^{\sigma/2}\Sum{a,b}{a+b=\sigma/2}\lambda'_{a}s_{23}^bs_{34}^a+\dotsb,
\end{align}
where the ellipsis denotes terms proportional to $s_{13}$. Matching the terms herein independent of $s_{13}$ to the expression~\eqref{inductionstep} for $A_4$ combined with eq.~\eqref{ansatzfinalb2}, we get the condition
\begin{align}
   (s_{12}+s_{14})^{\sigma/2} \Sum{a,b}{a+b=\sigma/2}\lambda'_{a}s_{23}^bs_{34}^a=c_0\Sum{a,b}{a+b=\sigma/2}\lambda_{a}s_{12}^as_{14}^bs_{23}^bs_{34}^a.
\end{align}
This equality can however only be satisfied if both sides vanish, which contradicts our assumption that $c_0(\dotsb)_\sigma$ is nonzero. 

This concludes the proof that all physically consistent exceptional four-point amplitudes in type $B_2$ theories take the simple form
\begin{align}
    A_4=\lambda_{\sigma}s_{13}^{\sigma}. 
\end{align}

%%%%%%%%%%%%%%%%%%%%%%%%%%%%%%%%%%%%%%%%%%%%%%%

\section{Quarton theory}
\label{app:B4}

In this appendix we work out the basic properties of the exotic \exotic{} theory briefly outlined in section~\ref{subsubsec:typeB4}. Let us start by writing down its Lie algebra in order to avoid multiple cross-references to the main text of the paper,
\begin{equation}
\begin{aligned}
[P_r,K_{sA}]&=\im g_{rs}Q_A,\qquad
&[K_{rA},K_{sB}]&=\im\eps_{AB}g_{rs}H,\\
[Q,K_{rA}]&=-\im\eps_A^{\phantom AB}K_{rB},\qquad
&[Q,Q_A]&=-\im\eps_A^{\phantom AB}Q_B,
\end{aligned}
\label{comm}
\end{equation}
where the indices $A,B$ take values from the set $\{1,2\}$. All the other commutators among the generators $J_{rs}$, $P_r$, $K_{rA}$, $Q$, $Q_A$ and $H$ are either fixed by rotational invariance or zero. What makes the \exotic{} theory exotic, and genuinely nonrelativistic, is the commutator $[K_{rA},K_{sB}]$, proportional to the Hamiltonian. This is the only place where the Lie algebra of the \exotic{} theory differs from that of the Schr\"odinger-Galileon theory, discussed in section~\ref{subsubsec:SchrodingerGalileon}. The former can therefore be considered a nontrivial deformation of the latter.

To convert the Lie algebra into a concrete effective action, we use the coset construction~\cite{Coleman1969a,Callan1969a} in a form applicable to spontaneously broken spacetime symmetries~\cite{Volkov1973a,Ogievetsky1974a}. We start by writing down a parameterization of the coset space, including all nonlinearly realized symmetries,
\begin{equation}
U(t,x,\t,\x)\equiv e^{\im tH}e^{\im x^r P_r}e^{\im\t^AQ_A}e^{\im\x^{rA}K_{rA}}.
\end{equation}
It is instructive to work out the symmetry transformations generated by the Lie algebra~\eqref{comm}. The generators $H$, $P_r$, $Q_A$ act by trivial shifts on $t$, $x^r$, $\t^A$, respectively. The generator $Q$ acts on $\t^A$ and $\x^{rA}$ by $\gr{SO}(2)$ rotations. The only nontrivial transformation is that generated by $K_{rA}$, with parameter $\beta^{rA}$, which acts as
\begin{equation}
t\to t-\tfrac12\eps_{AB}g_{rs}\beta^{rA}\x^{sB},\qquad
\t^A\to\t^A+\beta^A_rx^r,\qquad
\x^{rA}\to\x^{rA}+\beta^{rA}.
\label{Ktransfo}
\end{equation}
This again looks just like the symmetry transformation in the Schr\"odinger-Galileon theory, deformed by the nontrivial transformation of $t$.

With the Lie algebra and the coset space parameterization at hand, it is straightforward to work out the Lie-algebra-valued Maurer-Cartan (MC) form,
\begin{equation}
\om\equiv-\im U^{-1}\dd U\equiv\frac12\om_J^{rs}J_{rs}+\om_P^rP_r+\om_H H+\om_K^{rA}K_{rA}+\om_Q^AQ_A+\om_QQ.
\end{equation}
The components $\om_J^{rs}$ and $\om_Q$ vanish. The other, nonvanishing components take the values
\begin{equation}
\begin{aligned}
\om^r_P&=\dd x^r,\qquad
&\om_H&=\dd t+\tfrac12\eps_{AB}g_{rs}\x^{rA}\dd\x^{sB},\\
\om^{rA}_K&=\dd\x^{rA},\qquad
&\om^A_Q&=\dd\t^A-\x^A_r\dd x^r.
\end{aligned}
\label{MCform}
\end{equation}

%%%%%%%%%%%%%%%%%%%%%%%%%%%%%%%%%%%%%%%%%%%%%%%

\subsection{Wess-Zumino terms}

With the MC form at hand, we first have to check whether the symmetry of the \exotic{} theory admits some WZ terms in the Lagrangian. If present, these would be likely to dominate the low-energy expansion of the \exotic{} EFT. In $d$ spatial dimensions, WZ terms correspond to closed $(d+2)$-forms invariant under all the symmetries of the theory. In order to be able to test closedness, we write down the MC structure equations, satisfied by the MC form,
\begin{equation}
\dd\om^r_P=0,\qquad
\dd\om_H=\tfrac12\eps_{AB}g_{rs}\om^{rA}_K\w\om^{sB}_K,\qquad
\dd\om^{rA}_K=0,\qquad
\dd\om^A_Q=\om^r_P\w\om^A_{Kr}.
\label{MCequation}
\end{equation}
Except for $\dd\om_H$, these are identical to the MC equations of the Schr\"odinger-Galileon algebra. We can thus inspect possible presence of WZ terms along the line of argument of section~3.3 of ref.~\cite{Brauner2021a}. Out of the five different types of WZ terms discovered therein, three are not relevant for the \exotic{} theory due to the absence of any other spontaneously broken scalar generators but $Q_A$. Of the remaining two types of WZ terms, only one seems to respect the MC equations~\eqref{MCequation}. The corresponding closed $(d+2)$-form is proportional to
\begin{equation}
\eps_{AB}\eps_{r_1\dotsb r_d}\om_Q^A\w\om_Q^B\w\dd x^{r_1}\w\dotsb\w\dd x^{r_d}.
\end{equation}
This integrates to the following action in the physical $d+1$ spacetime dimensions,
\begin{equation}
S_{\text{WZ}}\propto\int\dd t\,\dd^d\vek x\,\eps_{AB}\t^A\de_0\t^B.    
\end{equation}
It is easy to check that under~\eqref{Ktransfo}, the integrand changes by a total time derivative.

%%%%%%%%%%%%%%%%%%%%%%%%%%%%%%%%%%%%%%%%%%%%%%%

\subsection{Invariant Lagrangians}

Let us now see how to build strictly invariant Lagrangian densities out of the components of the MC form~\eqref{MCform}. To that end, we need a way to contract spacetime indices, which is done with the help of the spacetime vielbein. The latter is extracted from $\om_P^r$ and $\om_H$,
\begin{equation}
e^s_0=0,\qquad
e^s_r=\delta^s_r,\qquad
n_0=1+\frac12\eps_{AB}g_{rs}\x^{rA}\de_0\x^{sB},\qquad
n_r=\frac12\eps_{AB}g_{st}\x^{sA}\de_r\x^{tB}.
\label{vielbein}
\end{equation}
Together, $e^s$ and $n$ define a basis of 1-forms on the $(d+1)$-dimensional spacetime. The corresponding basis of vectors, dual to the vielbein, then is
\begin{equation}
E^0_s=-\frac{n_s}{n_0},\qquad
E^r_s=\delta^r_s,\qquad
V^0=\frac1{n_0},\qquad
V^r=0.
\label{dualvielbein}
\end{equation}
The dual vielbein is used to project ordinary spacetime derivatives to covariant derivatives,
\begin{equation}
\nabla_0\equiv V^0\de_0+V^r\de_r=\frac1{n_0}\de_0,\qquad
\nabla_r\equiv E^0_r\de_0+E^s_r\de_s=\de_r-\frac{n_r}{n_0}\de_0.
\end{equation}

With the vielbein and the covariant derivatives at hand, we can now project out the spatial part of $\om_Q^A$. Setting this to zero provides us with an inverse Higgs constraint~\cite{Ivanov1975a} that can be used to eliminate the unphysical field $\x^{rA}$ in favor of the physical NG fields $\t^A$,
\begin{equation}
\x^A_r=\nabla_r\t^A.
\end{equation}
The only components of the MC form that are left unused are $\om_K^{rA}$ and the temporal part of $\om_Q^A$. These provide us with covariant building blocks for the construction of scalar Lagrangian densities,
\begin{equation}
\nabla_0\t^A=\frac1{n_0}\de_0\t^A,\qquad
\nabla_0\x^{rA}=\frac1{n_0}\de_0\x^{rA}=\frac1{n_0}\de_0\nabla^r\t^A,\qquad
\nabla_s\x^{rA}=\nabla_s\nabla^r\t^A.
\end{equation}
Whatever scalar Lagrangian we build out of these must be accompanied by the appropriate volume measure, $\dd t\,\dd^d\vek x\,n_0\det e=\dd t\,\dd^d\vek x\,n_0$.

%%%%%%%%%%%%%%%%%%%%%%%%%%%%%%%%%%%%%%%%%%%%%%%

\subsection{Discussion}

Let us see what we can construct out of our building blocks. We would like to have a kinetic term in the first place. But the only way to build an operator that does not contain temporal derivatives is to use  $\nabla_r\nabla_s\t^A$. This suggests a minimal action, based on the Lie algebra~\eqref{comm}, of the form
\begin{equation}
S=S_{\text{WZ}}-\frac12\int\dd t\,\dd^d\vek x\,n_0\,\delta_{AB}(\nabla_r\nabla^r\t^A)(\nabla_s\nabla^s\t^B).
\label{B4action}
\end{equation}
This is a type $B_4$ theory. There are other operators one can construct that contribute to the quadratic part of the Lagrangian, such as $\nabla_0\t^A\nabla_0\t^B$ or $\nabla_0\t^A\nabla_r\nabla^r\t^B$. But these will be subleading for energy proportional to the fourth power of momentum. Hence the minimal action~\eqref{B4action} seems to be natural in the technical sense.

It is interesting to compare this to theories based on polynomial shift symmetries, discussed in refs.~\cite{Hinterbichler2014a,Griffin2015a}. Therein, the authors constructed higher-order WZ terms which are enhanced in the sense that they admit a second-order polynomial shift symmetry (which ensures a kinetic term with four spatial derivatives) but contain less than three derivatives per field. Still, these WZ terms contain (with the exception of the kinetic term) more than two derivatives per field. In contrast, the leading part of the action~\eqref{B4action}, and of any interaction Lagrangian built solely out of $\nabla_r\nabla_s\t^A$, contains exactly two derivatives per field, and will therefore dominate over the WZ terms of refs.~\cite{Hinterbichler2014a,Griffin2015a} in the infrared.

%%%%%%%%%%%%%%%%%%%%%%%%%%%%%%%%%%%%%%%%%%%%%%%

\section{Relations among Noether currents and soft theorems}
\label{sec:Noetherrelations}

The main goal of this appendix is to derive the relation~\eqref{Rconstrainttext} for the remainder function $R^\mu(p)$, following closely the original argument of Cheung et al.~\cite{Cheung2017a}. We will start by reviewing the basics of relations among Noether currents of locally indistinguishable symmetries~\cite{Brauner2014a,Brauner2020a}. These provide a useful tool for analyzing the implications of redundant symmetry for the matrix elements of the broken current.

%%%%%%%%%%%%%%%%%%%%%%%%%%%%%%%%%%%%%%%%%%%%%%%

\subsection{Relations among Noether currents}

We consider a generic local theory of a single real scalar field $\theta$ whose classical action is invariant under a global transformation of the type
\begin{align}
    \label{16IHC}
    \theta'(x)=\theta(x)+\epsilon\xi[\theta,x](x).
\end{align}
Here $\eps$ is a constant infinitesimal parameter and the notation for $\xi[\theta,x]$ is chosen to indicate that this is a local function of $\theta$ and its derivatives, possibly explicitly depending on $x$. The corresponding Noether current $J^\mu[\theta,x]$ is defined by evaluating the variation of the action under a transformation with a coordinate-dependent parameter $\eps(x)$ such that it reduces to eq.~\eqref{16IHC} when $\eps(x)=\eps$ is a constant,
\begin{align}
    \delta S=\int\dd^{d+1}x\, J^{\mu}\partial_{\mu}\epsilon.
\end{align}
This definition leaves the Noether current ambiguous under addition of an arbitrary vector function whose divergence is identically zero. The corresponding conservation law is however unambiguous.

Suppose now that the action $S$ enjoys two sets of global symmetries characterized by infinitesimal parameters $\epsilon_1^{\alpha}$ and $\epsilon_2^{a}$. Suppose further that the two sets of transformations are locally indistinguishable in the sense that there is a set of coefficients $f^a_\alpha[\theta,x]$ such that setting $\eps_2^a(x)=f^a_\alpha[\theta,x](x)\eps_1^\alpha(x)$ makes the two local transformations identical. Then the variation of the action can be written in two equivalent ways,
\begin{align}
    \delta S=\int\dd^{d+1}x\, J_{2a}^{\mu}\partial_{\mu}\epsilon_2^a=\int\dd^{d+1}x\,J_{2a}^{\mu}(\epsilon_1^{\alpha}\partial_{\mu}f^a_{\alpha}+f^a_{\alpha}\partial_{\mu}\epsilon_1^{\alpha})=\int\dd^{d+1}x\,J_{1\alpha}^{\mu}\partial_{\mu}\epsilon_1^{\alpha}.
\end{align}
The last equality induces a constraint to be satisfied off-shell,
\begin{align}
    \label{osc}
    J_{2a}^{\mu}\partial_{\mu}f^a_{\alpha}=\partial_{\mu}N^{\mu}_{\alpha},
\end{align}
where $N^\mu_\alpha[\theta,x]$ is a set of local functions of the field, its derivatives and the coordinates. The two Noether currents are then related by
\begin{align}
    J^{\mu}_{1\alpha}=f^a_{\alpha}J^{\mu}_{2a}-N^{\mu}_{\alpha}.
    \label{currentrelations}
\end{align}
By taking a divergence and combining the result with eq.~\eqref{osc}, we obtain another off-shell identity,
\begin{align}
    \label{genconservation}
    \partial_{\mu}J^{\mu}_{1\alpha}=f^a_{\alpha}\partial_{\mu}J^{\mu}_{2a}.
\end{align}
This illustrates that on-shell conservation of $J_{1\alpha}^{\mu}$ is a consequence of conservation of $J_{2a}^{\mu}$. 

%%%%%%%%%%%%%%%%%%%%%%%%%%%%%%%%%%%%%%%%%%%%%%%

\subsection{Soft theorems from Noether current relations}

We shall now apply these general observations to the special case where $\eps^a_2$ is a one-parametric constant shift symmetry, $\theta\to\theta+\eps$, with Noether current $J^\mu[\theta]$. Here the notation indicates that the current does not depend explicitly on the coordinate simply because the action is assumed to be translationally invariant. The second set of transformations ($\eps^\alpha_1$) is taken to be a generalized shift symmetry,
\begin{align}
    \label{genshift}
    \theta'(x)=\theta(x)+\eps_j\bigl[\alpha^j(x)+\alpha_B^j(x)\mathcal{O}^B[\theta](x)\bigr],
\end{align}
where $\eps_j$ are infinitesimal parameters, $\alpha^j(x)$ and $\alpha_B^j(x)$ are fixed polynomials in the spacetime coordinates, and $\mathcal{O}^B[\theta]$ are local composite operators constructed out of $\theta$ and its derivatives. Promoting $\eps$ and $\eps_j$ to coordinate-dependent functions, we can see that the two transformations coincide if we set
\begin{align}
    \eps(x)&=\eps_j(x)f^j[\theta,x](x), \\
    f^j[\theta,x](x)&\equiv \alpha^j(x)+\alpha_B^j(x)\mathcal{O}^B[\theta](x).
\end{align}
The off-shell condition~(\ref{osc}) can then be written in the following form,
\begin{align}
    \label{lastclas}
    J^{\mu}[\theta]\partial_{\mu}\alpha^j=-\partial_{\mu}\bigl[\alpha_B^j\mathcal{O}^B[\theta]J^{\mu}[\theta]-N^{j\mu}[\theta,x]\bigr]+\alpha_B^j\mathcal{O}^B[\theta]\partial_{\mu}J^{\mu}[\theta].
\end{align}
The generalized shift symmetry implies the existence of a set of conserved currents, $J^{j\mu}[\theta,x]$, via eq.~\eqref{currentrelations}.

As the next step, we shall lift the above classical relations to the quantum level, assuming that they remain valid for renormalized quantum operators. See ref.~\cite{Cheung2017a} for a detailed justification of this step.\footnote{Everything that follows also remains valid without further assumptions for tree-level amplitudes.} The quantum version of eq.~\eqref{lastclas} reads
\begin{equation}
    \label{start}
    \begin{split}
    \bra{\beta}J^{\mu}[\theta](x)\ket{\alpha}\partial_{\mu}\alpha^j(x)=&-\partial_{\mu}\bra{\beta}\bigl[\alpha_B^j(x)\mathcal{O}^B[\theta](x)J^{\mu}[\theta](x)-N^{j\mu}[\theta,x](x)\bigr]\ket{\alpha}\\
    &+\alpha_B^j(x)\bra{\beta}\mathcal{O}^B[\theta](x)\partial_{\mu}J^{\mu}[\theta](x)\ket{\alpha},
    \end{split}
\end{equation}
where $\ket{\alpha}$ and $\ket{\beta}$ are an arbitrarily chosen initial and final state, respectively. By combining the Ward identities for the Noether currents,
\begin{equation}
    \bra{\beta}\de_\mu J^\mu[\theta](x)\ket{\alpha}=\bra{\beta}\de_\mu J^{j\mu}[\theta,x](x)\ket{\alpha}=0,
\end{equation}
with the relation~\eqref{genconservation}, we obtain
\begin{equation}
\begin{split}
    0=\bra{\beta}\de_\mu J^{j\mu}[\theta,x](x)\ket{\alpha}&=
    \bra{\beta}\alpha^j(x)\partial_{\mu}J^{\mu}[\theta](x)+\alpha^j_B(x)\mathcal{O}^B[\theta](x)\partial_{\mu}J^{\mu}[\theta](x)\ket{\alpha}\\
    &=\alpha^j_B(x)\bra{\beta}\mathcal{O}^B[\theta](x)\partial_{\mu}J^{\mu}[\theta](x)\ket{\alpha}.
\end{split}
\end{equation}
This lets us simplify eq.~(\ref{start}) to
\begin{equation}
    \label{untrans}
    \begin{split}
    \bra{\beta}J^{\mu}[\theta](x)\ket{\alpha}\partial_{\mu}\alpha^j(x)&=-\partial_{\mu}\bra{\beta}\bigl[\alpha_B^j(x)\mathcal{O}^B[\theta](x)J^{\mu}[\theta](x)-N^{j\mu}[\theta,x](x)\bigr]\ket{\alpha}\\
    &\equiv \partial_{\mu}\bra{\beta}M^{j\mu}[\theta, x](x)\ket{\alpha},
    \end{split}
\end{equation}
where the new object $M^{j\mu}[\theta,x]$ is defined just to simplify the notation.

Next we use the standard operator relations
\begin{equation}
J^\mu[\theta](x)=e^{\im P\cdot x}J^\mu[\theta](0)e^{-\im P\cdot x},\qquad
M^{j\mu}[\theta,x](x)=e^{\im P\cdot x}M^{j\mu}[\theta,x](0)e^{-\im P\cdot x},
\end{equation}
where $P^\mu$ is the momentum operator and the latter relation reminds us that this operator only acts on the coordinate dependence of the fields, not the explicit coordinate dependence of $M^{j\mu}[\theta,x]$. Then, eq.~(\ref{untrans}) becomes
\begin{align}
    \label{start2}
    e^{-\im p\cdot x}\bra{\beta}J^{\mu}[\theta](0)\ket{\alpha}\partial_{\mu}\alpha^j(x)&=\partial_{\mu}\left[\bra{\beta}M^{j\mu}[\theta, x](0)\ket{\alpha}e^{-\im p\cdot x}\right],
\end{align}
where $p\equiv p_\alpha-p_\beta$. The matrix element on the left-hand side has a pole corresponding to a single NG boson with energy $p^0=\omega(|\vek p|)$, cf.~eq.~\eqref{polecurrent}. The same pole must appear in the matrix element of $M^{j\mu}[\theta,x]$. Combining this with the usual rules of polology, we get
\begin{align}
    \label{start3}
    \bra{\beta}M^{j\mu}[\theta, x](0)\ket{\alpha}=\frac{\im}{p^0-\omega(\abs{\boldsymbol{p}})}\bra{0}M^{j\mu}[\theta, x](0)\ket{\theta(\boldsymbol{p})}\bra{\beta+\theta(\boldsymbol{p})}\ket{\alpha}+g^j_k(x)R^{k\mu}_M(p),
\end{align}
where the remainder function $R^{j\mu}_M(p)$ is regular in the limit $p^0\rightarrow \omega(\abs{\boldsymbol{p}})$, and the functions $g^j_k(x)$ in eq.~(\ref{start3}) arise from the coefficients of operators in $M^{j\mu}[\theta,x]$.
For the special choice $\ket{\alpha}=\ket{\theta(\boldsymbol{p})}$ and $\ket{\beta}=\ket{0}$, eq.~(\ref{start2}) yields 
\begin{align}
    e^{-\im p\cdot x}\bra{0}J^{\mu}[\theta](0)\ket{\theta(\boldsymbol{p})}\partial_{\mu}\alpha^j(x)=\partial_{\mu}\left[\bra{0}M^{j\mu}[\theta, x](0)\ket{\theta(\boldsymbol{p})}e^{-\im p\cdot x}\right].
\end{align}
This in combination with eqs.~(\ref{polecurrent}) and~(\ref{start3}) in turn gives the following relation between the remainder functions,
\begin{align}
    e^{-\im p\cdot x}\partial_{\mu}\alpha^j(x)R^{\mu}(p)=\partial_{\mu}\bigl[g^j_k(x)e^{-\im p\cdot x}\bigr]R^{k\mu}_M(p).
\end{align}
Integrating this over $x$ and imposing a final assumption that $R^{j\mu}_M(p)$ is regular in the limit $\boldsymbol{p}\rightarrow\vek0$,\footnote{This assumption could be violated if $M^{j\mu}[\theta, x]$ contains terms quadratic in fields.} we obtain in the sense of distributions~\cite{Cheung2017a} 
\begin{align}
    \Tilde{\alpha}^j(p)p_{\mu}R^{\mu}(p)=0.
\end{align}
This proves the relation~\eqref{Rconstrainttext} from the main text.

%%%%%%%%%%%%%%%%%%%%%%%%%%%%%%%%%%%%%%%%%%%%%%%

\bibliography{references}
\bibliographystyle{JHEP}

\end{document}